\pdfoutput=1
\documentclass[superscriptaddress,twocolumn,aps,letter,10pt]{revtex4-1}
\usepackage{amssymb, amsmath,amsthm}    
\usepackage{graphicx}   
\usepackage{verbatim}   
\usepackage{color}      
\usepackage[labelformat=simple]{subcaption}

\usepackage[linktocpage=true]{hyperref}  
\usepackage{cancel}
\usepackage{graphicx}
\usepackage{caption}
\usepackage{tikz}
\usepackage{tcolorbox}
\usepackage{pgflibraryarrows}
\usepackage{pgflibrarysnakes}
\usepackage{yfonts}
\usepackage{multirow}
\usetikzlibrary{decorations.markings}
\usetikzlibrary{decorations.footprints}
\definecolor{green}{RGB}{35,142,35}
\usetikzlibrary{decorations.pathmorphing}
\usepackage{float}

\usepackage{ulem}

\newtcolorbox{mymathbox}[1][]{colback=white, #1}

\hypersetup{
	colorlinks   = true,
	linkcolor = blue,
	citecolor    = red
}  
\renewcommand\theequation{{\color{blue}\arabic{equation}}}
\newcommand{\dd}{\mathrm{d}}

\renewcommand\thesection{{\color{purple}\bf{\Roman{section}}}}
\renewcommand\thesubsection{\thesection-{\color{purple}($\alph{subsection}$)}\hspace{0.2em}}
\renewcommand\thesubsubsection{{\color{purple}\arabic{subsubsection}}}

\makeatletter
\def\p@subsection{}
\makeatother

\newcommand{\bqa}{\begin{eqnarray}} 
\newcommand{\eqa}{\end{eqnarray}}
\newcommand{\nn}{\nonumber \\}

\usepackage{cleveref}

\allowdisplaybreaks

\begin{document}

\title{
 Noncommutativity between the low-energy limit and integer dimension limits in the $\boldsymbol{\epsilon}$-expansion:
a case study of the antiferromagnetic quantum critical metal
}

		\author{Andr\'es  Schlief}
		\affiliation{Perimeter Institute for Theoretical Physics, 31 Caroline St. N., Waterloo ON N2L 2Y5, Canada}
		\affiliation{Department of Physics \& Astronomy, McMaster University, 1280 Main St. W., Hamilton ON L8S 4M1, Canada}
		\author{Peter Lunts}
		\affiliation{Center for Computational Quantum Physics, Flatiron Institute, 162 5th Avenue, New York, NY 10010, USA}
		\author{Sung-Sik Lee}
		\affiliation{Perimeter Institute for Theoretical Physics, 31 Caroline St. N., Waterloo ON N2L 2Y5, Canada}
		\affiliation{Department of Physics \& Astronomy, McMaster University, 1280 Main St. W., Hamilton ON L8S 4M1, Canada}
		\date{\today}
		
			\begin{abstract}

We study the field theory for the SU($N_c$) symmetric antiferromagnetic quantum critical metal 
with a one-dimensional Fermi surface embedded 
in general space dimensions between two and three.
The asymptotically exact solution valid in this dimensional range provides 
an interpolation between the perturbative solution obtained from the $\epsilon$-expansion near three dimensions
and the nonperturbative solution in two dimensions.
We show that critical exponents are smooth functions of the space dimension.
However, physical observables 
exhibit subtle crossovers that
make it hard to access subleading scaling behaviors in two dimensions
from the low-energy solution obtained above two dimensions.  
These crossovers give rise to noncommutativities, 
where the low-energy limit does not commute with 
the limits in which the physical dimensions are approached.

			\end{abstract}
		
		\maketitle

	\section{\bf {Introduction}}

Quantum critical points (QCPs) host 
exotic quantum states that do not support well-defined 
single-particle excitations\cite{SUBIRBOOK}.
Universal long-distance physics of such critical states
are often described by interacting quantum field theories
that cannot be diagonalized in any known single particle basis.
In two space dimensions, 
strong quantum fluctuations make it hard 
to extract universal low-energy data 
from interacting theories.  
In the presence of supersymmetry\cite{SEIBERG1993469} or conformal symmetry\cite{1126-6708-2008-12-031,El-Showk2014},
kinematic constraints can be strong enough 
to fix some dynamical properties.
However, nonperturbative tools are  scarce
for strongly interacting non-relativistic quantum field theories (QFTs) in general.

For this reason, it has been theoretically challenging to 
understand non-Fermi liquid metals 
that arise near itinerant QCPs 
in two dimensions\cite{
HOLSTEIN,
HERTZ,
PLEE1,
REIZER,
PLEE2,
MILLIS,
ALTSHULER,
YBKIM,
NAYAK,
POLCHINSKI2,
ABANOV1,
STEWART,
ABANOV3,
ABANOV2,
LOH,
SENTHIL,
SSLEE,
MROSS,
MAX0,
MAX2,
HARTNOLL,
ABRAHAMS,
JIANG,
FITZPATRICK,
DENNIS,
LEESTRACK,
STRACK2,
SHOUVIK2,
PATEL2,
SHOUVIK,
RIDGWAY,
HOLDER,
PATEL,
VARMA2,
EBERLEIN,
MAIER,
SCHATTNER,
SHOUVIK3,
LIU,
XU,
HONGLIU,
CHOWDHURY,
VARMA3,
BERGLEDERER}.
Couplings between particle-hole excitations
and critical order parameter fluctuations  
present at QCPs 
invalidate Landau Fermi liquid theory
that is built on the quasiparticle paradigm\cite{LANDAU}.
As a result of abundant low-energy excitations 
that amplify infrared quantum fluctuations,
even perturbative expansions 
become subtle
in the presence of small parameters.
The $1/N$-expansion,
where $N$ is the number of flavors of fermions that form Fermi surfaces,
does not give a controlled expansion in two-dimensional non-Fermi liquids\cite{SSLEE,MAX2}.
The $\epsilon$-expansions pose different types of challenges.
In the dimensional regularization scheme 
which tunes the dimension of space 
with a fixed co-dimension of the Fermi surface\cite{FITZPATRICK,CHAKRAVARTY},
it is hard to access the physics in two dimensions
from higher dimensions 
because of a spurious ultraviolet (UV)/ infrared (IR) mixing caused by the size of Fermi surface\cite{IPSITA}.
If one tunes the co-dimension of the Fermi surface, 
one usually has to go beyond the one-loop order to capture the leading order physics  correctly\cite{SHOUVIK,Note1,LUNTS}\footnotetext[1]{For an earlier implementation of dimensional regularization that tunes the co-dimension of the Fermi surface for a superconducting QCP, see Ref. \cite{SENTHIL2}.}.
Although the $\epsilon$-expansion gives a controlled expansion,
extrapolating perturbative results obtained near the upper critical dimension
to strongly coupled theories in two spatial dimensions is a highly nontrivial task. 
For a brief review of recent progress in field theories of non-Fermi liquids, see Ref. \cite{SUNGSIKREVIEW}. For recent discussions on subtle issues in the $\epsilon$-expansion for relativistic QFTs\cite{WILSON1,WILSON2}, see Refs. \cite{LORENZO3,LORENZO2,LORENZO}.

 \begin{figure}[htbp]
 	\centering
 	\includegraphics[scale=0.9]{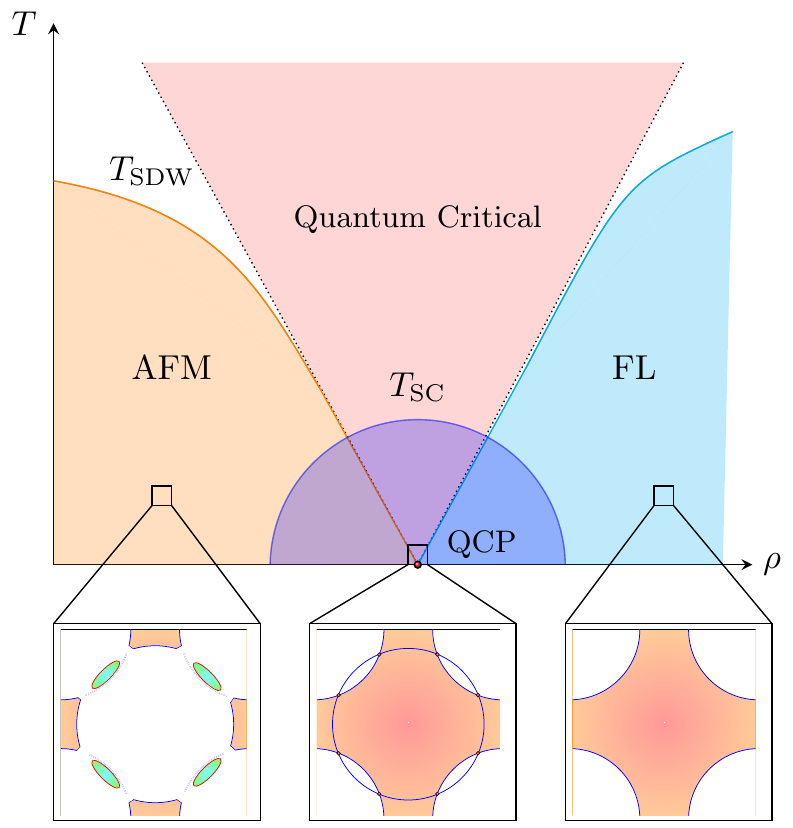}
 	\caption{
A schematic phase diagram for metals that undergo antiferromagnetic quantum phase transitions.
$T$ denotes temperature and $\rho$ denotes a tunning parameter that drives the transition 
from a paramagnetic Fermi liquid (FL) to an antiferromagnetically ordered Fermi liquid (AFM).
The dome near the critical point represents a superconducting phase.
The physics in the quantum critical region is dictated by the underlying quantum critical point (QCP).
\label{fig:PhaseDIag}}
 \end{figure}

In the past two decades, 
the non-Fermi liquids realized 
at the antiferromagnetic (AFM)  QCP
have been extensively studied 
both analytically\cite{ABANOV1,ABANOV3,ABANOV2,HARTNOLL,ABRAHAMS,LEESTRACK,VANUILDO,DECARVALHO,PATEL,PATEL2,VARMA2,MAIER,VARMA3,MAX2,SHOUVIK,SHOUVIK3} 
and numerically\cite{MAX1,LIHAI2,SCHATTNER2,GERLACH,WANG,LIHAI,WANG2}
because correlated metals 
such as electron doped cuprates\cite{HELM}, 
iron pnictides\cite{HASHIMOTO} 
and heavy fermion compounds\cite{PARK} exhibit strong AFM fluctuations.
In {\color{blue}Fig.} \ref{fig:PhaseDIag}, 
we show a schematic phase diagram 
for metals that exhibit AFM quantum phase transitions.  
Recently, the field theory that describes
the metallic AFM QCP
with the SU(2) symmetry and a $C_4$-symmetric Fermi surface has been solved 
both perturbatively near $d=3$ based on the $\epsilon$-expansion\cite{SHOUVIK,LUNTS}
and nonperturbatively in $d=2$\cite{SCHLIEF},
where $d$ is the space dimension.
The availability of both the perturbative solution valid near the upper critical dimension
and the nonperturbative  solution for the two-dimensional theory
provides a rare opportunity to test 
the extent to which the $\epsilon$-expansion is applicable
to strongly coupled theories in which $\epsilon \sim 1$.

In this paper, 
we test the dimensional regularization scheme (and the $\epsilon$-expansion) as a methodology
using the field theory for AFM quantum critical metals as a model theory.
We solve the theory
in general dimensions between two and three to understand
how the perturbative solution obtained from the $\epsilon$-expansion near the upper critical dimension
evolves as nonperturbative effects become stronger with decreasing dimension.
From this we expose both strengths and weaknesses of the dimensional regularization scheme.
On the one hand, the exact critical exponents are smooth functions of the space dimension,
and the $\epsilon$-expansion can provide a useful ansatz 
for the exact exponents in two dimensions.
On the other hand, it is difficult to capture full scaling behaviours in two dimensions
from the low-energy solution obtained above two dimensions
because the low-energy limit and the $d \rightarrow 2$ limit do not commute.

This paper is organized as follows. 
In Sec. \ref{sec:MODEL} we review the field theory
that describes AFM quantum critical metals in space dimensions 
between two and three\cite{SHOUVIK,LUNTS}.
In Sec. \ref{sec:COMMUTATIVITY}, 
we begin by summarizing the scaling forms of the low-energy Green's functions.
{\color{blue}Table} \ref{tb:Interpol}
encapsulates the main result of this paper:
{\it 
physical observables exhibit noncommutativities
in the sense that the low-energy limit and 
the limit in which physical dimensions are approached 
do not commute. }
After the summary, we provide details that lead to such scaling forms.
We first review the one-loop solution valid in $d=3$,
and discuss how the solution fails to capture the low-energy physics
in $d=3-\epsilon$ for any nonzero $\epsilon$.
This is caused by a noncommutativity
between the low-energy limit and the $d \rightarrow 3$ limit.
We then move on to the general solution valid in any $2 < d < 3$,  which shows how nonperturbative effects become important as the space dimension is lowered.
Finally, we compare this solution 
with the nonperturbative solution obtained at $d=2$.
While critical exponents vary smoothly in $d$,
the full low-energy Green's functions in $d=2$
\textit{cannot} be obtained by taking the $d \rightarrow 2$ limit
of the low-energy Green's function obtained in $d>2$
due to a noncommutativity between the $d\rightarrow 2$ limit and the low-energy limit.
We finish this paper by making some concluding remarks in Sec. \ref{sec:summary}.

	\section{\bf{Field Theory in $\boldsymbol{2\leq \lowercase{d} \leq 3}$}}\label{sec:MODEL}
	\begin{figure*}
	\centering
	\begin{subfigure}[b]{0.45\linewidth}
		\centering
		\includegraphics[scale=0.9]{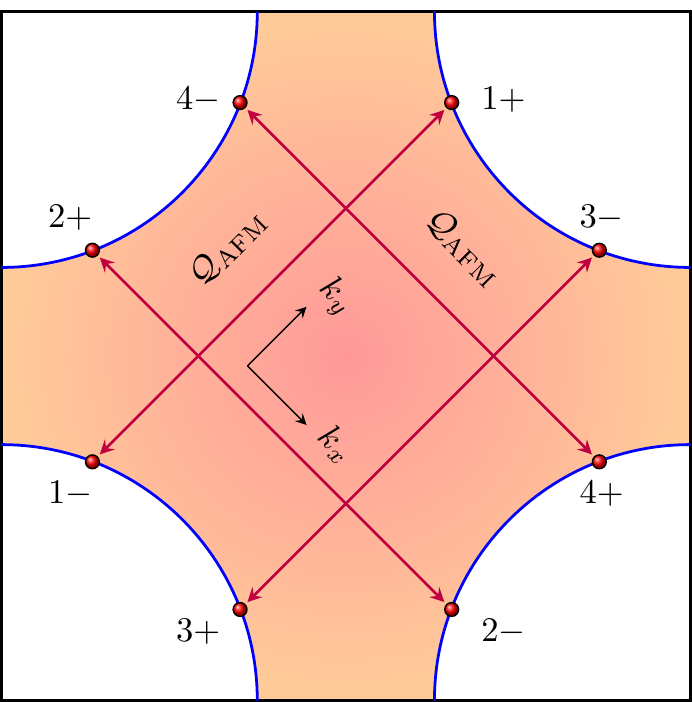}
		\caption{\hspace{0.5em} Setup in $d=2$}\label{fig:2dSetup}
	\end{subfigure}
	\begin{subfigure}[b]{0.45\linewidth}
		\centering
		\includegraphics[scale=0.6,trim={2cm 1.8cm 0 0},clip]{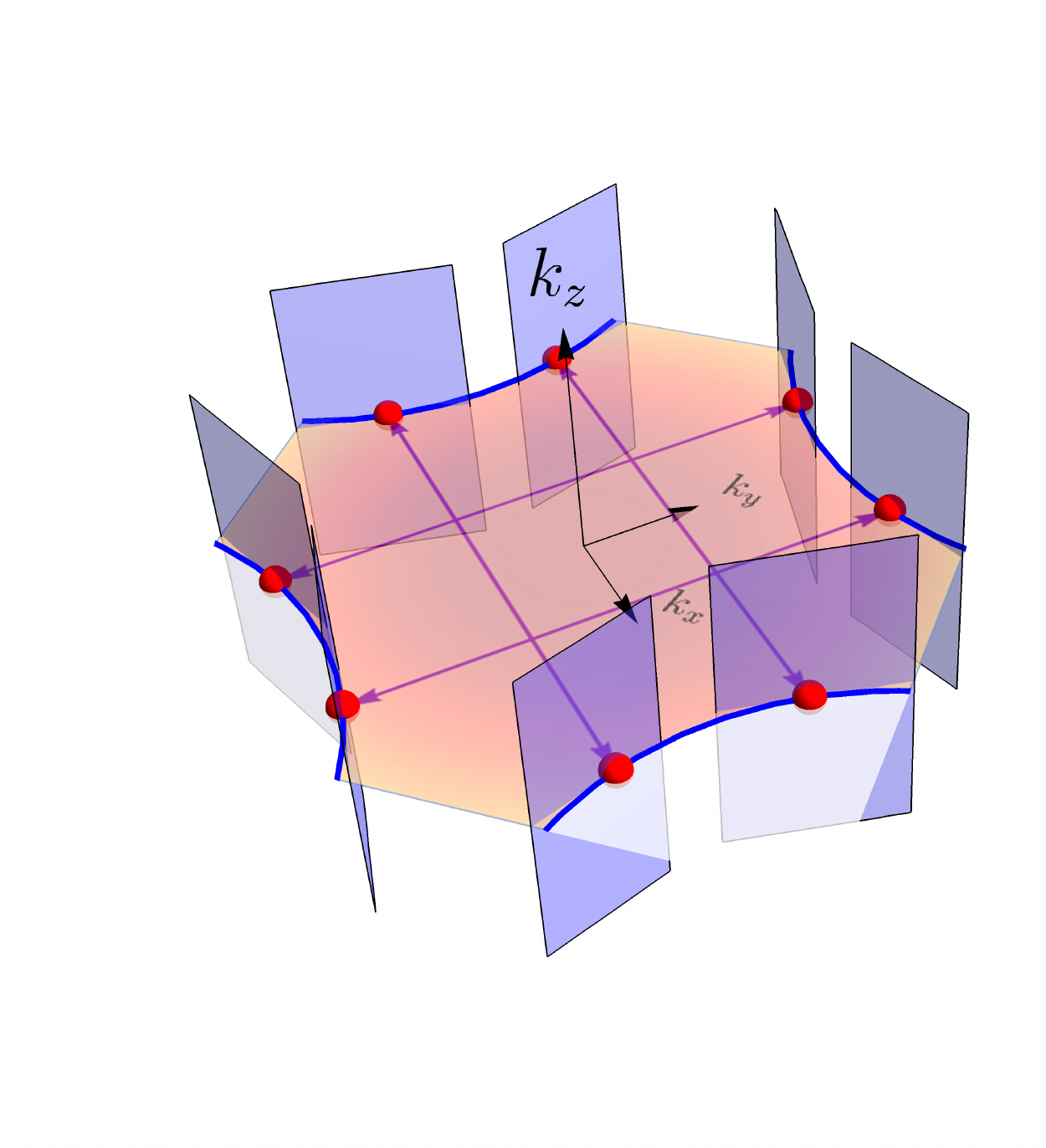}
		\caption{ \hspace{0.5em} Setup in $d=3$}\label{fig:3dSetup}
	\end{subfigure}
	\caption{({\color{blue}$a$}) 
A Fermi surface with $C_4$ symmetry in two dimensions. 
The (red) dots represent the hot spots connected by the AFM wave vector 
which is chosen to be $\mathcal{Q}_{\mathrm{AFM}}=\pm\sqrt{2}\pi\hat{k}_x$ or $\mathcal{Q}_{\mathrm{AFM}}=\pm\sqrt{2}\pi \hat{k}_y$ 
up to reciprocal lattice vectors $\sqrt{2}\pi(\hat{k}_x\pm\hat{k}_y)$. 
({\color{blue}$b$}) One-dimensional Fermi surface embedded in a three-dimensional momentum space. 
The (blue) planes correspond to locally flat patches
that include line nodes (blue lines) near the hot spots.
}
\end{figure*}

The minimal theory for the SU(2) symmetric AFM quantum critical metal in two dimensions is written as\cite{ABANOV1,MAX2,SHOUVIK,ABANOV3}
\begin{widetext}
		\begin{align}\label{eq:ActionTwoD}
		S_{d=2} &= \sum^{4}_{n=1}\sum_{\sigma=\uparrow, \downarrow}
\int\dd  k\overline{\Psi}_{n,\sigma}(k)\left(i\gamma_0k_0+i\gamma_1\varepsilon_{n}(\vec{k};v)\right)\Psi_{n,\sigma}(k)
+\frac{1}{4}\int\dd q\left(q^2_0+c^2|\vec{q}|^2\right)\mathrm{Tr}\left[\Phi(-q)\Phi(q)\right]\notag\\
		&+ ig \sum^{4}_{n=1}\sum_{\sigma,\sigma'=\uparrow, \downarrow} 
\int\dd k\int\dd q\overline{\Psi}_{\overline{n},\sigma}(k+q){\Phi}_{\sigma\sigma'}(q)\gamma_{1}\Psi_{n,\sigma'}(k)\\
		&+\frac{u}{4}\left[\prod^{3}_{i=1}\int\dd q_i\right]
\mathrm{Tr}\left[\Phi(q_1+q_3)\Phi(q_2-q_3)\right]\mathrm{Tr}\left[\Phi(-q_1)\Phi(-q_2)\right].\notag
		\end{align}
		\end{widetext}
Here, $k=(k_0,\vec{k})$ with $k_0$ denoting fermionic Matsubara frequency 
and $\vec{k}=(k_x,k_y)$, the two-dimensional  momentum  measured from each of the eight hot spots (points on the $C_4$-symmetric Fermi surface connected by the commensurate wave vector $\mathcal{Q}_{\mathrm{AFM}}$), 
as shown in {\color{blue}Fig.} \ref{fig:2dSetup}. 
We use a simplified notation, $\dd k=\frac{\dd k_0}{(2\pi)}\frac{\dd\vec{k}}{(2\pi)^2}$, for the integration measure.  
The four two-component spinors are given by
$\Psi_{1,\sigma}(k) = (\psi^{(+)}_{1,\sigma}(k),\psi^{(+)}_{3,\sigma}(k))^{\mathrm{T}}$,  
$\Psi_{2,\sigma}(k)=(\psi^{(+)}_{2,\sigma}(k),\psi^{(+)}_{4,\sigma})^{\mathrm{T}}$, 
$\Psi_{3,\sigma}(k)=(\psi^{(-)}_{1,\sigma}(k),-\psi^{(-)}_{3,\sigma}(k))^{\mathrm{T}}$ and 
$\Psi_{4,\sigma}(k)=(\psi^{(-)}_{2,\sigma}(k),-\psi^{(-)}_{4,\sigma}(k))^{\mathrm{T}}$, 
where $\psi^{(m)}_{n,\sigma}(k)$ is the Grassmanian field 
representing  electrons  near the hot spots labeled by $n=1,2,3,4$ and $m=\pm$, and  with spin $\sigma=\uparrow,\downarrow$.
$\gamma_0=\sigma_y$ and $\gamma_1=\sigma_x$ are the $2\times 2$ Pauli matrices,
and $\overline{\Psi}_{n,\sigma}(k) = \Psi^\dagger_{n,\sigma}(k)\gamma_0$.
 The energy dispersion relations of the fermions are given by 
 $\varepsilon_{1}(\vec{k};v) = vk_x+k_y$, 
$\varepsilon_{2}(\vec{k};v)= -k_x+vk_y$, 
$\varepsilon_{3}(\vec{k};v)=vk_x-k_y$ and 
$\varepsilon_{4}(\vec{k};v) = k_x+v k_y$, 
where $v$ measures the component of the Fermi velocity perpendicular to $\mathcal{Q}_{\mathrm{AFM}}$. 
The component of the Fermi velocity parallel to $\mathcal{Q}_{\mathrm{AFM}}$
is set to one. 
$q=(q_0,\vec{q})$ denotes the bosonic Matsubara frequency and two-dimensional momentum $\vec{q}=(q_x,q_y)$ 
measured relative to $\mathcal{Q}_{\mathrm{AFM}}$. 
The bosonic matrix field representing the collective spin fluctuations 
is written in the defining representation of SU(2):
 $\Phi(q) = \sum^{3}_{a=1}\phi^a(q)\tau^a$, where $\tau^a$ denotes the three generators of SU(2). 
 We choose the normalization of the generators as $\mathrm{Tr}[\tau^a\tau^b]=2\delta^{ab}$. 
 $\Phi(q)$ carries momentum $\vec{q}+\mathcal{Q}_{\mathrm{AFM}}$, and
$c$ denotes the velocity of the AFM spin fluctuations. 
The coupling between the collective mode and the fermions is denoted by $g$.
$\overline{n}$ denotes the hot spot connected to $n$ via 
$\mathcal{Q}_{\mathrm{AFM}}$,
that is, $\overline{1}=3,\overline{2}=4,\overline{3}=1$ and $\overline{4}=2$. 
Finally, $u$ sources the quartic interaction between the collective modes.

Now we write down the theory defined in $2\leq d\leq 3$, where $d$ is the space dimension\cite{SHOUVIK,DENNIS,LUNTS,SHOUVIK3}.
Here, the co-dimension of the Fermi surface is tuned while keeping its dimension fixed to be one.
This choice of dimensional regularization scheme maintains locality in real space, and avoids the UV/IR mixing 
that arises through couplings between different patches of the Fermi surface when its dimension is greater than one\cite{IPSITA}. The theory in ${2\leq d\leq 3}$ is written as
	  \begin{widetext}
	 \begin{align}\label{eq:ActionGeneralD}
	 	 	S_{d} &= \sum^{4}_{n=1}\sum^{N_c}_{\sigma=1}\sum^{N_f}_{j=1}\int \dd k\overline{\Psi}_{n,\sigma,j}(k)\left(i\boldsymbol{\Gamma}\cdot\mathbf{K}+i\gamma_{d-1}\varepsilon_{n}(\vec{k};v)\right)\Psi_{n,\sigma,j}(k)+\frac{1}{4}\int\dd q\left(|\mathbf{Q}|^2+c^2|\vec{q}|^2\right)\mathrm{Tr}\left[\Phi(-q)\Phi(q)\right]\notag\\
	 		&+\frac{ig}{\sqrt{N_f}}\sum^{4}_{n=1}\sum^{N_c}_{\sigma,\sigma'=1}\sum^{N_f}_{j=1}\int \dd k\int \dd q\overline{\Psi}_{\overline{n},\sigma,j}(k+q){\Phi}_{\sigma\sigma'}(q)\gamma_{d-1}\Psi_{n,\sigma',j}(k)\\
	 		&+\frac{1}{4}\left[\prod^{3}_{i=1}\int\dd q_i\right]\left[u_1\mathrm{Tr}\left[\Phi(q_1+q_3)\Phi(q_2-q_3)\right]\mathrm{Tr}\left[\Phi(-q_1)\Phi(-q_2)\right]+u_2\mathrm{Tr}\left[\Phi(q_1+q_3)\Phi(q_2-q_3)\Phi(-q_1)\Phi(-q_2)\right]\right].\notag
	 	 \end{align}
	 	\end{widetext}
Here, $k=(\mathbf{K},\vec{k})$, where
$\mathbf{K}$ denotes the $(d-1)$-dimensional vector 
composed of the Matsubara frequency and $(d-2)$ momentum components
that represent the extra space dimensions and $\vec{k}=(k_x,k_y)$.
The integration measure is denoted as $\dd k=\frac{\dd\mathbf{K}}{(2\pi)^{d-1}}\frac{\dd\vec{k}}{(2\pi)^2}$.
In $2 \leq d \leq 3$, the number of spinor components is fixed to be two.
$(\boldsymbol{\Gamma},\gamma_{d-1})=(\gamma_0,\gamma_1,\cdots,\gamma_{d-2},\gamma_{d-1})$ 
denotes $2\times 2$ gamma matrices that satisfy the Clifford algebra: 
$\{\gamma_\mu,\gamma_\nu\}=2\delta_{\mu\nu}\mathbb{I}_{2\times 2}$ 
with $\mathbb{I}_{2\times 2}$ being the identity matrix. The fermionic kinetic term in Eq. (\ref{eq:ActionGeneralD}) describes a metal with a one-dimensional Fermi surface embedded in a $d$-dimensional momentum space.
We choose $\gamma_0=\sigma_y$ and $\gamma_{d-1}=\sigma_x$ without loss of generality. 
For completeness, the fermion flavor is promoted to $j=1,2,\dots,N_f$.
We also generalize the SU(2) spin group to SU($N_c$) such that ${\sigma=1,2,\dots, N_{c}}$.
Accordingly, the boson field is written as
 $\Phi(q) = \sum^{N^2_c-1}_{a=1}\phi^a(q)\tau^a$, 
where $\tau^a$ denotes the $(N^2_c-1)$ generators of  SU($N_c$) subject to the normalization ${\mathrm{Tr}[\tau^a\tau^b]=2\delta^{ab}}$. $u_1$ and $u_2$ source two possible quartic interactions
which are independent from each other for $N_c\geq 4$. In what follows, we consider the theory for general $N_c\geq 2$ and $N_{f}\geq 1$. However, the validity of the solution presented in this paper does not rely on $N_{c}$ or $N_{f}$ being large.
Finally, we note that the field theory in $2\leq d\leq 3$ inherits the underlying $C_4$ symmetry of the Fermi surface.

	\renewcommand\thesubsection{{\color{purple}($\alph{subsection}$)}\hspace{0.2em}}
\renewcommand\thesubsubsection{{\color{purple}\arabic{subsubsection}}}


\section{\bf{Noncommutativity between the Low-Energy and the Physical Dimension Limits}}\label{sec:COMMUTATIVITY}

\begin{table*}
		\centering
		\begin{tabular}{cccc}
			\hline
			\hline
			 & $d=2$& $2<d<3$ & $d=3$\\
			\hline
			\hline
			&&&\\[-1.5ex]
			$F_{z}(\mu)$&$ \exp\left(2\sqrt{N^2_c-1}\dfrac{\ell^\frac{1}{2}}{\log(\ell)}\right)$&$\exp\left((d-2)\mathfrak{F}_z(d)(N^2_c-1)^\frac{1}{d}\ell^\frac{1}{d}\right)$&$\ell^{\dfrac{(N^2_c+N_cN_f-1)}{2(N^2_c+N_cN_f-3)}}$\\[3ex]
			$F_{\Psi}(\mu)$&$\displaystyle\ell^{\frac{3}{8}} $&$\displaystyle\sqrt{\ell}$& $\displaystyle\sqrt{\log(\ell)}$\\[3ex]
			$F_{\Phi}(\mu)$&$\displaystyle  \exp\left(\frac{2\ell^\frac{1}{2}}{\sqrt{N^2_c-1}}\right)$& $\displaystyle\exp\left(-\frac{\mathfrak{F}_\Phi(d)}{2}\frac{((d-2)N^2_c-d)}{(N^2_c-1)^\frac{d-1}{d}}\ell^\frac{1}{d}\right)$&$\displaystyle \log(\ell)$\\[4ex]
			$v(\mu)$&$\displaystyle\frac{\pi^2 N_c N_f }{2(N^2_c-1)\ell\log(\ell)}$&$\displaystyle \frac{\pi N_c N_f(d-2)}{4(N^2_c-1)\zeta(d)(d-1)\ell}$&$\displaystyle \frac{\pi N_cN_f (N^2_c-N_cN_f-3)}{4(N^2_c-1) (N^2_c+N_cN_f-1)\log(\ell)}$\\[3ex]
			$c(\mu)$&$\displaystyle\frac{\pi}{4\sqrt{N^2_c-1}}\frac{1}{\sqrt{\ell}}$&$\displaystyle \left[\frac{\pi \beta^4_{d}\mathfrak{B}(d)}{(3-d)(d-1)\zeta(d)(N^2_c-1)}\right]^\frac{1}{d}\frac{1}{\ell^\frac{1}{d}}$&$\displaystyle \frac{\pi (N^2_c-N_cN_f-3)}{4(N^2_c-1+N_cN_f)}\frac{1}{\log(\ell)}$\\[4ex]
			\hline
			\hline
		\end{tabular}
		\caption{
Scale-dependent universal crossover functions and renormalized velocities in the low-energy limit for each fixed $d$.
Here $\ell\equiv\log(\Lambda/\mu)$ is a logarithmic length scale
associated with a running energy scale $\mu$,
and a UV cutoff $\Lambda$.
$\beta_{d},\zeta(d),\mathfrak{F}_z(d),\mathfrak{F}_\Phi(d)$ and $\mathfrak{B}(d)$ are 
smooth and positive functions defined in Eqs.  (\ref{eq:Betad}), (\ref{eq:zetad}), (\ref{eq:Fzz}), (\ref{eq:Fphi}), and (\ref{eq:DefinitionofB}) respectively. It is noted that
$\beta_2 = \sqrt{\pi/2},\zeta(2)=(2\pi)^{-1},\mathfrak{F}_{z}(2) =\sqrt{2},\mathfrak{F}_{\Phi}(2)=2\sqrt{2}$ and $\mathfrak{B}(2) = (4\pi^2)^{-1}$ in $d=2$,
and 
$\beta_{3-\epsilon}=\sqrt{4\pi\epsilon}$, $\zeta(3-\epsilon)=\epsilon/2$, $\mathfrak{F}_z(3)=3/(2^{14}h^{*}_5)^\frac{1}{3}$, $\mathfrak{F}_{\Phi}(3)=3/(2^{8}h^{*}_5)^\frac{1}{3}$ and $\mathfrak{B}(3)=2 h^{*}_5$ with $h^{*}_{5} \approx 5.7 \times 10^{-4}$\cite{LUNTS} in $d=3-\epsilon$ to leading order in $\epsilon\ll 1$.
}
\label{tb:Interpol}
\end{table*}

In this section, we first summarize the main results of the paper without derivation.
The scaling form of the Green's functions in $2\leq d\leq 3$ is given by
\begin{widetext}
\begin{align}
	G_{1}(k;d) &= \frac{1}{iF_\Psi(|\mathbf{K}|)}
\frac{1}{F_z(|\mathbf{K}|)\boldsymbol{\Gamma}\cdot\mathbf{K}+\gamma_{d-1}(v(|\mathbf{K}|)k_x+k_y)},\label{eq:GeneralFermionGreensFunction}\\
	D(k;d) &=\frac{1}{F_{\Phi}(|\mathbf{K}|)}\frac{1}{F_z(|\mathbf{K}|)^{d-1}|\mathbf{K}|^{d-1}+c(|\mathbf{K}|)^{d-1}(|k_x|^{d-1}+|k_y|^{d-1})},
	\label{eq:GeneralBosonGreensFunction}
\end{align}
\end{widetext}
for $\frac{\vec{k}}{|\mathbf{K}|F_z(|\mathbf{K}|)}\sim 1$. $G_1(k;d)$ denotes the fermion Green's function at hot spot $n=1$
in $d$ space dimensions.
The Green's functions at other hot spots are related to $G_1(k;d)$ through the $C_4$ symmetry of the theory. 
$D(k;d)$ is the Green's function for the AFM collective mode.
 Here, the condition $\vec{k}/(|\mathbf{K}|F_z(|\mathbf{K}|))\sim 1$
is chosen so that the forms of the Green's functions are invariant 
(up to the weak scale dependence of the velocities)
under scale transformations 
in which momentum and frequency are simultaneously taken to zero.
If the dynamical critical exponent was fixed,
the scale invariance of the Green's function would be manifest
under the rescaling
in which $\vec k/|{\bf K}|^{1/z}$ is fixed,
where $z$ is the dynamical critical exponent.
In the present case, the dynamical critical exponent depends  weakly on the scale,
and it flows to $z=1$ in the low-energy limit, 
as will be shown later.
At finite energy scales, the Green's functions are invariant under the scale transformation
in which $\vec{k}/(|\mathbf{K}|F_z(|\mathbf{K}|))$ is fixed,
where $F_z(|{\bf K}|)$ is a function 
that encodes the scale dependence of the dynamical critical exponent.
The leading power-law dependences of the Green's functions in energy and momentum reflect
the dynamical critical exponent $(z=1)$,
and the scaling dimensions of the fermion $\Bigl( [\Psi(k)]=-(d+2)/2 \Bigr)$ 
and the collective mode $\Bigl( [\Phi(k)]=-d \Bigr)$  at the fixed point.
%
The full Green's functions deviate from the perfect power-law behaviors 
due to a scale dependence of marginally irrelevant operators.
In $d<3$, the ratio between velocities, 
\begin{equation}
w(\mu) \equiv v(\mu) /c (\mu)
\end{equation}
controls quantum corrections,
where  $v(\mu)$ and $c(\mu)$ are the renormalized velocities that depend on the energy scale $\mu$.
As will be shown later, a slow flow of $w(\mu)$ generates super-logarithmic corrections 
captured by $F_z(\mu), F_\Psi(\mu)$ and $F_\Phi(\mu)$,
that is, corrections that are smaller than a power-law
but larger than any fixed power of a logarithm in energy.
$F_\Psi(\mu)$ ($F_\Phi(\mu)$) represents 
the correction to the scaling dimension of the fermion (boson) field. 
In $d=3$,  quantum corrections are controlled by $g^2/v$, 
which yield logarithmic corrections to the power-law scalings.
The scale dependences of $v(\mu), c(\mu), F_z(\mu), F_\Psi(\mu)$ and $F_\Phi(\mu)$
in each dimension
are summarized in {\color{blue}Table} \ref{tb:Interpol}. 

Although the critical exponents that characterize the fixed point are smooth functions of $d$,
$v(\mu), c(\mu), F_z(\mu), F_\Psi(\mu)$ and $F_\Phi(\mu)$,
evaluated in the small $\mu$ limit,
are not,
as is shown in {\color{blue}Table} \ref{tb:Interpol}. 
This leads to discontinuities of $\lim_{k\rightarrow 0} G_{1}(k;d)$ and $\lim_{k\rightarrow 0} D(k;d)$
as functions of $d$.
The discontinuities are caused by a lack of commutativity 
between the low-energy limit and the limits in which $d$ approaches the physical dimensions,
\bqa
\lim_{k \rightarrow 0} \lim_{d \rightarrow 2,3} G_1(k;d) \neq
\lim_{d \rightarrow 2,3} \lim_{k \rightarrow 0} G_1(k;d),
\label{NCM}
\eqa
and similarly for $D(k;d)$. Since the Green's functions diverge at $k=0$,
Eq. (\ref{NCM}) makes sense only if
the small $k$ limit is viewed 
as the asymptotic limit of the Green's functions.
In other words, 
$\lim_{k\rightarrow 0}G_1(k;d)$ 
should be understood as the asymptote of $G_1(k;d)$ in the small $k$ limit, 
that is, the
$k$-dependent function that 
$G_1(k;d)$ asymptotically approaches
in the small $k$ limit at a fixed $d$
rather than $G_1(0;d)$.
With this, Eq. (\ref{NCM}) implies that
the low-energy asymptote of 
$G_1(k;d)$ at $d=2,3$
can not be reproduced
by taking the $d \rightarrow 2,3$ limits
of the low-energy asymptotes of $G_1(k;d)$ 
obtained in $2<d<3$.

\begin{figure}
	\centering
	\includegraphics[scale=0.95]{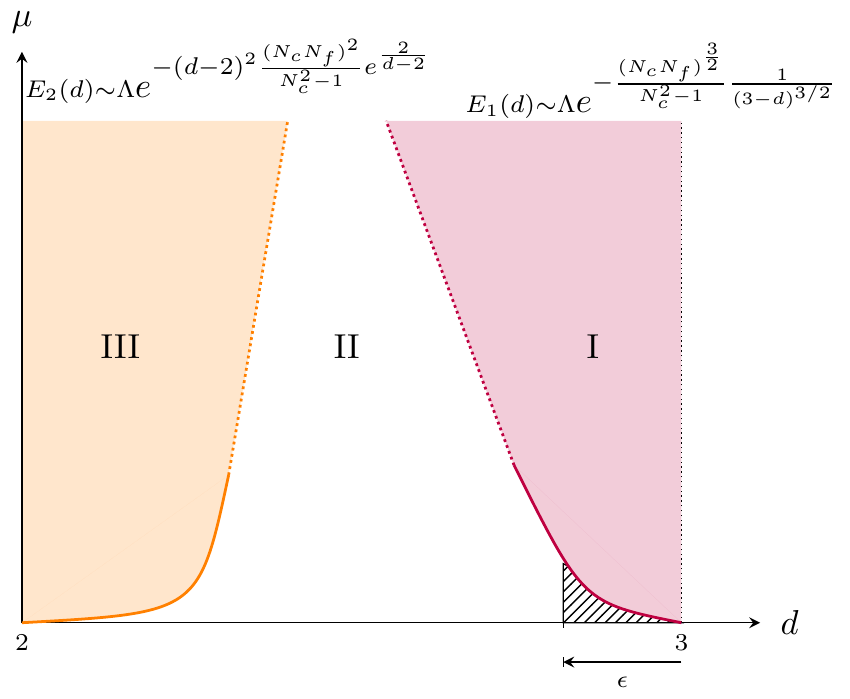}	
	\caption{
		Two crossover energy scales that divide the plane of spatial dimension ($d$) and energy scale ($\mu$) into three regions. At low energies,
		$w(\mu)$ flows to an order one number in region I, 
		while it flows to zero in regions II and III.
		Region III  is distinguished from region II by the fact that 
		physical observables receive additional logarithmic corrections.
		\label{fig:DimPlots}
	}
\end{figure}

The expressions in {\color{blue}Table} \ref{tb:Interpol}
are obtained by taking the low-energy limit at a fixed dimension.
Because of the noncommutativity in Eq. (\ref{NCM}),
$\lim_{k \rightarrow 0} G_1(k;d)$ is not  a continuous function of $d$ at $d=2$ and $d=3$. The noncommutativity arises because of 
the existence of crossover energy scales
that vanish in the $d \rightarrow 2, 3$ limits.
In the plane of spatial dimension and energy scale,
there are three distinct regions 
divided by these crossover energy scales as is shown in {\color{blue}Fig.} \ref{fig:DimPlots}.
The first crossover energy scale is given by $E_1(d) \sim \Lambda e^{-\frac{(N_cN_f)^\frac{3}{2}}{(N^2_c-1)}(3-d)^{-\frac{3}{2}}}$ 
which vanishes exponentially as $d$ approaches three,
where $\Lambda$ is a UV energy scale.
The second scale, $E_2(d) \sim \Lambda  e^{-(d-2)^2\frac{(N_cN_f)^2}{(N^2_c-1)}e^{\frac{2}{d-2}}}$ vanishes
in a doubly exponential fashion as $d$ approaches two.
The three regions divided by $E_1(d)$ and $E_2(d)$ 
are governed by different physics.

In region I of {\color{blue}Fig.} \ref{fig:DimPlots} (${\mu > E_1(d)}$), 
the low-energy physics is described at the one-loop order by a quasi-local marginal Fermi liquid,
where $v$ and $c$ flow to zero as $1/\log (\log(\Lambda/\mu))$  with ${w \sim \mathcal{O}(1)}$\cite{SHOUVIK}. 
Because the velocities flow to zero, the magnitude of higher-loop diagrams is not only determined by the number of vertices, 
but also by enhancement factors of $1/v$ and $1/c$ 
that originate from the fact that modes become dispersionless at low energies. 
In particular,  the one-loop fixed point is controlled only when $g^2$ flows to zero faster than $v$ and $c$.
In $d=3$, the one-loop results become asymptotically exact at low energies
because $\lambda\equiv g^2/v$ flows to zero much faster than any power of the velocities.
While the quasi-local marginal Fermi liquid behavior persists
down to the zero energy limit in $d=3$,
the low-energy physics becomes qualitatively different below three dimensions.
In $d=3-\epsilon$ with $\epsilon>0$, $\lambda$ becomes order of ${\epsilon}$, while $v$ and $c$ still flow to zero logarithmically at the one-loop order.
Due to the enhanced quantum fluctuations associated with the vanishing velocities and non-vanishing $\lambda$,
higher-loop effects become qualitatively important at energies below the crossover energy scale $E_1(d)$\cite{LUNTS,SHOUVIK}.
For any nonzero $\epsilon<1$, the theory flows into a new region (region II)
in which leading order quantum fluctuations are no longer contained within the one-loop order.
The noncommutativity
between the $d \rightarrow 3$ and $\mu \rightarrow 0$ limits 
arises because $E_1(d)$ vanishes as $d \rightarrow 3$.

It turns out that it is sufficient to include a two-loop quantum correction 
in addition to the one-loop quantum corrections
to the leading order in ${\epsilon\ll 1}$ 
because all other higher-loop corrections are suppressed by $\epsilon$ 
in the shaded area of region II shown in {\color{blue}Fig.} \ref{fig:DimPlots}\cite{SHOUVIK3,LUNTS}.
The physics below $E_1(d)$ is qualitatively different from that of region I.
In particular, $w$ flows to zero in the low-energy limit in ${d=3-\epsilon}$
due to the two-loop effect that modifies the flow of the velocities. 
The fact that quantum corrections are not organized by the number of loops
even close to the upper critical dimension 
is a feature caused by the emergent quasi-locality where
velocities flow to zero in the low-energy limit.

As $d$ decreases further away from three, 
an infinite set of diagrams, which are suppressed by higher powers of $\epsilon$ near three dimensions, becomes important.
Although it is usually hopeless to include all higher-order quantum corrections, 
in the present case one can use $w$ as a control parameter 
since $w$ dynamically flows to zero in the low-energy limit.
In the small $w$ limit, 
only the diagrams in {\color{blue}Figs.} \ref{fig:SDDiagramA}, \ref{fig:SDDiagramB} and \ref{fig:Diagrams} 
remain important even when $\epsilon \sim 1$\cite{LUNTS}. In there, the double wiggly line represents the renormalized boson propagator which is self-consistently dressed with the diagrams in {\color{blue}Figs.} \ref{fig:SDDiagramA} and \ref{fig:SDDiagramB}.
The propagator of the collective mode becomes 
$D(q) = \frac{1}{|{\bf Q}|^{d-1} + c(v)^{d-1} ( |q_x|^{d-1} + |q_y|^{d-1} )}$,
where $c(v)$ is the velocity of the incoherent collective mode given by
$c(v)^{d} \sim \frac{v}{d-2}$.

The behavior in region II 
does not extend smoothly to $d=2$ 
because of another crossover 
set by an energy scale $E_2(d)$
that vanishes in the $d \rightarrow 2$ limit.
The existence of the crossover is expected from the fact that 
the relation, $c(v)^{d} \sim \frac{v}{d-2}$
valid in region II becomes ill-defined in $d=2$.
The UV divergence in the $d \rightarrow 2$ limit
is caused by the incoherent nature of the AFM collective mode
which has significant low-energy spectral weight 
even at large momenta. 
At $d=2$, the divergence gives rise to a logarithmic enhancement
of $c(v)$ as $c(v)^{2} \sim v \log (1/w(v))$.
The extra logarithmic correction 
causes the additional set of diagrams in {\color{blue}Fig}. \ref{fig:Diagrams2}
to become important in region III.
This gives rise to a lack of commutativity between
the $d \rightarrow 2$ limit and the low-energy limit.

In what follows, 
we elaborate on the points summarized in this section 
starting from $d=3$.
The subsections \ref{sec:threeD} and \ref{sec:twoD} are mostly summaries of
Refs. \cite{SHOUVIK,LUNTS} and \cite{SCHLIEF}
for regions I and III, respectively.
The subsection \ref{sec:ansatz} is devoted to region II, 
which is the main new result of the present paper.

\begin{figure*}
	\centering
	\begin{subfigure}[b]{0.31\linewidth}
		\includegraphics[scale=1]{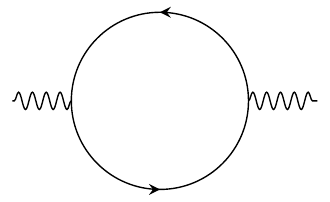}
		\caption{\label{fig:1LBSE}\label{fig:oneloop1}}
	\end{subfigure}
	\begin{subfigure}[b]{0.31\linewidth}
		\includegraphics[scale=0.8,trim={0 -.5cm 0 0}]{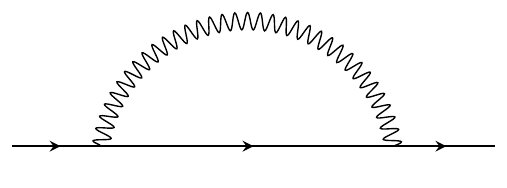}
		\caption{\label{fig:1LFSE}\label{fig:oneloop2}}
	\end{subfigure}
	\begin{subfigure}[b]{0.31\linewidth}
		\includegraphics[scale=1]{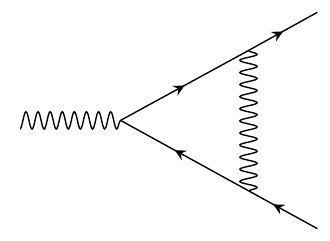}
		\caption{\label{fig:oneloop3}}
	\end{subfigure}
	\begin{subfigure}[b]{0.31\linewidth}
		\includegraphics[scale=1]{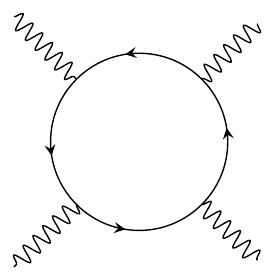}
		\caption{\label{fig:oneloop4}}
	\end{subfigure}
	\begin{subfigure}[b]{0.31\linewidth}
		\includegraphics[scale=0.65,trim={0 -.5cm 0 0}]{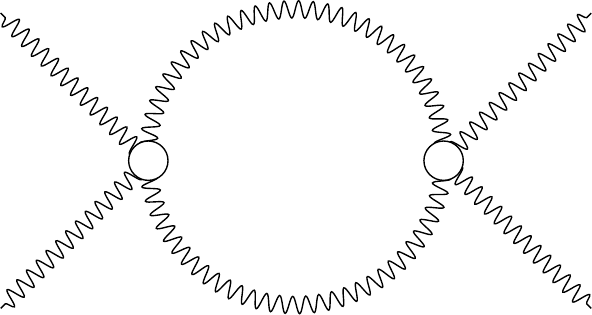}
		\caption{\label{fig:1L4P2}\label{fig:oneloop5}}
	\end{subfigure}
	\begin{subfigure}[b]{0.31\linewidth}
		\includegraphics[scale=1,trim={0 -0.5cm 0 0}]{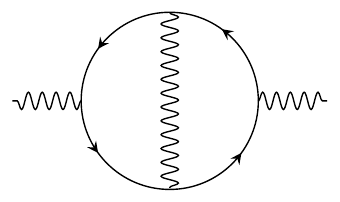}
		\caption{\label{fig:2LBSE}}
	\end{subfigure}
	\caption{Quantum corrections at the modified one-loop order. \label{fig:DiagramsM1L}}
\end{figure*} 

\subsection{Region I : from $\boldsymbol{d=3}$ to $\boldsymbol{d=3-\epsilon}$ }\label{sec:threeD}

In three dimensions, the Yukawa coupling is marginal under the Gaussian scaling.
The one-loop quantum corrections shown in {\color{blue}Figs.} \ref{fig:oneloop1} to \ref{fig:oneloop5} drive
all parameters of the theory ($g,v,c,u_i$) 
to flow to zero in such a way that the ratios defined by $\lambda=g^2/v$, $w=v/c$ and $\kappa_i=u_i/c^2$ 
become\cite{SHOUVIK}
\begin{align}
\lambda^{*}=0,\quad c^{*}=0, \quad w^{*}=\frac{N_cN_f}{N^2_c-1}, \quad \& \quad \kappa^{*}_{i}=0
\label{1loopft}
\end{align}
in the low-energy limit.
As is shown in {\color{blue}Table} \ref{tb:Interpol}, 
the velocities flow to zero as
$v(\ell), c(\ell) \sim 1/\log(\ell)$ 
in the logarithmic length scale $\ell\equiv\log(\Lambda/\mu)$,
while the rescaled coupling flows to zero as $\lambda(\ell)\sim 1/\ell$.
Because $\lambda$ flows to zero faster than both $v$ and $c$, 
the ratios $g^{2n}/c^m$ and $g^{2n}/v^m$,
which control the perturbative expansion, flow to zero for any $n,m>1$. 
This implies that all higher-order corrections are suppressed at low energies. 
The physical observables receive only logarithmic quantum corrections
compared to the Gaussian scaling.
The crossover functions that capture the corrections are given by
\begin{align}
	F_{z}(|\mathbf{K}|) &= (\log(\Lambda/|\mathbf{K}|))^{\frac{(N^2_c+N_cN_f-1)}{2(N^2_c+N_cN_f-3)}}, \\
	F_{\Psi}(|\mathbf{K}|)&= \sqrt{\log(\log(\Lambda/|\mathbf{K}|))}, \\
	F_{\Phi}(|\mathbf{K}|) &= \log(\log(\Lambda/|\mathbf{K}|)),
\end{align}
in the small $|\mathbf{K}|$ limit with $\frac{\vec{k}}{|\mathbf{K}|F_z(|\mathbf{K}|)}\sim 1$ fixed.
See Appendix  \ref{sec:PhysicalObservablesThreeD} for details.

For $d<3$, $\lambda= g^2/v$ no longer flows to zero, although $v$ and $c$ still do under the one-loop renormalization group (RG) flow. 
This puts the control of the one-loop analysis in peril even close to three dimensions.
Due to the enhanced infrared quantum fluctuations caused by the modes that become increasingly dispersionless at low energies,
some higher-loop diagrams, albeit suppressed by powers of $\epsilon$, diverge at the one-loop fixed point.
The divergence is cured only after the two-loop correction in {\color{blue}Fig.} \ref{fig:2LBSE} is included. The energy scale below which the two-loop effect becomes qualitatively important marks the crossover energy scale $E_1(d)\sim \Lambda \exp\left(-\frac{(N_cN_f)^\frac{3}{2}}{(N^2_c-1)}\epsilon^{-\frac{3}{2}}\right)$. At energies below $E_1(d)$, 
the two-loop self-energy speeds up the collective mode 
such that $v$ and $c$ flow to zero with a hierarchy, $c \gg v$, with ${c^3\sim \epsilon v/N_c N_f}$\cite{LUNTS}.
The low-energy fixed point is characterized by
\begin{align}\label{eq:FixedPointEpsilon}
\hspace{-0.5cm}	\lambda^{*}=4\pi\epsilon,\quad x^{*}=\frac{N_cN_f}{16\pi \mathfrak{B}(3)},\quad w^{*}=0,\quad \&\quad \kappa^{*}_i=0,
\end{align}
where $x\equiv g^2/c^3$ and $\mathfrak{B}(3) \approx 0.0012434$
(See Appendix \ref{Sec:Quantum} for details).
It can be shown that 
all other higher-loop corrections remain finite and they are suppressed by $\epsilon$
at the modified one-loop (M1L) fixed point where
the two-loop effect is taken into account
in addition to the one-loop corrections\cite{LUNTS}. 
The shaded area of region II in {\color{blue}Fig.} \ref{fig:DimPlots} is where the M1L description is valid
at low energies.

Comparing these results with those obtained in three dimensions shows a qualitative change in the low energy physics. 
Especially, the fixed point value of $w$ is not a continuous function of ${d}$.
In region I, the one-loop effect causes $w$ to flow to the $\mathcal{O}(1)$ value given in Eq. (\ref{1loopft}).
Below the crossover energy scale $E_1(d)$,
$w$ flows to zero as $w(\ell)=\frac{N_cN_f}{2^{\frac{10}{3}}\mathfrak{B}(3)^\frac{1}{3}(N^2_c-1)^\frac{2}{3}}{\epsilon}^{-1}\ell^{-\frac{2}{3}}$\cite{LUNTS}. 
For small but nonzero ${\epsilon}$, the M1L description is controlled, and $w$ flows to zero at sufficiently low energies. 
Thus, the low-energy fixed point below three dimensions 
is qualitatively different from the fixed point that the theory flows into in three dimensions.
This discrepancy shows that the low-energy limit does not commute with the $d \rightarrow 3$ limit.
The change in the flow of $w$ is responsible for the disparity between the low-energy physical observables in $d=3-\epsilon$ in the $\epsilon\rightarrow 0$ limit and those in $d=3$. 

\begin{figure}[H]
	\centering
	\includegraphics[scale=0.46]{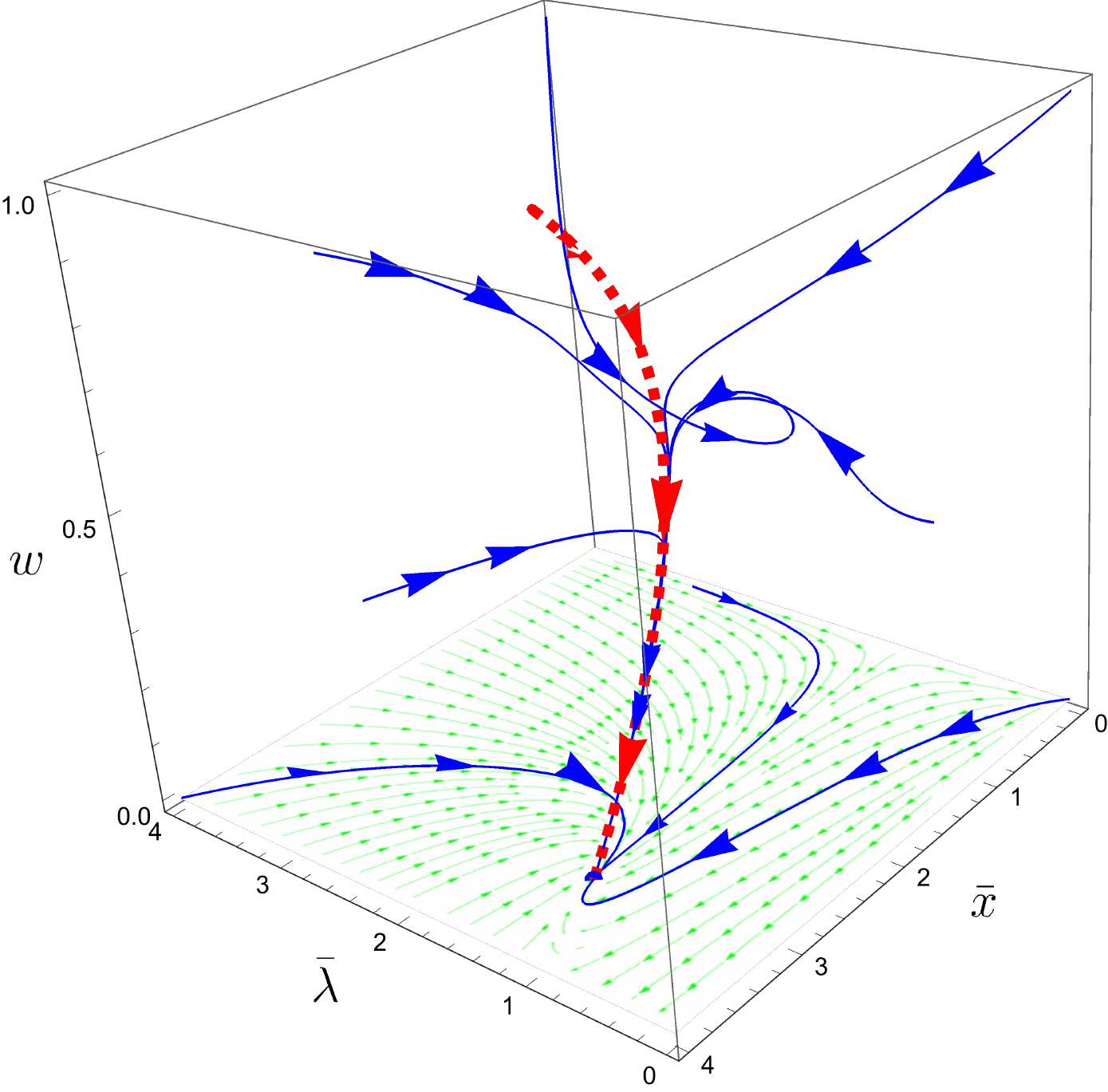}
	\caption{
		RG flow projected in the space of $(\overline{\lambda},\overline{x},w)$ 
		for $N_c=2,N_f=1$ and $\epsilon=0.01$ with $\kappa_i=0$.
		The axes are scaled as $\overline{\lambda}=10\lambda$ and $\overline{x}=x/10$. 
		The dashed (red) line corresponds to the one-dimensional manifold 
		towards which the RG flow is rapidly attracted
		before a slow flow along the manifold takes 
		the couplings to the low-energy fixed point 
		located on the $w=0$ plane. The three trajectories that do not seem to converge to the universal one-dimensional manifold lay on the $w=0$ plane.	\label{fig:RGFlow} }	
\end{figure}

There are two relatively well separated stages of the RG flow
in the space of $\lambda, x, w$ and $\kappa_i$ for $\epsilon>0$ and $\mu<E_1(d)$.
In the first stage, the RG flow converges 
towards a one-dimensional manifold,
where deviations away from the manifold
die out as a power-law in the energy scale.
The one-dimensional manifold can be parameterized by one of the parameters, say $w$,
where $\lambda, x$ and $\kappa_i$ take $w$-dependent values.
Once the RG flow converges to the one-dimensional manifold,
all couplings are controlled by a slow sub-logarithmic flow of $w$\cite{LUNTS}.
This is shown in {\color{blue}Fig.} 
\ref{fig:RGFlow}. 
At low energies, we can keep only one coupling, 
although the microscopic theory has five independent parameters.

At the IR fixed point, the fermion keeps the Gaussian scaling dimension, $[\Psi(k)]=-(d+2)/2$
while the collective mode acquires an anomalous dimension which gives $[\Phi(k)]=-d$.
Interestingly, the scaling dimensions of the fields are set such that
the fermion kinetic term and the Yukawa coupling are marginal
while the boson kinetic term and the quartic coupling are irrelevant. A similar protection of the scaling exponents arises in the $1/N$-expansion for the nematic QCP in $d$-wave superconductors\cite{HUH}.
Physically, the collective mode is strongly dressed by particle-hole excitations,
while its feedback to fermions remains small.
This provides a crucial hint in constructing
a nonperturbative ansatz for regions II and III.

\subsection{Region II: $\boldsymbol{2 < d < 3}$}
\label{sec:ansatz}

As the dimension approaches two, quantum fluctuations become progressively stronger, 
and the perturbative expansion no longer works.
In the following we describe a nonperturbative approach that captures the universal low-energy physics for 
any $0<\epsilon \leq 1$\cite{SCHLIEF}.

	 \subsubsection{Tree Level Scaling: Gaussian vs. Interaction-driven}

Under the Gaussian scaling, 
which prioritizes the kinetic terms over the interactions in Eq. (\ref{eq:ActionGeneralD}),
the scaling dimensions of $g$ and $u_i$ are $\epsilon/2$ and $\epsilon$, respectively. 
For $\epsilon \sim 1$, quantum corrections to the Gaussian scaling are expected to be $\mathcal{O}(1)$
and the $\epsilon$-expansion breaks down.  
For strongly coupled theories, it is better to start with a scaling 
which takes into account the interaction upfront rather than perturbatively.
The interaction-driven scaling\cite{SHOUVIK2} is 
a scaling that treats the interaction ahead of some kinetic terms.
Here we use the information obtained from the $\epsilon$-expansion 
to construct a scaling ansatz for general $\epsilon$.
In particular, we choose a scaling in which the fermion kinetic term and the fermion-boson interaction
are treated as marginal operators at the expense of treating the boson kinetic and quartic terms as irrelevant.
This uniquely fixes the scaling dimensions of the fields as in  {\color{blue}Table} \ref{tb:Scalings}.

The ansatz is consistent with the result from the $\epsilon$-expansion 
which suggests that the collective mode is likely to acquire an $\mathcal{O}(1)$ anomalous dimension near $d=2$. 
Since the boson dynamics is dominated by particle-hole excitations,
treating the boson kinetic term as an irrelevant operator is natural.
Dropping those terms that are irrelevant under the interaction-driven scaling, 
we write down the minimal action as
    \begin{widetext}
	 \begin{align}\label{eq:MinimalLocalAction}
	 \begin{split}
	 S_{d} &= \sum^{4}_{n=1}\sum^{N_c}_{\sigma=1}\sum^{N_f}_{j=1}\int\dd k\overline{\Psi}_{n,\sigma,j}(k)\left(i\boldsymbol{\Gamma}\cdot\mathbf{K}+i\gamma_{d-1}\varepsilon_{n}(\vec{k};v)\right)\Psi_{n,\sigma,j}(k)\\
	 &+\frac{i\beta_d\sqrt{v}}{\sqrt{N_f}}\sum^{4}_{n=1}\sum^{N_c}_{\sigma,\sigma'=1}\sum^{N_f}_{j=1}\int \dd k\int\dd q\overline{\Psi}_{\overline{n},\sigma,j}(k+q){\Phi}_{\sigma\sigma'}(q)\gamma_{d-1}\Psi_{n,\sigma',j}(k),
	 \end{split}
	 \end{align}
	 \end{widetext}
where 
	 \begin{align}\label{eq:Betad}
	 \beta_{d} = \frac{\pi^\frac{d-1}{4}}{\Gamma\left(\frac{d}{2}\right)}\sqrt{\frac{\Gamma(d)\Gamma\left(\frac{d-1}{2}\right)
	 \cos\left(\frac{\pi(d+2)}{2}\right)}{2^{3-d}}}
	 \end{align}
is a positive constant in $2\leq d<3$. 
The freedom in choosing the overall scale of the boson field is used to fix the Yukawa coupling in terms of $v$ such that $g^2/v \sim (3-d)$.
The choice of $\beta_d$ is such that the one-loop boson self-energy becomes order of one.
Roughly speaking, the fermion-boson coupling is replaced by $\sqrt{v}$ 
as the interaction is screened in such a way that $g^2$ and $v$ balance with each other in the low-energy limit\cite{SCHLIEF,SHOUVIK,LUNTS}. 
Since the $\epsilon$-expansion is organized in powers of $g^2/v$, 
the theory with $g^2/v \sim 1$ is a strongly coupled theory 
that cannot be accessed perturbatively in $\epsilon$.

  \begin{table}[htbp]
	\begin{tabular}{ccc}
		\hline
		\hline
		{\hspace{0.5em}\bf{Quantity}\hspace{0.5em}} & {\hspace{0.5em}\bf{Gaussian}\hspace{0.5em}} & {\hspace{0.5em}\bf{ID}\hspace{0.5em}}\\[1ex]
		\hline
		\hline
		&&\\[-1ex]	 		
		$[\Psi(k)]$ & $\displaystyle -\left(\frac{d+2}{2}\right)$ & $\displaystyle-\left(\frac{d+2}{2}\right)$\\[2ex]
		$[\Phi(k)]$ & $\displaystyle -\left(\frac{d+3}{2}\right)$ &$\displaystyle -d$\\[2ex]
		$[g]$&$\displaystyle \frac{3-d}{2}$&$\displaystyle 0 $\\[2ex]
		$[u_i]$&$\displaystyle 3-d$&$\displaystyle -(3-d)$\\[2ex]
		\hline 
		\hline	  		
	\end{tabular}
	\caption{
Comparison between the scaling dimensions of fields and couplings
deduced from the Gaussian and interaction-driven (ID) scalings.  
}
\label{tb:Scalings}
\end{table}

The five parameters ($v, c, g, u_1,u_2$) in the original theory are  now reduced to one ($v$) in the minimal theory.
The velocity $v$ specifies the low-energy effective theory
within the one-dimensional manifold shown in {\color{blue}Fig.} \ref{fig:RGFlow}.
The minimal theory is valid at energy scales low enough 
that the five parameters of the theory have already flown to the
one-dimensional manifold, and all renormalized couplings are 
tied to one leading irrelevant parameter.

 \subsubsection{Schwinger-Dyson Equation for the Boson Dynamics}
 
  \begin{figure}[htbp]
 	\centering
 	\includegraphics[scale=1.2]{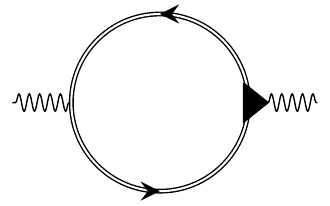}
 	\caption{Schwinger-Dyson (SD) equation for the exact boson self-energy. The double line represents the fully dressed fermion propagator and the triangle represents the fully dressed vertex.
\label{fig:SD}}
 \end{figure}

In the absence of the bare kinetic term for the boson, its dynamics is entirely generated from the self-consistent Schwinger-Dyson (SD) equation shown in {\color{blue}Fig.} \ref{fig:SD}. 
The SD equation for the exact boson self-energy is given by  
 \bqa
  && D(q)^{-1} = m_{\mathrm{C.T.}} -2\beta^2_{d}v \sum^{4}_{n=1}\int \dd k\nn
 && \times\mathrm{Tr}\left[\gamma_{d-1}G_{\overline{n}}(k+q)\varGamma^{(2,1)}_n(k,q)G_n(k)\right].  \label{eq:SDEquation}
 \eqa
Here $m_{\mathrm{C.T.}}$ is a counter term that tunes the mass to zero in order to keep the theory at criticality. 
$\varGamma^{(2,1)}_n(k,q)$ denotes the fully dressed vertex function.
$D(q)$ and  $G_n(k)$ denote the fully dressed boson and fermion propagators, respectively. 
%
%

We proceed following the scheme used in Ref. \cite{SCHLIEF}:
%
 \begin{itemize}
 	\item[{\color{blue}1.}] 
	We first assume that $v \ll 1$ and solve the SD equation in the small $v$ limit 
	to obtain the boson dynamics to the leading order in $v$.
 	\item[{\color{blue}2 .}] 
	By using the dressed boson propagator obtained under the assumption that $v$ is small, 
	we show that $v$ indeed flows to zero in the low-energy limit.
 \end{itemize}
 

\begin{figure*}
	\centering
	\begin{subfigure}[b]{0.31\linewidth}
		\includegraphics[scale=1.1]{SE_boson.pdf}
		\caption{\label{fig:SDDiagramA}}
	\end{subfigure}
	\begin{subfigure}[b]{0.31\linewidth}
		\includegraphics[scale=1.1]{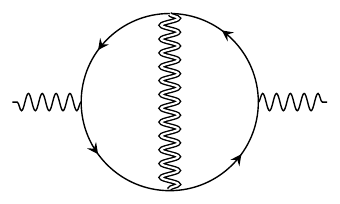}
		\caption{\label{fig:SDDiagramB}}
	\end{subfigure}
	\begin{subfigure}[b]{0.31\linewidth}
		\includegraphics[scale=1.1]{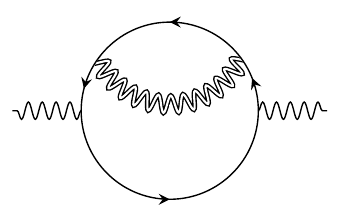}
		\caption{\label{fig:SDDiagramD}}
	\end{subfigure}
	\caption{Leading order corrections to the boson self-energy in the small $v$ limit. 
The solid line represents the  bare fermion propagator and the double wiggly line denotes the fully dressed self-consistent propagator in Eq. (\ref{eq:BosonPropagator}). 
	\label{fig:SDDiagrams}}
\end{figure*} 
 

We start with an ansatz for the fully dressed boson propagator in the small $v$ limit:
 \begin{align}\label{eq:BosonPropagator}
 D(q)^{-1} = |\mathbf{Q}|^{d-1}+c(v)^{d-1}(|q_x|^{d-1}+|q_y|^{d-1}),
 \end{align}
 \noindent where $c(v)$ is the `velocity' of the damped AFM collective mode
 that is to be determined as a function of $v$ from the SD equation. 
This ansatz is consistent with the interaction-driven scaling and the symmetries of the theory. However, the ultimate justification for the ansatz comes from the fact that Eq. (\ref{eq:BosonPropagator}) satisfies the SD equation as will be shown below.

Assuming that $v\ll c(v)\ll 1$, one can show that a general $L$-loop diagram with $L_f$ fermion loops and $E$ external legs scales at most as  
 \begin{align}\label{eq:UpperBound}
 \mathcal{G}(L,L_f,E)\sim  v^\frac{E-2}{2}\left(\frac{v}{c(v)}\right)^{L-L_f}
 \end{align}  
 up to logarithmic corrections.
 The proof closely follows the one given in Refs. \cite{SCHLIEF,LUNTS}. 
 In Appendix \ref{sec:SCALING}, we provide a brief review of the proof.
%
 The magnitude of general diagrams is not determined solely by the number of interaction vertices 
 since $v$ appears not only in the interaction term, but also in the fermion kinetic term. 
In the presence of the assumed hierarchy between velocities  ($v \ll c(v) \ll 1$) 
there is a systematic suppression of diagrams with $L>L_f$.

 
To the zeroth order in $v$, only the one-loop diagram in 
{\color{blue}Fig.} \ref{fig:SDDiagramA} survives. 
However, the leading order graph is independent of the spatial momentum. 
To determine  such a dependence of the boson propagator, 
one has to go to the next order in $v$ shown in {\color{blue}Figs.} \ref{fig:SDDiagramB} and \ref{fig:SDDiagramD}.
{\color{blue}Fig.}  \ref{fig:SDDiagramD} is again independent of the spatial momentum, 
and only  {\color{blue}Fig.} \ref{fig:SDDiagramB} remains important to the next leading order in $v$.
This is shown in Appendix \ref{Sec:Quantum}. 
{\color{blue}Fig.} \ref{fig:SDDiagramA} and
{\color{blue}Fig.} \ref{fig:SDDiagramB}
give rise to the SD equation:\\

\begin{widetext}
\begin{align}\label{eq:SDEquationSimplified}
\begin{split}
D(q)^{-1}&=m'_{\mathrm{C.T.}}+|\mathbf{Q}|^{d-1}-\frac{4\beta^4_d v^2}{N_c N_f}\sum^{4}_{n=1}\int\dd p\int\dd k\\
&\times \mathrm{Tr}\left[\gamma_{d-1}G^{(0)}_{\overline{n}}(k+p)\gamma_{d-1}G^{(0)}_{n}(k+q+p)\gamma_{d-1}G^{(0)}_{\overline{n}}(k+q)\gamma_{d-1}G^{(0)}_{n}(k)\right]D(p),
\end{split}
\end{align}
\end{widetext}
\noindent where
\begin{align}\label{eq:BareFermionPropagator}
G^{(0)}_n(k) = \frac{1}{i}\left(\frac{\boldsymbol{\Gamma}\cdot\mathbf{K}+\gamma_{d-1}\varepsilon_{n}(\vec{k};v)}{\mathbf{K}^2+\varepsilon_n(\vec{k};v)^2}\right)
\end{align}
\noindent denotes the bare fermion propagator and $m'_{\mathrm{C.T.}}$ is a two-loop mass counter term.
The term $|\mathbf{Q}|^{d-1}$ in Eq. (\ref{eq:SDEquationSimplified}) is the contribution from the one-loop self-energy. 
Explicit computation of the two-loop boson self-energy with Eq. (\ref{eq:BosonPropagator}) in the small $v$ limit 
indeed yields the boson propagator of the form in Eq. (\ref{eq:BosonPropagator}) with a self-consistent equation for $c(v)$
 (see Appendix {\color{purple}\ref{Sec:Quantum}} for details), 
 \begin{align}\label{eq:cgeneral}
	c(v)^{d-1} = \frac{4\beta^4_d\mathfrak{B}(d)}{(3-d)N_c N_f}\frac{v}{c(v)}\textswab{S}\left(d-2; \frac{v}{c(v)}  \right),
\end{align}
where $\textswab{S}(d-2;w(v))$ is 
defined in Eq. (\ref{eq:SFunction}).
It has the following limiting behaviors: 
$\lim_{w(v) \rightarrow 0}\textswab{S}(d-2;w(v))= 1/(d-2)$ 
and $\lim_{d\rightarrow 2}\textswab{S}(d-2;w(v)) = \log(1/w(v))$.
$\mathfrak{B}(d)$, defined in Eq. (\ref{eq:DefinitionofB}), is positive and finite in $2\leq d\leq  3$. 
Here we consider the low-energy limit at a fixed $d>2$.
If $w(v) \ll 1$, an assumption that needs to be checked later,
we can use $\lim_{w(v) \rightarrow 0}\textswab{S}(d-2;w(v))= 1/(d-2)$ 
to solve Eq. (\ref{eq:cgeneral})  and obtain
 \begin{align}\label{eq:CFunctionV}
 	c(v)  = \left(\frac{4\beta^4_d\mathfrak{B}(d)}{(3-d)(d-2)N_c N_f}\right)^{\frac{1}{d}}v^\frac{1}{d}.
 \end{align}
This general expression reduces to $c(v)^3/v=64\pi^2\mathfrak{B}(3)\epsilon/N_cN_f$
near three dimensions, which matches the result from the $\epsilon$-expansion in Ref. \cite{LUNTS}.
Finally we note that  $v\ll c(v)\ll 1$ and, thus, the assumed hierarchy of velocities ($w(v)\ll1$) is satisfied if $v \ll 1$. 
This gives the first consistency check of the scaling ansatz.

    \subsubsection{Low-energy fixed point}

     \begin{figure}[htbp]
	\centering
	\begin{subfigure}[b]{0.49\linewidth}
		\centering
		\includegraphics[scale=1]{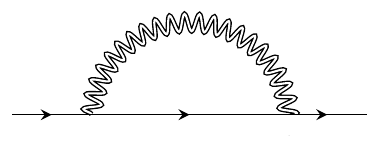}
		\caption{\label{fig:FSE1}}
	\end{subfigure}
	\begin{subfigure}[b]{0.49\linewidth}
		\centering
		\includegraphics[scale=1]{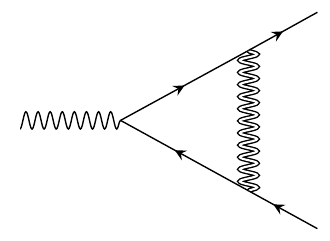}
		\caption{\label{fig:YUK}}
	\end{subfigure}
	\caption{
Leading order quantum corrections to the minimal local action.
In $2<d<3$, all other diagrams are strictly subleading in $v$.
	\label{fig:Diagrams}}
\end{figure}

The remaining question is whether $v$ flows to zero in the low-energy limit.
The beta function for $v$ is determined by the fermion self-energy, 
and the vertex correction determines the $\mathcal{O}(w(v))$ correction to the scaling dimension of the collective mode.
Because the Yukawa coupling remains marginal in any $2 \leq d \leq 3$
according to the interaction-driven scaling,
the quantum corrections are logarithmically divergent in all $2 \leq d \leq 3$. This is in contrast to the conventional perturbative approaches where logarithmic divergences arise only at the critical dimensions.
We determine local counter terms by requiring that physical observables 
are independent of UV cutoff scales (See Appendix \ref{Sec:RG} for details on the RG scheme).

According to Eq. (\ref{eq:UpperBound}), 
the contribution of the diagrams in {\color{blue} Fig.} \ref{fig:FSE1}
to the beta function of $v$ is at most $\mathcal{O}\left( w(v) \right) v$.
An explicit computation in Appendix \ref{Sec:Quantum}
shows that the contribution is actually  suppressed further by $c(v)$.
The reason for the additional suppression by $c(v)$ 
is that the external momentum can be directed to flow only through the boson propagator. 
As a result, the self-energy depends on the external spatial momentum through $c(v)\vec{k}$. 
According to Eq. (\ref{eq:UpperBound}), higher order diagrams are suppressed by at least one more power of $w(v)$.
Because
\bqa
w(v)  \sim v^{(d-1)/d} \ll c(v) \sim v^{1/d}
\label{eq:comp}
\eqa 
for $d>2$,
higher order diagrams remain smaller than
{\color{blue} Fig.} \ref{fig:FSE1} despite its additional suppression by $c(v)$.
In the small $v$ limit, 
{\color{blue} Fig.} \ref{fig:FSE1}  determines the beta function for $v$ 
(See Appendix \ref{Sec:Fixed} for a derivation),
   \begin{align}\label{eq:BetafunctionV}
   	\beta_v \equiv\frac{\dd v}{\dd\log\mu}= \frac{4(N_c^2-1)}{\pi N_c N_f}\frac{(d-1)\zeta(d)}{d-2}v^2
   	\end{align}
   \noindent 
   to the leading order in $v$ in $2<d<3$,
   where  
   \begin{align}\label{eq:zetad}
\zeta(d)&=-\frac{\cos \left(\frac{\pi  d}{2}\right) \Gamma \left(\frac{2d-3}{d-1}\right) \Gamma \left(\frac{1}{d-1}\right) \Gamma
	\left(\frac{d-1}{2}\right)}{2^{3-d}\pi ^{3/2} \Gamma \left(\frac{d}{2}\right)}
\end{align}
is positive in $2\leq d<3$. The beta function indeed shows that $v$ flows to zero at low energies in any $2< d<3$.
This completes the proof that the 
theory flows to the fixed point described by the ansatz introduced in the previous section 
if the bare value of $v$ is small.

At the low-energy fixed point with $v=0$, 
the dynamical critical exponent ($z$)
and the corrections to the interaction-driven scaling dimensions of the fields 
($ \eta_\Psi$ and $\eta_\Phi$)
in {\color{blue}Table} \ref{tb:Scalings}
are given by:
	\begin{align}\label{eq:CriticalFixedPoint}
	z=1,\qquad \eta_{\Psi}=0, \qquad \eta_{\Phi}=0.
	\end{align}
It is noted that $ \eta_\Psi=\eta_\Phi = 0$ does not mean 
that the fixed point is the Gaussian fixed point
because $ \eta_\Psi$, $\eta_\Phi$ 
denote the correction to the
interaction-driven scaling, which already includes
the $\mathcal{O}(1)$ anomalous dimension for the collective mode
compared to the noninteracting  theory.

\subsubsection{Green's Functions}

Defining the logarithmic length scale $\ell=\log(\Lambda/\mu)$, Eqs. (\ref{eq:CFunctionV}) and (\ref{eq:BetafunctionV}) imply that $w(v)$ flows to zero as 
 \begin{align}\label{eq:wofl}
 \hspace{-0.5cm}	w(\ell) = \frac{\pi^\frac{d-1}{d} N_cN_f(d-2)}{4((d-1)\zeta(d)(N^2_c-1))^\frac{d-1}{d}}\left[\frac{(3-d)}{\beta^4_d\mathfrak{B}(d)}\right]^\frac{1}{d}\frac{1}{\ell^{\frac{d-1}{d}}}
 \end{align}
for $\ell\gg\ell_0$ with $\ell_0\equiv\frac{1}{v_0}\frac{N_c N_f}{N^2_c-1}\textswab{S}\left(d-2;v^{\frac{d-1}{d}}_0\right)^{-1}\sim \frac{(d-2)}{v_0}\frac{N_cN_f}{N^2_c-1}$ and $v_0\ll 1$ denoting the bare value of $v$ (See Appendix \ref{Sec:Fixed} for details).
Even though $w(\ell)=0$ is a stable low-energy fixed point,
$w(\ell)$ is nonzero at intermediate energy scales
unless one starts with a fine tuned theory with a  perfectly nested Fermi surface. 
This gives rise to corrections to the scaling form of physical observables.
While critical exponents are well defined only at fixed points,
it is useful to introduce `scale-dependent critical exponents'
that determine the scaling forms of physical observables
in the presence of a slowly running irrelevant coupling, 
	\begin{align}
	z(\ell) &=1+\frac{(N^2_c-1)\zeta(d)}{N_cN_f} w(\ell),\label{eq:zexponent}\\
	\eta_\Psi(\ell) &=-\frac{(N^2_c-1)(d-1){\zeta}(d)}{2N_cN_f} w(\ell) ,\label{eq:EtaPsi}\\
	\eta_\Phi(\ell) &=-\frac{\left[(d-2)N^2_c-d+1\right](d-1){\zeta}(d)}{N_cN_f(d-2)} w(\ell). \label{eq:EtaPhi}
	\end{align}
Their derivation can be found in Appendix \ref{Sec:UNIV}.
Had $w(\ell)$ flown to a nonzero value at the fixed point,
the $\mathcal{O}(w(\ell))$ corrections would have modified the critical exponents  in Eq. (\ref{eq:CriticalFixedPoint}).
Since $w(\ell)$ flows to zero, the exponents predicted by the interaction-driven scaling are exact,
and the corrections introduce only subleading scalings in the physical observables.

The scaling form of the fermion Green's function is given by 
Eq.  (\ref{eq:GeneralFermionGreensFunction}) with 
	\begin{align}
	\hspace{-0.5cm}F_{z}(|\mathbf{K}|) &= \exp\left[(d-2)\mathfrak{F}_z(d)(N^2_c-1)^\frac{1}{d} \log\left[\frac{\Lambda}{|\mathbf{K}|}\right]^\frac{1}{d}\right],\label{eq:FZ}\\
	\hspace{-0.5cm}F_{\Psi}(|\mathbf{K}|)&=\sqrt{\log\left[\frac{\Lambda}{|\mathbf{K}|}\right]}, \label{eq:FPSI}
	\end{align}
and
	\begin{align}\label{eq:Fzz}
		\mathfrak{F}_{z}(d)&=\frac{\pi d}{4(d-1)}\left(\frac{(3-d)(d-1)\zeta(d)}{\pi\beta^4_d\mathfrak{B}(d)}\right)^\frac{1}{d}.
	\end{align}
It is noted that $F_{z}(|\mathbf{K}|)$ and $F_{\Psi}(|\mathbf{K}|)$ introduce corrections that are not strong enough to modify the exponents in the power-law behavior,
yet $F_{z}(|\mathbf{K}|)$ is stronger than logarithmic corrections of marginal Fermi liquids\cite{VARMALI,VARMA}. 
Similarly the crossover function for the bosonic Green's function in Eq. (\ref{eq:GeneralBosonGreensFunction}) is given by
	\begin{align}
	\hspace{-0.5cm}	F_{\Phi}(|\mathbf{Q}|) = \exp\left[ \frac{\mathfrak{F}_{\Phi}(d)\log\left[\frac{\Lambda}{|\mathbf{Q}|}\right]^\frac{1}{d}}{2(N^2_c-1)^\frac{d-1}{d}}\left(d-(d-2)N^2_c\right)\right],
		\end{align}
with 
\begin{align}\label{eq:Fphi}
\mathfrak{F}_{\Phi}(d)= \frac{d \pi^\frac{d-1}{d}}{2}\left(\frac{(d-1)(3-d)\zeta(d)}{\beta^4_d\mathfrak{B}(d)}\right)^\frac{1}{d}.
\end{align}
In Appendix \ref{Sec:UNIV} we provide the derivation of these results.
Compared to the bare boson propagator, the physical propagator describing the low-energy dynamics of the AFM collective mode is highly damped and incoherent. 
We note that the deviation of fermion Green's function from that of Fermi liquids 
as well as the incoherent nature of the AFM collective mode become stronger as $d$ is lowered.  
This is expected because the effect of interactions is stronger in lower dimensions.

\subsection{Region III : from $\boldsymbol{d>2}$  to $\boldsymbol{d=2}$}\label{sec:twoD}

In this section, we discuss how the results obtained in $2 < d < 3$ 
are connected to the solution in $d=2$\cite{SCHLIEF}.
We note that the expression in Eq. (\ref{eq:CFunctionV}), which is divergent  in  $d=2$, is valid only for $d > 2$.
This is because the $d \rightarrow 2$ limit and the $w(v) \rightarrow 0$ limit do not commute in Eq. (\ref{eq:cgeneral}).
In order to access the physics in $d=2$, we have to take the $d \rightarrow 2$ limit 
before the low-energy limit is taken.
In $d=2$, the $1/(d-2)$ divergence in Eq. (\ref{eq:CFunctionV}) is replaced by $\log(1/w(v))$,
and the solution to  Eq. (\ref{eq:cgeneral}) is given by
\begin{align}
	c(v) = \sqrt{\frac{1}{8N_cN_f}v\log\left(\frac{1}{v}\right)}
\end{align}
to the leading order in $v$\cite{SCHLIEF}.
Notice that the hierarchy $v \ll c(v) $ still holds if $v \ll 1$,
and general diagrams still obey Eq. (\ref{eq:UpperBound}) up to logarithmic corrections.

\begin{figure}[htbp]
	\begin{subfigure}[b]{0.49\linewidth}
		\centering
		\includegraphics[scale=0.9]{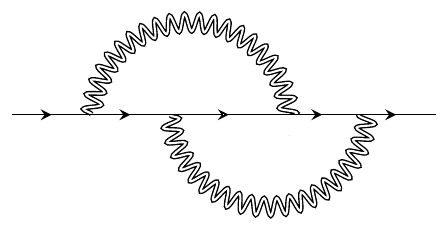}
		\caption{\label{fig:FSE2}}
	\end{subfigure}
	\begin{subfigure}[b]{0.49\linewidth}
		\centering
		\includegraphics[scale=0.8,trim={0 -0.5cm 0 0}]{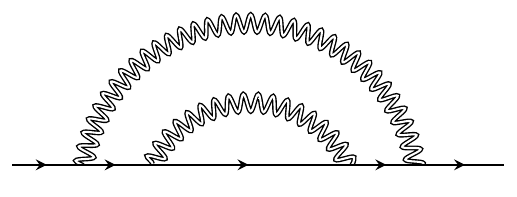}
		\caption{\label{fig:FSE22}}
	\end{subfigure}
	\caption{Two-loop fermion self-energies. As explained in the text, the two-loop diagram ({\color{blue}$a$}) is of the same order as the one-loop diagram in {\color{blue}Fig.} \ref{fig:FSE1}. The diagram in ({\color{blue}$b$}) is subleading due to an additional suppression by $c(v)$. 
	\label{fig:Diagrams2}}
\end{figure}

Another complication that arises  in $d=2$ is that
the inequality in Eq. (\ref{eq:comp}) no longer holds.
This means that the two-loop fermion self-energies shown in {\color{blue}Fig.} \ref{fig:Diagrams2}
can be as important as the one-loop graph in {\color{blue}Fig.} \ref{fig:FSE1}. {\color{blue}Fig.} \ref{fig:FSE22} is also additionally suppressed by $c(v)$ for the same reason that {\color{blue}Fig.} \ref{fig:FSE1} is further suppressed by $c(v)$.
However, this extra suppression is absent in {\color{blue}Fig.} \ref{fig:FSE2} 
because the external momentum cannot be directed to flow only through the boson lines. 
As a result,  {\color{blue}Fig.} \ref{fig:FSE2}  is of the same order as the one-loop fermion self-energy in $d=2$.
Taking into account the contribution from  {\color{blue}Fig.} \ref{fig:FSE1} and  {\color{blue}Fig.} \ref{fig:FSE2}, 
we obtain the beta function for $v$ in $d=2$\cite{SCHLIEF},
\begin{align}
	\beta_v= \frac{2}{\pi^2}\frac{(N^2_c-1)}{N_cN_f}v^2\log\left(\frac{1}{v}\right).
\end{align}
It again predicts that $v$ flows to zero if $v$ is small to begin with.

In $d=2$, the scale-dependent critical exponents are given by
\begin{align}
	z(\ell) &= 1+\frac{(N^2_c-1)}{2\pi N_c N_f} w(\ell) ,\label{eq:zexponent2}\\
	\eta_{\Psi}(\ell) &=-\frac{(N^2_c-1)}{4\pi N_cN_f}w(\ell),\\
	\eta_\Phi(\ell) &=\frac{1}{2\pi N_c N_f} w(\ell) \log\left(\frac{1}{w(\ell)}\right),
	\label{eq:EtaPhi2}
\end{align}
where $w(\ell)$ flows to zero as
\begin{align}\label{eq:W2d}
	w(\ell) = \frac{2\pi N_c N_f}{\sqrt{N_c^2-1}}\frac{1}{\sqrt{\ell}\log(\ell)}
\end{align}
for $\ell\gg \ell_0$ with $\ell_0\equiv  \lim_{d\rightarrow 2}\frac{1}{v_0}\frac{N_c N_f}{N^2_c-1}\textswab{S}\left(d-2;v^{\frac{d-1}{d}}_0\right)^{-1}\sim \frac{2}{v_0\log(1/v_0)}\frac{N_cN_f}{N^2_c-1}$ and $v_0\ll 1$ denoting the bare value of $v$ (See Appendix \ref{Sec:Fixed} for details). 

Comparing Eq. (\ref{eq:W2d}) with the $d\rightarrow 2$ limit of Eq. (\ref{eq:wofl}) shows that the flow of $w(\ell)$ in $d>2$ does not smoothly extend to $d=2$. 
This is due to the existence of a crossover energy scale, $E_2(d)$ 
that vanishes in the $d\rightarrow 2$ limit. 
As the energy scale is lowered, the crossover from region III to region II 
occurs at a scale where
$\lim_{w(v)\rightarrow 0}\textswab{S}(d-2;w(v))= 1/(d-2) \sim \lim_{d\rightarrow 2}\textswab{S}(d-2;w(v)) = \log(1/w(v))$ in Eq. (\ref{eq:cgeneral}). 
From Eq. (\ref{eq:W2d}),  the crossover energy scale is obtained to be $E_{2}(d)\sim \Lambda\exp\left(-(d-2)^2\frac{(N_cN_f)^2}{(N^2_c-1)}e^{2/(d-2)}\right)$. 
The double exponential dependence 
originates from the fact that $w(v)$ needs to be exponentially small in $-(d-2)^{-1}$ for the crossover to happen,
and, up to sublogarithmic corrections, $w(v)^2$ itself flows to zero logarithmically in two dimensions. The sublogarithmic correction to the flow of $w(\ell)$ is responsible for the extra factor of $(d-2)^2$ in the exponential. For $\mu>E_{2}(d)$ (region III), $w(\ell)$ flows to zero according to Eq. (\ref{eq:W2d}), while for $\mu<E_{2}(d)$ (region II), the flow is dictated by Eq. (\ref{eq:wofl}). 
Thus, unless $d=2$, the theory will always flow into region II at sufficiently low energies.

Finally, the corrections to the exponents predicted by the interaction-driven scaling go to zero in the long distance limit
because $w(\ell)$ flows to zero.
The Green's functions at intermediate energy scales
receive super-logarithmic corrections given by the crossover functions\cite{SCHLIEF},
%
%
%
 \begin{align}
 	F_z(k_0)&=  \exp\left(2\sqrt{N^2_c-1}\frac{\log(\Lambda/|k_0|)^\frac{1}{2}}{\log(\log(\Lambda/|k_0|))}\right),\\
 	F_\Psi(k_0)&=\left (\log\left(\Lambda/|k_0|\right)\right)^\frac{3}{8},\\
 	F_\Phi(k_0) &=\exp\left(\frac{2(\log(\Lambda/|k_0|))^\frac{1}{2}}{\sqrt{N^2_c-1}}\right) .
 \end{align}
The crossover functions in $d=2$ are different from 
the $d \rightarrow 2$ limit of the crossover functions obtained in $d>2$.  
This is due to the fact that the low-energy limit and the $d \rightarrow 2$ limit do not commute.

	\section{\bf{Summary}}
\label{sec:summary}

In this paper, we solved the low-energy effective theory
for the commensurate AFM quantum critical metal 
with a $C_4$-symmetric one-dimensional Fermi surface 
embedded in space dimensions between two and three.
The exact critical exponents and 
the subleading  corrections generated 
from the leading irrelevant perturbation are obtained 
by extending the  nonperturbative approach 
based on an interaction-driven scaling\cite{SCHLIEF}.
The solution in $2\leq d\leq 3$ provides
an interpolation between the perturbative solution
obtained based on the $\epsilon$-expansion near the upper critical dimension
and the nonperturbative solution for the two-dimensional theory.
The general solution exposes both merits and subtle issues 
of RG schemes based on dimensional regularization. 
On the one hand, the critical exponents that characterize the low-energy fixed point
are smooth functions of the space dimension.
This allows one to make an educated guess on the critical exponents 
in two dimensions from the solution obtained in higher dimensions.
On the other hand, the full scaling behaviors in two dimensions
are not correctly captured by the low-energy solutions obtained above two dimensions.
A crossover scale that vanishes in the $d \rightarrow 2$ limit
makes it difficult to access the full scaling forms of physical observables in $d=2$ 
from solutions obtained in the low-energy limit in $d>2$.
These crossovers give rise to
emergent noncommutativities,
where the low-energy limit
and the limits in which physical dimensions
are approached do not commute.

	\section{\bf{Acknowledgements}} 
The research was supported by
the Natural Sciences and Engineering Research Council of Canada.
Research at the Perimeter Institute is supported
in part by the Government of Canada
through Industry Canada,
and by the Province of Ontario through the
Ministry of Research and Information. 
The Flatiron Institute is a division of the Simons Foundation.

\onecolumngrid

	\bibliographystyle{apsrev4-1}

		\bibliography{references}


	\appendix

	\renewcommand\thesection{{\color{purple}\bf{\Alph{section}}}}
	\renewcommand\thesubsection{{\color{purple}($\alph{subsection}$)}\hspace{0.2em}}
	\renewcommand\thesubsubsection{{\color{purple}\arabic{subsubsection}}}
	
	\renewcommand\theequation{{\color{purple}\thesection.\arabic{equation}}}

	\section{\bf{Physical Observables in Three Dimensions}}
	\label{sec:PhysicalObservablesThreeD}
	
Here we derive the scaling form of the Green's functions in $d=3$. 
We first summarize the regularization and renormalization group (RG) prescription\cite{SHOUVIK},
and proceed to compute the scaling form of the low-energy Green's functions.
	 
	\subsection{Regularization and RG Scheme in $\boldsymbol{d=3}$}\label{Sec:RGPrescription3d}
	
Since $d=3$ is the upper critical dimension of the theory, 
every term in Eq. (\ref{eq:ActionGeneralD}) is marginal under the Gaussian scaling,
and quantum corrections are expected to be logarithmically divergent. 
We regulate the theory by introducing two UV cutoffs : 
$\Lambda$ in the frequency and co-dimensional momentum space that is $\mathrm{SO}(d-1)$ symmetric, 
and $\widetilde{\Lambda}$ in the original two-dimensional momentum subspace. 
We assume that they are comparable in magnitude. 
To make sure that physical observables are independent of the UV energy scales, 
we add the following counter terms to the action 
	\begin{align}\label{eq:3dCT}
		S^{\mathrm{C.T}}_{d=3} &= \sum^{4}_{n=1}\sum^{N_c}_{\sigma=1}\sum^{N_f}_{j=1}\int \dd k\overline{\Psi}_{n,\sigma,j}(k)\left(i\mathcal{A}_1\boldsymbol{\Gamma}\cdot\mathbf{K}+i\gamma_{2}\widetilde{\varepsilon}_{n}(\vec{k};v)\right)\Psi_{n,\sigma,j}(k)+\frac{1}{4}\int\dd q\left(\mathcal{A}_4|\mathbf{Q}|^2+\mathcal{A}_5c^2|\vec{q}|^2\right)\mathrm{Tr}\left[\Phi(-q)\Phi(q)\right]\notag\\
	&+\frac{ig\mathcal{A}_6}{\sqrt{N_f}}\sum^{4}_{n=1}\sum^{N_c}_{\sigma,\sigma'=1}\sum^{N_f}_{j=1}\int \dd k\int \dd q\overline{\Psi}_{\overline{n},\sigma,j}(k+q){\Phi}_{\sigma\sigma'}(q)\gamma_{2}\Psi_{n,\sigma',j}(k)\\
	&+\frac{1}{4}\left[\prod^{3}_{i=1}\int\dd q_i\right]\left[\mathcal{A}_7u_1\mathrm{Tr}\left[\Phi(q_1+q_3)\Phi(q_2-q_3)\right]\mathrm{Tr}\left[\Phi(-q_1)\Phi(-q_2)\right]+\mathcal{A}_8u_2\mathrm{Tr}\left[\Phi(q_1+q_3)\Phi(q_2-q_3)\Phi(-q_1)\Phi(-q_2)\right]\right].\notag
	\end{align}
	
	\noindent Here, $\gamma_2=\sigma_x$ is the first Pauli matrix, $\widetilde{\varepsilon}_{1}(\vec{k};v) = \mathcal{A}_2vk_x+\mathcal{A}_3k_y$, 
	$\widetilde{\varepsilon}_{2}(\vec{k};v) = -\mathcal{A}_3 k_x+\mathcal{A}_2vk_y$,  
	$\widetilde{\varepsilon}_{3}(\vec{k};v) =\mathcal{A}_2 vk_x-\mathcal{A}_3k_y$, 
	and $\widetilde{\varepsilon}_{4}(\vec{k};v) = \mathcal{A}_3 k_x + \mathcal{A}_2vk_y$. The $\mathcal{A}_i$'s are momentum-independent counter term coefficients.	Adding this counter term action to Eq. (\ref{eq:ActionGeneralD}) in $d=3$ yields the renormalized action,
		\begin{align}
	S^{\mathrm{Ren}}_{d=3} &= \sum^{4}_{n=1}\sum^{N_c}_{\sigma=1}\sum^{N_f}_{j=1}\int \dd k_{B}\overline{\Psi}_{n,\sigma,j;B}(k_{B})\left(i\boldsymbol{\Gamma}\cdot\mathbf{K}_{B}+i\gamma_{2}{\varepsilon}_{n}(\vec{k}_{B};v_{B})\right)\Psi_{n,\sigma,j;B}(k_{B})\notag.\\
	&+\frac{1}{4}\int\dd q_{B}\left(|\mathbf{Q}_{B}|^2+c^2_{B}|\vec{q}_{B}|^2\right)\mathrm{Tr}\left[\Phi_{B}(-q_{B})\Phi_{B}(q_{B})\right]\notag\\
	&+\frac{ig_{B} }{\sqrt{N_f}}\sum^{4}_{n=1}\sum^{N_c}_{\sigma,\sigma'=1}\sum^{N_f}_{j=1}\int \dd k_{B}\int \dd q_{B}\overline{\Psi}_{\overline{n},\sigma,j;B}(k_{B}+q_{B}){\Phi}_{B;\sigma\sigma'}(q_{B})\gamma_{2}\Psi_{n,\sigma',j;B}(k_{B})\\
	&+\frac{1}{4}\left[\prod^{3}_{i=1}\int\dd q_{i;B}\right]\left\{u_{1;B}\mathrm{Tr}\left[\Phi_{B}(q_{1;B}+q_{3;B})\Phi_{B}(q_{2;B}-q_{3;B})\right]\mathrm{Tr}\left[\Phi_{B}(-q_{1;B})\Phi_{B}(-q_{2;B})\right]\right.\notag\\
	&\left.+u_{2;B}\mathrm{Tr}\left[\Phi(q_{1;B}+q_{3;B})\Phi_{B}(q_{2;B}-q_{3;B})\Phi_{B}(-q_{1;B})\Phi_{B}(-q_{2;B})\right]\right\}.\notag
	\end{align}
	\noindent The renormalized frequency, momenta, fields, velocities and couplings are related to the bare ones through
	\begin{align}
	\mathbf{K}_{B}&=\frac{\mathcal{Z}_1}{\mathcal{Z}_3}\mathbf{K},\qquad \vec{k}_{B} =\vec{k},\qquad v_{B} = \frac{\mathcal{Z}_2}{\mathcal{Z}_3}v, \qquad c_B = \sqrt{\frac{\mathcal{Z}_5}{\mathcal{Z}_4}}\left(\frac{\mathcal{Z}_1}{\mathcal{Z}_3}\right)c,\\
	\quad g_{B} &=\frac{\mathcal{Z}_6}{\mathcal{Z}_3\sqrt{\mathcal{Z}_4}} g,\qquad u_{1;B} = \frac{\mathcal{Z}_7}{\mathcal{Z}^2_4}\left(\frac{\mathcal{Z}_1}{\mathcal{Z}_3}\right)^2u_1, \qquad u_{2;B} =\frac{\mathcal{Z}_8}{\mathcal{Z}^2_4}\left(\frac{\mathcal{Z}_1}{\mathcal{Z}_3}\right)^2u_{2}, \\
\Psi_{B}(k_{B}) &= \mathcal{Z}^\frac{1}{2}_{\Psi}\Psi(k),\qquad \& \qquad 	\Phi_{B}(k_{B}) = \mathcal{Z}^\frac{1}{2}_{\Phi}\Phi(k),
	\end{align}
	\noindent where $\mathcal{Z}_i=1+\mathcal{A}_i$, $\mathcal{Z}_\Psi = \mathcal{Z}_3(\mathcal{Z}_3/\mathcal{Z}_1)^2$ , $\mathcal{Z}_{\Phi} =\mathcal{Z}_4(\mathcal{Z}_3/\mathcal{Z}_1)^4$, and the field indices have been suppressed.	The renormalized action gives rise to the quantum effective action that can be expanded as
	\begin{align}\label{eq:QEA1}
	\begin{split}
	&\boldsymbol{\Gamma}[\{\overline{\Psi},\Psi,\Phi\},v,c,g,u_i;\mu]=\sum^{\infty}_{m=0}\sum^{\infty}_{n=0}\boldsymbol{\Gamma}^{(2m,n)}[\{\overline{\Psi},\Psi,\Phi\},v,c,g,u_i;\mu],\qquad\qquad \qquad \text{where}\\
	&\boldsymbol{\Gamma}^{(2m,n)}[\{\overline{\Psi},\Psi,\Phi\},v,c,g,u_i;\mu]= \left(\prod^{2m+n}_{j=1}\int\dd k_{j}\right)(2\pi)^{d+1}\delta\left(\sum^{m}_{j=1}k_{j}-\sum^{2m+n}_{j=1+m}k_{j}\right)\\
	&\times\varGamma^{(2m,n)}\left(k_1,\dots,k_{2m+n-1}, v,c,g,u_i;\mu\right)\overline{\Psi}(k_1)\cdots\overline{\Psi}(k_{m})\Psi(k_{1+m})\cdots\Psi(k_{2m})\Phi(k_{1+2m})\cdots\Phi(k_{n+2m}).
	\end{split}
	\end{align}
	\noindent Here, $\varGamma^{(2m,n)}\left(k_1,\dots,k_{2m+n-1}, v,c,g,u_i;\mu\right)$ denote the one-particle irreducible (1PI) vertex functions that implicitly depend on all discrete indices. The summation over these indices has been left implicit. 
The counter-term coefficients in Eq. (\ref{eq:3dCT}) are determined according to a minimal subtraction scheme 
which imposes the following renormalization conditions on the vertex functions, 
	\begin{align}
	&	\lim_{\widetilde{\Lambda}\rightarrow \infty}\lim_{\Lambda\rightarrow \infty}	\frac{1}{2i}\frac{\partial}{\partial\mathbf{K}^2}\mathrm{Tr}\left[(\mathbf{K}\cdot\boldsymbol{\Gamma})\varGamma^{(2,0)}_{n}(k)\right]\bigg|_{|\mathbf{K}|=\mu,\vec{k}=0}=1+E_1(v,c,g,u_i),\qquad n=1,2,3,4,\label{eq:RGConditionFermionFrequency13d}\\
	&\lim_{\widetilde{\Lambda}\rightarrow \infty}\lim_{\Lambda\rightarrow \infty}\frac{1}{2i}\frac{\partial}{\partial k_x}\mathrm{Tr}\left[\gamma_{2}\varGamma^{(2,0)}_{n=1}(k)\right]\bigg|_{|\mathbf{K}|=0,k_x=\mu,k_y=0}=v(1+E_2(v,c,g,u_i)),\label{eq:RGConditionKx3d}\\
	&\lim_{\widetilde{\Lambda}\rightarrow \infty}\lim_{\Lambda\rightarrow \infty}\frac{1}{2i}\frac{\partial}{\partial k_y}\mathrm{Tr}\left[\gamma_{2}\varGamma^{(2,0)}_{n=1}(k)\right]\bigg|_{|\mathbf{K}|=0,k_x=0,k_y=\mu}=1+E_3(v,c,g,u_i),\label{eq:RGConditionKy3d}\\
	&	\lim_{\widetilde{\Lambda}\rightarrow \infty}\lim_{\Lambda\rightarrow \infty}\frac{\partial}{\partial\mathbf{Q}^2}\left[\varGamma^{(0,2)}(q)\right]\bigg|_{|\mathbf{Q}|=\mu,\vec{q}=0}= 1+E_4(v,c,g,u_i),\label{eq:RGConditionQ2}\\
&	\lim_{\widetilde{\Lambda}\rightarrow \infty}\lim_{\Lambda\rightarrow \infty}\left[\frac{\partial}{\partial q_j^2}	\varGamma^{(0,2)}(q)\right]\bigg|_{|\mathbf{Q}|=0, \vec q=(\mu,\mu)} = c^2(1+E_{5}(v,c,g,u_i)),\qquad j=x,y,\label{eq:RGConditionq2}\\	
&	\lim_{\widetilde{\Lambda}\rightarrow \infty}\lim_{\Lambda\rightarrow \infty}\frac{1}{2}\mathrm{Tr}\left[\gamma_{2}\varGamma^{(2,1)}_{n}(k,q)\right]\bigg|_{q=0,|\mathbf{K}|=\mu,\vec{k}=0}=1+E_6(v,c,g,u_i),\qquad n=1,2,3,4\label{eq:RGConditionVertex3d},\\
&\lim_{\widetilde{\Lambda}\rightarrow \infty}\lim_{\Lambda\rightarrow \infty}	\varGamma^{(0,4)}_{abcd}(k_1,k_2,k_3) \bigg|_{|\mathbf{K}_{i}|=\mu,\vec{k}_i=\vec{0}}= \frac{1}{4}(u_1\mathrm{Tr}[\tau^a\tau^b]\mathrm{Tr}[\tau^c\tau^d]+u_2\mathrm{Tr}[\tau^a\tau^b\tau^c\tau^d])+E_7(v,c,g,u_i).\label{eq:RG4Point}
	\end{align}
	\noindent 
Here $\mu$ is an energy scale at which the physical observables are measured. 
$E_i(v,c,g,u_i)$'s are finite functions of the renormalized couplings.
They vanish in the $u_i\rightarrow 0$ and $g\rightarrow 0$ limits. 
$\tau^{a}$ denote the generators of $\mathrm{SU}(N_c)$ with $a=1,2,\dots, N^2_c-1$.
The conditions in Eqs. (\ref{eq:RGConditionFermionFrequency13d}) to (\ref{eq:RGConditionKy3d}) fix the fermion two-point function at the $n=1$ hot spot and, by virtue of the $C_4$ symmetry of the theory, they also fix the two-point function at the other three hot spots. 
The renormalization conditions in Eqs. (\ref{eq:RGConditionQ2}) and (\ref{eq:RGConditionq2}) fix the bosonic two-point function.
Eqs. (\ref{eq:RGConditionVertex3d}) and (\ref{eq:RG4Point}) fix the Yukawa vertex and the bosonic four-point function, respectively. 
	
	Under the Gaussian scaling, the 1PI vertex functions have scaling dimension $[\varGamma^{(2m,n)}(\{k_i\};v,c,g,u_i;\mu)]=4-n-3m$ and the renormalized vertex functions are related to the bare ones via,
	\begin{align}
	\varGamma^{(2m,n)}_{B}(\{k_{i;B}\};v_{B},c_{B},g_{B},u_{i;B};\Lambda,\widetilde{\Lambda}) = \left(\frac{\mathcal{Z}_3}{\mathcal{Z}_1}\right)^{2(2m+n-1)}\mathcal{Z}^{-m}_{\Psi}\mathcal{Z}^{-\frac{n}{2}}_{\Phi}	\varGamma^{(2m,n)}(\{k_i\};v,c,g,u_i;\mu).
	\end{align}
	\noindent Since the bare vertex functions are independent of the running energy scale $\mu$, 
the vertex functions satisfy the RG equation,
	\begin{align}
	\begin{split}\label{eq:RGEquation3d}
	\left[
	\sum^{2m+n-1}_{i=1}\left(z\mathbf{K}_{i}\cdot\nabla_{\mathbf{K}_i}+\vec{k}_i\cdot\nabla_{\vec{k}_i}\right)-\beta_v\frac{\partial}{\partial v}-\beta_c\frac{\partial}{\partial c}-\beta_g\frac{\partial}{\partial g}-\beta_{u_{1}}\frac{\partial}{\partial u_{1}}-\beta_{u_{2}}\frac{\partial}{\partial u_{2}}\right.\\
	\left.+2m\left(\eta_{\Psi}-\frac{5}{2}\right)+n(\eta_{\Phi}-3)+2(2m+n-1)(z+1)\right]	\varGamma^{(2m,n)}(\{ k_i\}, v,c,g,u_i;\mu)=0,
	\end{split}
	\end{align}
	
	\noindent where the critical exponents and beta functions of the velocities and couplings are given by
	\begin{align}
	z &= 1-\frac{\dd}{\dd\log\mu}\log\left(\frac{\mathcal{Z}_3}{\mathcal{Z}_1}\right),\label{eq:z3d}\\
	\eta_{\Psi(\Phi)} &= \frac{1}{2}\frac{\dd}{\dd\log\mu}\log \mathcal{Z}_{\Psi(\Phi)}\label{eq:eta3d}\\
	\beta_{\mathsf{A}}&=\frac{\dd \mathsf{A}}{\dd\log\mu},\qquad \mathsf{A}= v,c,g,u_1,u_2. 
	\end{align}
	\noindent Here, $z$ denotes the dynamical critical exponent and $\eta_{\Psi}$ ($\eta_{\Phi}$) denotes the anomalous scaling dimension of the fermion (boson) field with respect to the Gaussian scaling.
	
	The one-loop counter term coefficients in $d=3$ are given by\cite{SHOUVIK}
	\begin{align}
	\mathcal{Z}_1 &= 1-\frac{(N^2_c-1)}{4\pi^2N_cN_f}\frac{g^2}{c}h_1(v,c)\log\left(\frac{\Lambda}{\mu}\right),\\
	\mathcal{Z}_2&=1+\frac{(N^2_c-1)}{4\pi^2 N_cN_f}\frac{g^2}{c}h_2(v,c)\log\left(\frac{\Lambda}{\mu}\right),\\
	\mathcal{Z}_3&=1-\frac{(N^2_c-1)}{4\pi^2 N_cN_f}\frac{g^2}{c}h_2(v,c)\log\left(\frac{\Lambda}{\mu}\right),\\
	\mathcal{Z}_4&=1-\frac{1}{4\pi}\frac{g^2}{v}\log\left(\frac{\Lambda}{\mu}\right),\\
	\mathcal{Z}_5&=0,\\
	\mathcal{Z}_6&=1-\frac{1}{8\pi^3N_cN_f}\frac{g^2}{c}h_3(v,c)\log\left(\frac{\Lambda}{\mu}\right),\\
	\mathcal{Z}_7 &=1+\frac{1}{2\pi^2 c^2}\left[(N^2_c+7)u_1+2\left(\frac{2N^2_c-3}{N_c}\right)u_2+3\left(\frac{N^2_c+3}{N^2_c}\right)\frac{u^2_2}{u_1}\right]\log\left(\frac{\Lambda}{\mu}\right),\\
	\mathcal{Z}_8&=1+\frac{1}{2\pi^2c^2}\left[12u_1+2\left(\frac{N^2_c-9}{N_c}\right)u_2\right]\log\left(\frac{\Lambda}{\mu}\right).
	\end{align}
	\noindent Here, $h_i(v,c)$ are finite functions of $v$ and $c$ defined in Ref. \cite{SHOUVIK}.
They have the following limiting behaviors: $\lim_{c\rightarrow 0}h_1(w c,c)=\frac{\pi}{2}, \lim_{c\rightarrow 0}h_2(w c,c)=2c,$ and $\lim_{c\rightarrow 0}h_3(w c,c)=2\pi^2/(1+w)$, with $w=v/c$ fixed. 
In the low-energy limit,
all $g$, $v$, $c$, $u_i$ flow to zero such that 
$\lambda \equiv g^2/v \sim 1/l$, 
$\kappa_i \equiv u_i/c^2 \sim 1/l$,
$v \sim c \sim 1/\log(l)$,
where $l$ is the logarithmic length scale\cite{SHOUVIK}.
The quasi-local marginal Fermi liquid fixed point is stable.
While the leading scaling behaviors are characterized by the Gaussian critical exponents,
there exist logarithmic corrections generated from 
the marginally irrelevant couplings.
Below, we discuss those corrections in the two-point functions.
For simplicity, we set $u_i=0$,
and focus on the corrections from the Yukawa coupling.
	
	\subsection{Fermionic and Bosonic Green's Functions}
	
	The scaling form of the two-point functions is governed by 
	\begin{align}\label{eq:RGThreeD}
	\begin{split}
	\left[z\mathbf{K}\cdot\nabla_{\mathbf{K}}+\vec{k}\cdot\nabla_{\vec{k}}-\beta_w\frac{\partial}{\partial w}-\beta_\lambda\frac{\partial}{\partial\lambda}-\beta_c\frac{\partial}{\partial c}+\widetilde{D}_{\mathsf{a}}\right]	\varGamma^{(2)}_{\mathsf{a}}(k, \lambda,w,c;\mu)=0,
	\end{split}
	\end{align}
	Here, $\mathsf{a}=\mathsf{b},\mathsf{f}$ labels the bosonic and fermionic two-point functions, respectively. 
	We write the RG equation in terms of 
$c$, $\lambda\equiv g^2/v$ and $w\equiv v/c$.
In particular, $\lambda$ controls the perturbative expansion in three dimensions\cite{SHOUVIK}.
	$\widetilde{D}_{\mathsf{a}}$ denotes the total scaling dimension of the two-point vertex functions,
	\begin{align}
	\widetilde{D}_{\mathsf{f}}&=2(\eta_\Psi+z)-3,\\
		\widetilde{D}_{\mathsf{b}} &=2(\eta_\Phi+z-2),
	\end{align}
	\noindent where the dynamical critical exponent and the anomalous dimensions of the fields are defined in Eq. (\ref{eq:z3d}) and (\ref{eq:eta3d}), respectively.
 Eq. (\ref{eq:RGThreeD}) can be rewritten as
	\begin{align}\label{eq:RGEqFinalThreeD}
	\left[\mathbf{K}\cdot\nabla_{\mathbf{K}}+\frac{\vec{k}}{z(l)}\cdot\nabla_{\vec{k}}+\frac{\dd}{\dd l}+\frac{\widetilde{D}_{\mathsf{a}}(l)}{z(l)}\right]	\varGamma^{(2)}_{\mathsf{a}}(k, \lambda(l),w(l),c(l))=0,
	\end{align}
	where the scale-dependent couplings obey
	\begin{align}
	\begin{split}\label{eq:BProbs}
	\frac{\dd w(l)}{\dd l} &= -\frac{\beta_w}{z(l)},\quad
	\frac{\dd\lambda (l)}{\dd l} = -\frac{\beta_\lambda}{z(l)},\quad \quad 
	\frac{\dd c(l)}{\dd l} = -\frac{\beta_c}{z(l)} \qquad \text{with}\qquad (w(0),\lambda(0),c(0))=(w_0,\lambda_0,c_0),
	\end{split}
	\end{align}
	and $l$ is the logarithmic length scale.
	The solution to Eq. (\ref{eq:RGEqFinalThreeD}) is given by
	\begin{align}\label{eq:GeneralSol3D}
	\varGamma^{(2)}_{\mathsf{a}}(\mathbf{K},\vec{k},\lambda_0,w_0,c_0) = \exp\left(\int^{l}_{0}\dd \ell \frac{\widetilde{D}_{\mathsf{a}}(\ell)}{z(\ell)}\right)\varGamma^{(2)}_{\mathsf{a}}\left(e^{l}\mathbf{K},\exp\left(\int\limits^{l}_{0}\frac{\dd\ell}{z(\ell)}\right)\vec{k},\lambda(l),w(l),c(l)\right).
	\end{align}
	The boundary problems in Eq. (\ref{eq:BProbs}) are solved by following the results of Ref. \cite{SHOUVIK},
	\begin{align}
	\lambda(l) &= \frac{4\pi(N^2_c-1+N_cN_f)}{N^2_c+N_cN_f-3}\frac{1}{l},\label{eq:lambdal}\\
	w(l) &= \frac{N_cN_f}{N^2_c-1}+\mathcal{O}\left(\frac{1}{\log(l)}\right),\label{eq:wl}\\
	c(l) &=\frac{\pi(N^2_c+N_cN_f-3)}{4(N^2_c-1+N_cN_f)}\frac{1}{\log l},\label{eq:cl}
	\end{align}
	\noindent in the large $l$ limit.

	The integrations over the length scale in Eq. (\ref{eq:GeneralSol3D}) are straightforward to perform in both the bosonic and fermionic cases after separating the contributions from the dynamical critical exponent and the net anomalous dimension of the fields. Setting $l=\log(\Lambda/|\mathbf{K}|)$ in Eq. (\ref{eq:GeneralSol3D}) for the fermion two-point function,
	we obtain the scaling form,
	\begin{align}\label{eq:F2P3D}
	\varGamma^{(2,0)}_{n}(\mathbf{K},\vec{k}) =\varGamma^{(2)}_{f}(\mathbf{K},\vec{k}) = {F}_{\Psi}(|\mathbf{K}|)\left(i{F}_{z}(|\mathbf{K}|)\boldsymbol{\Gamma}\cdot\mathbf{K}+i\gamma_2\varepsilon_{n}(\vec{k};v_{|\mathbf{K}|}) \right),
	\end{align}
where
	\begin{align}
	{F}_{z}(|\mathbf{K}|) &=\exp\left(\frac{(N^2_c+N_cN_f-1)}{2(N^2_c+N_cN_f-3)}\log(\log(\Lambda/|\mathbf{K}|))\right),\\
	{F}_{\Psi}(|\mathbf{K}|) &=\sqrt{\log(\log(\Lambda/|\mathbf{K}|))}.
	\end{align}
	Moreover, $v_{|\mathbf{K}|} = v \left(  
	\log(\Lambda/|\mathbf{K}|)
	\right)$
	with
	\begin{align}
	v(l) = w(l)c(l)\approx \frac{\pi
		N_c N_f
		(N^2_c+N_cN_f-3)}{4
		(N^2_c-1)
		(N^2_c+N_cN_f-1)}\frac{1}{\log l}
	\end{align}
	in the low-energy limit and Eq. (\ref{eq:F2P3D}) is obtained by keeping $\frac{\vec{k}}{(|\mathbf{K}|F_{z}(|\mathbf{K}|))}\sim 1$ fixed. \\

	%

	Similarly, the boson two-point function takes the form,
	\begin{align}\label{eq:2PB3D}
	\varGamma^{(0,2)}(\mathbf{Q},\vec{q}) =\varGamma^{(2)}_{b}(\mathbf{Q},\vec{q}) = {F}_{\Phi}(|\mathbf{Q}|)\left({F}_{z}(|\mathbf{Q}|)^2|\mathbf{Q}|^2+c_{|\mathbf{Q}|}^2|\vec{q}|^2\right),
	\end{align}
	\noindent where
	\begin{align}
	{F}_\Phi(|\mathbf{Q}|) = \log(\log(\Lambda/|\mathbf{Q}|)),
	\end{align}
	and $c_{|\mathbf{Q}|} 
	= c \left(
	\log(\Lambda/|\mathbf{Q}|)
	\right)$.  Eq. (\ref{eq:2PB3D}) is obtained by setting $l = \log(\Lambda/|\mathbf{Q}|)$ while  keeping $\frac{\vec{q}}{(|\mathbf{Q}|F_{z}(|\mathbf{Q}|))}\sim 1$ fixed and $c(l)$ is given by Eq. (\ref{eq:cl}).

	\section{\bf{Upper bound for higher-loop diagrams in $\boldsymbol{d<3}$}}\label{sec:SCALING}
	
Here we sketch the proof of the upper bound in Eq. (\ref{eq:UpperBound}).
Since the proof is essentially identical to the one given in Ref.  \cite{SCHLIEF}, 
here we only highlight the important steps without a full derivation.
We assume that the completely dressed boson propagator is given by  Eq. (\ref{eq:BosonPropagator}) in the limit in which the hierarchy of velocities, $v\ll c(v)\ll 1$, is satisfied. 
A general diagram with $L$ loops, $L_{f}$ fermion loops, $E$ external legs, and $V=2L-2+E$ vertices is given by
	\begin{align}\label{eq:DiagramMagnitude}
\mathcal{G}(L,L_f,E)\sim v^\frac{V}{2}
\int\prod^{L}_{r=1}\dd p_r\left[\prod^{I_f}_{l=1}
\frac{1}{\boldsymbol{\Gamma}\cdot\mathbf{K}_{l}+\gamma_{d-1}\varepsilon_{n_l}(\vec{k}_l;v)}
\right]\left[\prod^{I_b}_{s=1}\frac{1}{|\mathbf{Q}_s|^{d-1}+c(v)^{d-1}(|q_{s,x}|^{d-1}+|q_{s,y}|^{d-1})}\right].
\end{align}
	\noindent Here  $p_r=(\mathbf{P}_{r},\vec{p}_r)$ denotes the $(d+1)$-dimensional frequency and momentum that runs in the $r$-th loop.
	$k_{l}=(\mathbf{K}_{l},\vec{k}_l)$ ($q_{s}=(\mathbf{Q}_s,\vec{q}_s)$), 
	which is a linear combination of the loop momenta and external momenta,
	 denotes the frequency and momentum vector of the $l$-th fermion ($s$-th boson) propagator.  $I_f$ ($I_b$) denotes the number of internal fermion (boson) propagators and $n_l$ symbolizes the hot spot index for the $l$-th fermion propagator. 
	
	In the small $v$ limit, patches of the Fermi surface become locally nested, 
	and the AFM collective mode becomes dispersionless. 
	For a small but finite $v$, the integrations over internal fermion (boson) spatial momenta are cut off at momentum scales proportional to $1/v$ ($1/c(v)$).
	This gives rise to enhancement factors of $1/v$ ($1/c(v)$). 
	The enhancement of a diagram becomes maximal when the diagram contains only fermions belonging to patches of the Fermi surface that become locally nested in the small $v$ limit. 
	Because of this we consider Eq. (\ref{eq:DiagramMagnitude}) for $n_l=1,3$, without loss of generality.

	 Since the enhancement factor comes
	 from the integrations over the $x$ and $y$ components of the momenta,
	 we focus on the $2L$-dimensional integration over those components.
Through a change of variables of the $2L$ spatial loop momenta described in Ref. \cite{SCHLIEF},
Eq. (\ref{eq:DiagramMagnitude}) can be rewritten as
	\begin{align}\label{eq:Mag3}
\hspace{-0.5cm}\mathcal{G}(L,L_f,E)\sim v^{\frac{V}{2}-L_f}c(v)^{-(L-L_{f})}
\int\prod^{2L}_{r=1}\dd {p}'_i\left[\prod^{2L}_{l=L-L_{f}+1}
\frac{1}{\cdots+\gamma_{d-1}p'_l}
\right]\left[\prod^{L-L_{f}}_{s=1}\frac{1}{\cdots+|p'_{s}|^{d-1}+\mathcal{O}(c(v)^{d-1})}\right]{R}(p').
\end{align}
Here 
$p'_i$ denotes the new $2L$ variables for the $x$ and $y$ components of the internal momenta. The ellipsis denote the frequency and co-dimensional momenta that play no role in determining the enhancement factor.
${R}(p')$ denotes the product of all the remaining propagators. 
The point of the change of basis is to make it manifest that there is at least one propagator
that guarantees that the integrand decays in the UV at least as $1/p'_i$ in each of the internal momenta
once a factor of $v^{-1}$ or $c(v)^{-1}$ is scaled out from each loop.
Each fermion loop contributes 
 $v^{-1}$ because the $x$-momentum component becomes unbounded
in the small $v$ limit.  Each of the remaining ${L-L_{f}}$ loops contribute 
 a factor of $c(v)^{-1}$ because the $x$-momentum  component in the loop necessarily runs through a boson propagator and is cut off at a scale proportional to $1/c(v)$, since $c(v)\gg v$.
%
It follows from this that the magnitude of a generic $L$-loop diagram with $L_{f}$ fermionic loops is at most
\begin{align}\label{eq:VScaling}
\mathcal{G}(L,L_f,E)\sim v^\frac{E-2}{2}\left(\frac{v}{c(v)}\right)^{L-L_{f}}
\end{align}
up to a potential logarithmic correction in the small $v$ limit. 
We note that Eq. (\ref{eq:VScaling}) is independent of the space dimension because the fully dressed boson propagator in Eq. (\ref{eq:BosonPropagator}) depends on $q_x$ and $q_y$ only through ${c(v)\vec{q}}$ and the velocities along the extra co-dimensions are fixed to be one. 
	
	\section{\bf{Regularization and RG Scheme}}\label{Sec:RG}
	
Here we briefly explain the RG scheme used in $d<3$.
The main difference from the case with $d=3$ is that 
we start with the interaction-driven scaling in $d<3$.
As a result, the minimal action only includes the fermion kinetic term
and the Yukawa interaction.
Quantum corrections are computed with the self-consistent boson propagator in Eq. (\ref{eq:BosonPropagator}).
Under the  interaction-driven scaling, the Yukawa vertex is marginal in any dimension between two and three. 
As a result, we expect logarithmic divergences in this dimensional range. 
We regularize the theory with the same prescription as the one given in Sec. \ref{Sec:RGPrescription3d} of Appendix 	\ref{sec:PhysicalObservablesThreeD} and follow a similar RG scheme.
We add the following local counter term to the action in Eq. (\ref{eq:MinimalLocalAction}) such that low-energy physical observables are  independent of the UV cutoff scales:
	\begin{align}\label{eq:CT}
		\begin{split}
		S^{\mathrm{C.T.}}_{d} &= \sum^{4}_{n=1}\sum^{N_c}_{\sigma=1}\sum^{N_f}_{j=1}\int\dd k\overline{\Psi}_{n,\sigma,j}(k)\left(iA_1\boldsymbol{\Gamma}\cdot\mathbf{K}+i\gamma_{d-1}\widetilde{\varepsilon}_{n}(\vec{k};v)\right)\Psi_{n,\sigma,j}(k)\\
		&+A_6\frac{i\beta_d\sqrt{v}}{\sqrt{N_f}}\sum^{4}_{n=1}\sum^{N_c}_{\sigma,\sigma'=1}\sum^{N_f}_{j=1}\int\dd k\int\dd q\overline{\Psi}_{\overline{n},\sigma,j}(k+q){\Phi}_{\sigma\sigma'}(q)\gamma_{d-1}\Psi_{n,\sigma',j}(k).
		\end{split} 
		\end{align}
Here,
		 $\widetilde{\varepsilon}_{1}(\vec{k};v) = A_2vk_x+A_3k_y$, 
		 $\widetilde{\varepsilon}_{2}(\vec{k};v) = -A_3 k_x+A_2vk_y$,  
		 $\widetilde{\varepsilon}_{3}(\vec{k};v) =A_2 vk_x-A_3k_y$ ,
		 and $\widetilde{\varepsilon}_{4}(\vec{k};v) = A_3 k_x + A_2vk_y$. The $A_i's$ are momentum-independent counter term coefficients.
 Adding this counter term action to Eq. (\ref{eq:MinimalLocalAction}) yields the renormalized action,
	\begin{align}
		\begin{split}
		S^{\mathrm{Ren}}_{d} &= \sum^{4}_{n=1}\sum^{N_c}_{\sigma=1}\sum^{N_f}_{j=1}\int\dd k_{B}\overline{\Psi}_{n,\sigma,j;B}(k_B)\left(i\boldsymbol{\Gamma}\cdot\mathbf{K}_{B}+i\gamma_{d-1}{\varepsilon}_{n}(\vec{k}_{B};v_B)\right)\Psi_{n,\sigma,j,B}(k_B)\\
		&+\frac{i\beta_d\sqrt{v_B}}{\sqrt{N_f}}\sum^{4}_{n=1}\sum^{N_c}_{\sigma,\sigma'=1}\sum^{N_f}_{j=1}\int\dd k_{B}\int \dd q_{B}\overline{\Psi}_{\overline{n},\sigma,j;B}(k_{B}+q_{B}){\Phi}_{B;\sigma\sigma'}(q_{B})\gamma_{d-1}\Psi_{n,\sigma',j;B}(k_{B}).
		\end{split} 
		\end{align}
The renormalized frequency, momenta, fields and velocity are related to the bare ones via the multiplicative relations:
	\begin{align}\label{eq:RenormalizedBare}
	\begin{split}
	\mathbf{K}_{B} = \frac{Z_1}{Z_3}\mathbf{K},\quad   \vec{k}_{B}=\vec{k},\quad v_{B} = \frac{Z_2}{Z_3}v,\quad 
	\Psi_{B}(k_{B}) = Z^\frac{1}{2}_{\Psi}\Psi(k),\quad \& \quad  \Phi_{B}(k_{B})=Z^\frac{1}{2}_{\Phi}\Phi(k),\\
	\end{split}
	\end{align}
	\noindent where $Z_i=1+A_i$, $Z_{\Psi}=Z_3(Z_3/Z_1)^{d-1}$, $Z_\Phi=\frac{Z^2_6}{Z_3Z_2}(Z_3/Z_1)^{2(d-1)}$ and the field indices are suppressed.
	It is noted that the expression for $Z_\Phi$ is different from the one used in $d=3$ 
	because here we are using the interaction-driven scaling.
	 The renormalized action gives rise to the quantum effective action in Eq. (\ref{eq:QEA1}). However, the dependences on $v,c,g,u_1$ and $u_2$ of the latter and the 1PI vertex functions are now replaced by a single parameter: $v$. The counter term coefficients  in Eq. (\ref{eq:CT}) are fixed by the renormalization conditions imposed over the vertex functions:
			\begin{align}
		\lim_{\widetilde{\Lambda}\rightarrow \infty}\lim_{\Lambda\rightarrow \infty}	\frac{1}{2i}\frac{\partial}{\partial\mathbf{K}^2}\mathrm{Tr}\left[(\mathbf{K}\cdot\boldsymbol{\Gamma})\varGamma^{(2,0)}_{n}(k)\right]\bigg|_{|\mathbf{K}|=\mu,\vec{k}=0}&=1+F_1(v),\qquad n=1,2,3,4,\label{eq:RGConditionFermionFrequency1}\\
		\lim_{\widetilde{\Lambda}\rightarrow \infty}\lim_{\Lambda\rightarrow \infty}\frac{1}{2i}\frac{\partial}{\partial k_x}\mathrm{Tr}\left[\gamma_{d-1}\varGamma^{(2,0)}_{n=1}(k)\right]\bigg|_{|\mathbf{K}|=0,k_x=\mu,k_y=0}&=v(1+F_2(v)),\label{eq:RGConditionKx}\\
		\lim_{\widetilde{\Lambda}\rightarrow \infty}\lim_{\Lambda\rightarrow \infty}\frac{1}{2i}\frac{\partial}{\partial k_y}\mathrm{Tr}\left[\gamma_{d-1}\varGamma^{(2,0)}_{n=1}(k)\right]\bigg|_{|\mathbf{K}|=0,k_x=0,k_y=\mu}&=1+F_3(v),\label{eq:RGConditionKy}\\
		\lim_{\widetilde{\Lambda}\rightarrow \infty}\lim_{\Lambda\rightarrow \infty}\frac{1}{2}\mathrm{Tr}\left[\gamma_{d-1}\varGamma^{(2,1)}_{n}(k,q)\right]\bigg|_{q=0,|\mathbf{K}|=\mu,\vec{k}=0}&=1+F_4(v),\qquad n=1,2,3,4\label{eq:RGConditionVertex},
		\end{align}
	\noindent which follow from a minimal subtraction scheme. Here, we have left implicit the dependence of the vertex function on $v$.
 $\mu$ is an energy scale at which the physical observables are measured and $F_i(v)$ are functions that vanish in the small $v$ limit. 

	Since the bare quantities are independent of the running energy scale $\mu$, 
	the 1PI vertex functions obey the RG equation:
	   	\begin{align}\label{eq:RGEquation}
	\begin{split}
	\left[
	\sum^{2m+n-1}_{i=1}\left(z\mathbf{K}_{i}\cdot\nabla_{\mathbf{K}_i}+\vec{k}_i\cdot\nabla_{\vec{k}_i}\right)-\beta_v\frac{\partial}{\partial v}+m\left(2\eta_\Psi-(d+2)\right)+n\left(\eta_{\Phi}-d\right)\right.
	\\
\left. \frac{}{}	 +(2m+n-1)(2+z(d-1))
	\right]	\varGamma^{(2m,n)}(\{ k_i\}, v;\mu)=0,
	\end{split}
	\end{align}
	\noindent which is obtained  by combining the fact that, under the interaction-driven scaling, the vertex functions have engineering scaling dimension $[\varGamma^{(2m,n)}(\{k_i\},v;\mu)]=-md-n+d+1$ and that the bare vertex functions are related to the renormalized ones via
	  	\begin{align}
	  	\varGamma^{(2m,n)}_{B}[\{k_{i,B},v_{B};\Lambda,\widetilde{\Lambda}\}] =\left(\frac{Z_3}{Z_1}\right)^{(d-1)(2m+n-1)}Z^{-m}_\Psi Z^{-\frac{n}{2}}_\Phi\varGamma^{(2m,n)}(\{k_i\},v;\mu).
	  	\end{align} 
The dynamical critical exponent, the beta function for $v$, and the anomalous scaling dimensions of the fields are given by
	  \begin{align}
	  z&=1-\frac{\dd}{\dd\log\mu}\log\left(\frac{Z_3}{Z_1}\right),\label{eq:z1}\\
	  \beta_v&= \frac{\dd v}{\dd\log\mu},\label{eq:Betav}\\
	  \eta_{\Psi(\Phi)} &=\frac{1}{2}\frac{\dd\log Z_{\Psi(\Phi)}}{\dd\log\mu}\label{eq:etapsiphi},
	  \end{align} 
	  \noindent respectively.
Here $\eta_\Psi$ and $\eta_\Phi$ denote the deviations of the scaling dimensions of the fields
from the ones predicted by the interaction-driven scaling
(not the Gaussian scaling).

	\section{\bf{Quantum Corrections }}\label{Sec:Quantum}

	\noindent Here we provide details on the computations of the quantum corrections to the minimal local action depicted in {\color{blue}Figs.} \ref{fig:SDDiagramA}, \ref{fig:SDDiagramB}, \ref{fig:SDDiagramD}, \ref{fig:FSE1}, \ref{fig:YUK} and \ref{fig:FSE2}.


	\subsection{One-loop boson self-energy}\label{sec:ONELOOPBOSON}

\noindent The one-loop correction that generates dynamics of the boson is shown in {\color{blue}Fig.} \ref{fig:SDDiagramA}. Its contribution to the  quantum effective action reads
\begin{align}
\delta \boldsymbol{\Gamma}^{(0,2)}_{\mathrm{1L}} = \frac{1}{4}\int\dd q\Pi^{\mathrm{1L}}(q)\mathrm{Tr}\left[\Phi(-q)\Phi(q)\right],
\end{align}

\noindent where the one-loop boson self-energy is given by
	\begin{align}\label{eq:SEBoson}
	\Pi^{\mathrm{1L}}(q) = -2v\beta^2_{d}\sum^{4}_{n=1}\int\dd k\mathrm{Tr}\left[\gamma_{d-1}G^{(0)}_{\overline{n}}(k+q)\gamma_{d-1} G^{(0)}_{n}(k)\right].
	\end{align}
	
	\noindent Here $G^{(0)}_{n}(k)$ is the bare fermion propagator given in Eq. (\ref{eq:BareFermionPropagator}) and $\beta_d$ is defined in Eq. (\ref{eq:Betad}).
	Taking the trace over the spinor indices and integrating over the spatial momenta $\vec{k}$, yields
	\begin{align}
	\Pi^{\mathrm{1L}}(q) = -2\beta^2_d\int\limits_{\mathbb{R}^{d-1}}\frac{\dd \mathbf{K}}{(2\pi)^{d-1}}\frac{\mathbf{K}\cdot(\mathbf{K}+\mathbf{Q})}{|\mathbf{K}||\mathbf{K}+\mathbf{Q}|}.
	\end{align}
	\noindent 
	Subtracting the mass renormalization, we focus on the momentum dependent self-energy : 
	$\Delta\Pi^{\mathrm{1L}}(q) = \Pi^{\mathrm{1L}}(q)-\Pi^{\mathrm{1L}}(0)$. 
	Integration over $\mathbf{K}$ is done after imposing a cutoff $\Lambda$ in the UV. 
	In the $\Lambda/|\mathbf{Q}|\gg1$ limit this becomes
	\begin{align}
	\Delta\Pi^{\mathrm{1L}}(q) =  \frac{\beta^2_d\Gamma\left(\frac{5-d}{2}\right)\Gamma\left(\frac{d}{2}\right)}{2^{2d-5}\pi^\frac{d}{2}\Gamma\left(\frac{d+1}{2}\right)}|\mathbf{Q}|^{d-1}\left(\frac{1}{3-d}-\frac{1}{3-d}\left[\frac{2\cos\left(\frac{\pi d}{2}\right)}{\pi (d-3)}\right]\left(\frac{\Lambda}{|\mathbf{Q}|}\right)^{d-3}\right).
	\end{align}
	\noindent  
	While the expression is logarithmically divergent in $d=3$, it is UV finite for $d<3$.
In $d < 3$, the one-loop boson self-energy is given by 
	\begin{align}
	\Delta \Pi^{\mathrm{1L}}(q) = |\mathbf{Q}|^{d-1}.
	\end{align}

	\subsection{Two-loop boson self-energy}\label{sec:TWOLOOPBOSON}
	
We first compute the two-loop boson self-energy shown in {\color{blue}Fig.} \ref{fig:SDDiagramB}, and
then  comment on the contribution arising from {\color{blue}Fig.} \ref{fig:SDDiagramD}. The contribution of {\color{blue}Fig.} \ref{fig:SDDiagramB}  to the quantum effective action is given by 
		\begin{align}
	\delta\boldsymbol{\Gamma}^{(0,2)}_{\mathrm{2L}} = \frac{1}{4}\int\dd q \Pi^{\mathrm{2L}}(q)\mathrm{Tr}\left[\Phi(q)\Phi(-q)\right]
		\end{align}
		with
		\begin{align}\label{eq:QuantumCorrection}
		\Pi^{\mathrm{2L}}(q) = -\frac{4\beta^4_{d}v^2}{N_cN_f}\sum^{4}_{n=1}\int\dd k\int\dd p\mathrm{Tr}\left[\gamma_{d-1}G^{(0)}_{\overline{n}}(k+p)\gamma_{d-1}G^{(0)}_{n}(k+q+p)\gamma_{d-1}G^{(0)}_{\overline{n}}(k+q)\gamma_{d-1}G^{(0)}_{n}(k)\right]D(p).
		\end{align}
		\noindent 
		Here $\beta_d$ is defined in Eq. (\ref{eq:Betad}) and $D(p)$ is given by the self-consistent propagator in Eq. (\ref{eq:BosonPropagator}). 
		The frequency-dependent part of the two-loop self-energy is subleading with respect to the one-loop boson self-energy by a factor of $w(v)=v/c(v)$.
		Therefore, we focus on the momentum dependent part by setting $\mathbf{Q}=\mathbf{0}$. Taking the trace over the spinor indices, changing variables to $k_{+}=\varepsilon_{n}(\vec{k};v)$ and $k_{-}=\varepsilon_{\overline{n}}(\vec{k}+\vec{q};v)$, and noting that the latter has a Jacobian of $1/(2v)$, the spatial part of the two-loop boson self-energy takes the form,
	\begin{align}
	\Pi^{\mathrm{2L}}(\mathbf{0},\vec{q}) &= -\frac{4v\beta^4_{d}}{N_cN_f}\sum^{4}_{n=1}\int\dd k \int\dd p\left\{\frac{1}{(\mathbf{K}^2+k^2_{+})(\mathbf{K}^2+k^2_{-})((\mathbf{K}+\mathbf{P})^2+(k_{+}+\varepsilon_{n}(\vec{p}+\vec{q};v))^2)}\right. \\
	&\left.\times\frac{1}{((\mathbf{K}+\mathbf{P})^2+(k_{-}+\varepsilon_{\overline{n}}(\vec{p}-\vec{q};v))^2)}\left[\left(\mathbf{K}^2-k_{+}k_{-}\right)\left((\mathbf{K}+\mathbf{P})^2-(k_{+}+\varepsilon_{n}(\vec{p}+\vec{q};v))(k_{-}+\varepsilon_{\overline{n}}(\vec{p}-\vec{q};v))\right)\right.\right.\notag\\
	&\left.\left.-\mathbf{K}\cdot(\mathbf{K}+\mathbf{P})(k_{+}+k_{-}+\varepsilon_{n}(\vec{p}+\vec{q};v)+\varepsilon_{\overline{n}}(\vec{p}-\vec{q};v))(k_{+}+k_{-}) \right]\right\}D(p).\notag
	\end{align}
					\noindent This expression can be written as a sum of the contributions from the four hot spots,
		\begin{align}\label{eq:ALLHOTS}
		\Pi^{\mathrm{2L}}(\mathbf{0},\vec{q}) =\sum^{4}_{n=1}\Pi^{\mathrm{2L}}_{n}(\vec{q}).
		\end{align}
		\noindent Let us first  consider the contribution from the $n=1$ hot spot.
		Since the self-energy depends on the external momentum component $q_x$ only through $vq_x$, the first hot spot gives rise to the self-energy that depends on $q_y$ to the leading order in the small $v$ limit.
	After setting $q_x=0$, we perform a change of variables $p_x\rightarrow p_x/v$ to write the  the two-loop boson self-energy as
		
		\begin{align}\label{eq:PrePX}
		\Pi^{\mathrm{2L}}_{1}(\vec{q}) &= -\frac{4w(v)^{d-1}\beta^4_{d}}{N_cN_f}\int\dd k \int\dd p\left\{\frac{1}{(\mathbf{K}^2+k^2_{+})(\mathbf{K}^2+k^2_{-})((\mathbf{K}+\mathbf{P})^2+(k_{+}+p_x+p_y+q_y)^2)}\right.\\
		&\left. \times \frac{1}{((\mathbf{K}+\mathbf{P})^2+(k_{-}+p_x-p_y+q_y)^2)}\left[\left(\mathbf{K}^2-k_{+}k_{-}\right)\left((\mathbf{K}+\mathbf{P})^2-(k_{+}+p_x+p_y+q_y)(k_{-}+p_x-p_y+q_y)\right)\right.\right.\notag\\
		&\left.\left. -\mathbf{K}\cdot(\mathbf{K}+\mathbf{P})(k_{+}+k_{-}+2p_x+2q_y)(k_{+}+k_{-})\right]\left(\frac{1}{w(v)^{d-1}|\mathbf{P}|^{d-1}+|p_x|^{d-1}+v^{d-1}|p_y|^{d-1}}\right)\right\}.\notag
		\end{align}
		
		\noindent  
		We can neglect $|vp_y|^{d-1}$ in the boson propagator in the small $v$ limit.
The integration over $p_x$ is divided into two regimes: 
$p_x\in(-\lambda,\lambda)$ and $p_x\in \mathbb{R}\setminus(-\lambda,\lambda)$ where 
$\lambda \sim \min(k_{+},k_{-},\mathbf{P},\mathbf{K},p_y)$ is a momentum scale below which the $p_x$ dependence in the fermion propagators can be ignored. 
The exact form of $\lambda$ is unimportant in the small $w(v)$ limit.
		  The integration over the first regime is divergent in the small $w(v)$ limit
		  due to the infrared singularity that is cut off by $w(v) |\mathbf{P}|$.
On the other hand, the contribution from the second regime is regular. 
To the leading order in $w(v)\ll 1$, we can keep only the first contribution to write the $p_x$ integration as
		\begin{align}\label{eq:IntPX}
	|\mathbf{P}|^{2-d}	\overline{\textswab{S}}\left(d-2;w(v);\frac{\lambda}{|\mathbf{P}|}\right) \equiv \frac{\pi }{(d-2)}\frac{1}{\Gamma\left(\frac{d-2}{d-1}\right)\Gamma\left(\frac{d}{d-1}\right)}\int\limits^{\lambda}_{-\lambda}\frac{\dd p_x}{(2\pi)} \left(\frac{w(v)^{d-2}}{w(v)^{d-1}|\mathbf{P}|^{d-1}+|p_x|^{d-1}}\right).
		\end{align} 
		\noindent In the $w(v)\rightarrow0$ and in the $d\rightarrow 2$ limits,  $\overline{\textswab{S}}(d-2;w(v);\lambda/|\mathbf{P}|)$ becomes independent of $\lambda/|\mathbf{P}|$ because it  has the following 
		limiting behaviors:
			\begin{align}\label{eq:LimitsSchlief}
			\lim_{d \rightarrow 2}\overline{\textswab{S}}\left(d-2;w(v);\frac{\lambda}{|\mathbf{P}|}\right) = -\log\left(w(v)\right), \qquad  \lim_{w(v)\rightarrow 0}\overline{\textswab{S}}\left(d-2;w(v);\frac{\lambda}{|\mathbf{P}|}\right) = \frac{1}{d-2}.
		\end{align} 
		\noindent Since we are mainly interested in these limits, we can replace $\overline{\textswab{S}}(d-2;w(v);\lambda/|\mathbf{P}|)$ with
		\begin{align}\label{eq:SFunction}
	\textswab{S}(d-2;w(v)) \equiv	\overline{\textswab{S}}(d-2;w(v);1) = \frac{1}{d-2}\left[1-w(v)^{d-2}\right],
		\end{align}
		\noindent where the last equality comes from explicitly computing Eq. (\ref{eq:IntPX}) 
at $\lambda/|\mathbf{P}|=1$ in the small $w(v)$ limit.
The $p_x$, $p_y$ and $k_{+}$ integrations in Eq. (\ref{eq:PrePX}) result in
		\begin{align}
		\begin{split}
		\Pi^{\mathrm{2L}}_{1}(\vec{q}) &= -\frac{4(d-2)\beta^4_{d}w(v)}{\pi N_cN_f}{\Gamma\left(\frac{d-2}{d-1}\right)}\Gamma\left(\frac{d}{d-1}\right)\textswab{S}(d-2;w(v))\int\limits_{\mathbb{R}^{d-1}}\frac{\dd \mathbf{K}}{(2\pi)^{d-1}}\int\limits_{\mathbb{R}}\frac{\dd k_{-}}{(2\pi)} \int\limits_{\mathbb{R}^{d-1}} \frac{\dd\mathbf{P}}{(2\pi)^{d-1}}\\
		&\times\frac{|\mathbf{P}|^{2-d}}{|\mathbf{K}||\mathbf{K}+\mathbf{P}|}\left[\frac{4\mathbf{K}^2(\mathbf{K}+\mathbf{P})^2-2q_y k_{-}\mathbf{K}\cdot(\mathbf{K}+\mathbf{P})-\mathbf{K}\cdot(\mathbf{K}+\mathbf{P})k^2_{-}}{(4\mathbf{K}^2+k^2_{-})(4(\mathbf{K}+\mathbf{P})^2+(k_{-}+2q_y)^2)}\right],
		\end{split}
		\end{align}
		
\noindent to leading order in $v\ll 1$.	Subtracting the mass renormalization, 
	the momentum dependent self-energy (defined as $\Delta\Pi^{\mathrm{2L}}_{1}(\vec{q})\equiv \Pi^{\mathrm{2L}}_{1}(\vec{q})-\Pi^{\mathrm{2L}}_{1}(\vec{0})$) is obtained to be
		\begin{align}
		\begin{split}
		\Delta\Pi^{\mathrm{2L}}_{1}(\vec{q}) &= -\frac{4(d-2)\beta^4_{d}w(v)}{\pi N_cN_f}{\Gamma\left(\frac{d-2}{d-1}\right)}\Gamma\left(\frac{d}{d-1}\right)\textswab{S}(d-2;w(v))\int\limits_{\mathbb{R}^{d-1}}\frac{\dd \mathbf{K}}{(2\pi)^{d-1}}\int\limits_{\mathbb{R}}\frac{\dd k_{-}}{(2\pi)} \int\limits_{\mathbb{R}^{d-1}} \frac{\dd\mathbf{P}}{(2\pi)^{d-1}}\\
		&\times\frac{|\mathbf{P}|^{2-d}}{|\mathbf{K}||\mathbf{K}+\mathbf{P}|} \left[\frac{\mathcal{F}(\mathbf{P},\mathbf{K},k_{-},q_y) }{(4\mathbf{K}^2+k^2_{-})(4(\mathbf{K}+\mathbf{P})^2+(k_{-}+2q_y)^2)(4(\mathbf{K}+\mathbf{P})^2+k^2_{-})}\right],
		\end{split}
		\end{align}		
		\noindent where 
		\begin{align}
		\begin{split}
		\mathcal{F}(\mathbf{P},\mathbf{K},k_{-},q_y) &= \left[4\mathbf{K}^2(\mathbf{K}+\mathbf{P})^2-2q_y k_{-}\mathbf{K}\cdot(\mathbf{K}+\mathbf{P})-\mathbf{K}\cdot(\mathbf{K}+\mathbf{P})k^2_{-}\right](4(\mathbf{K}+\mathbf{P})^2+k^2_{-})\\
		&-\left[4\mathbf{K}^2(\mathbf{K}+\mathbf{P})^2-\mathbf{K}\cdot(\mathbf{K}+\mathbf{P})k^2_{-}\right](4(\mathbf{K}+\mathbf{P})^2+(k_{-}+2q_y)^2).
		\end{split}
		\end{align}

		\begin{figure*}
			\centering
			
			\begin{subfigure}{0.45\linewidth}
				\includegraphics[scale=0.625]{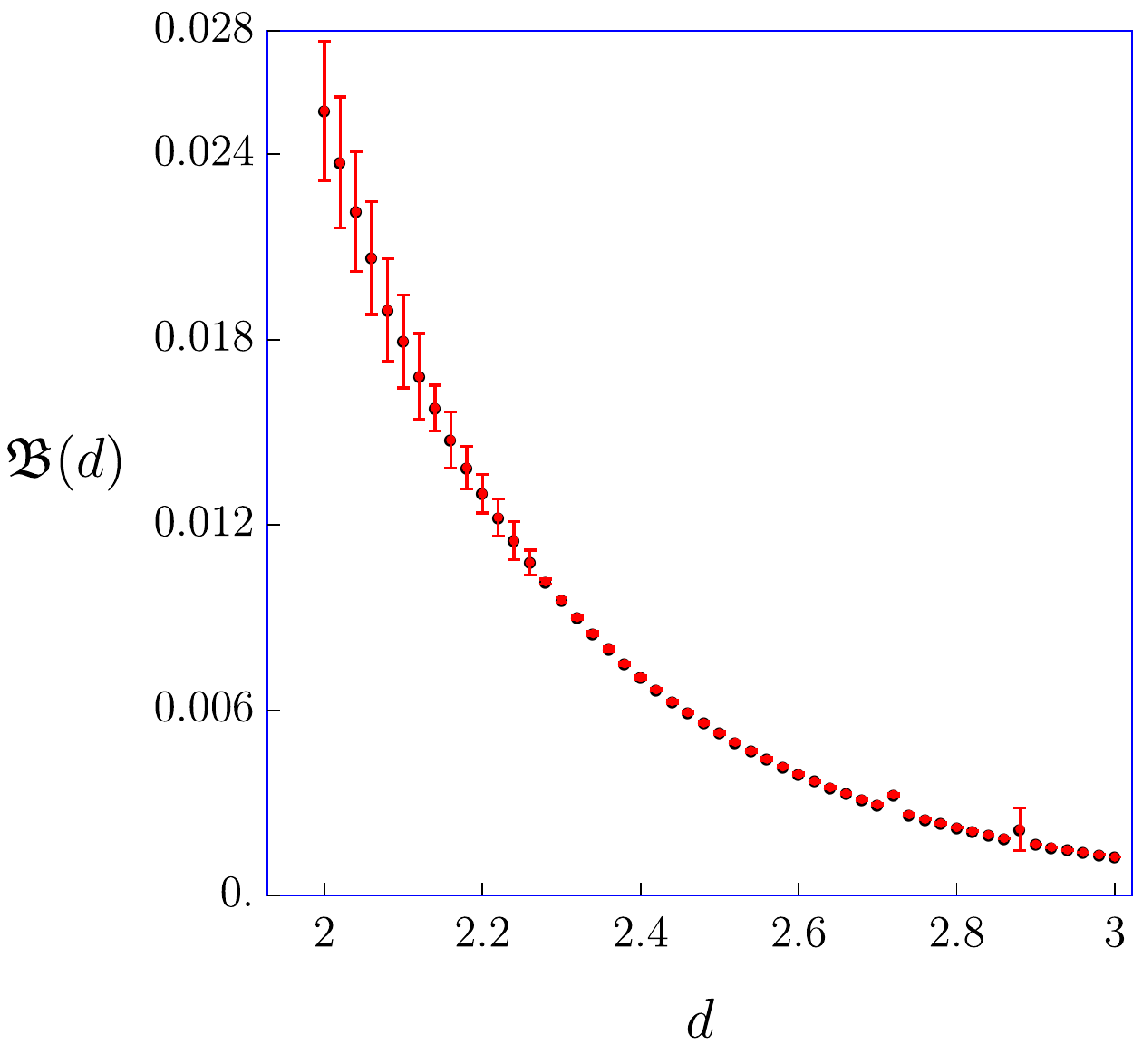}
				\caption{\hspace{0.5em}  $\mathfrak{B}(d)$ for $2\leq d\leq 3$.}
			\end{subfigure}
			\begin{subfigure}{0.45\linewidth}
				\includegraphics[scale=0.6]{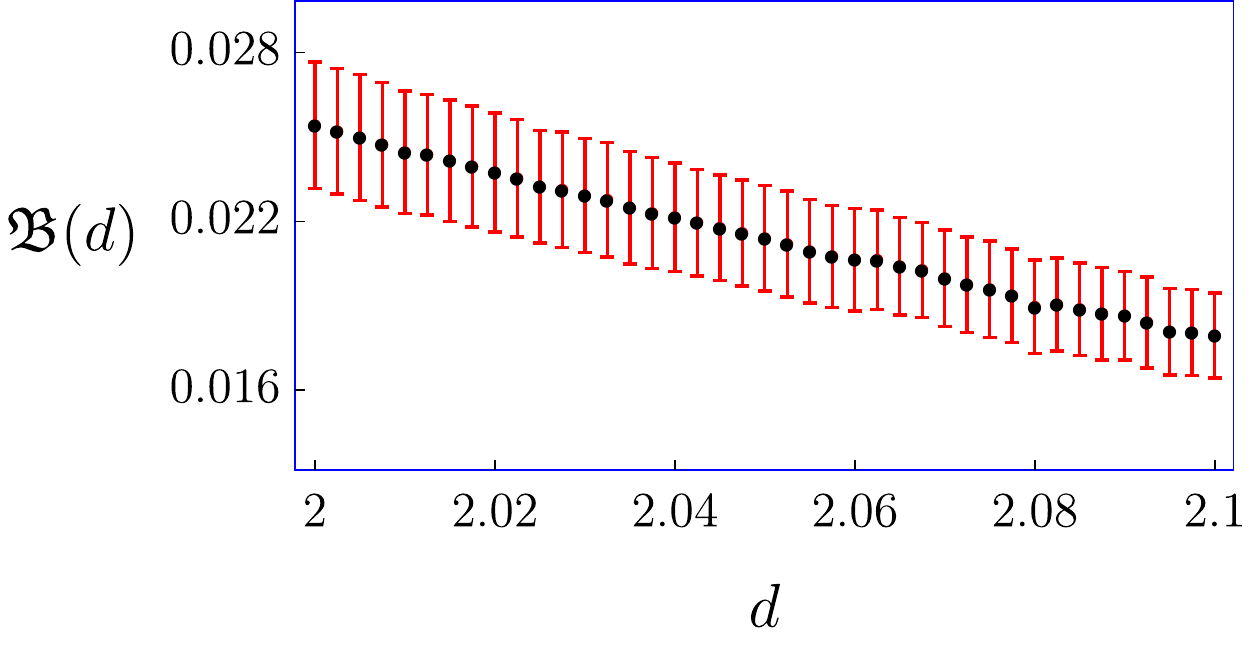}\\\vfill
				\includegraphics[scale=0.6]{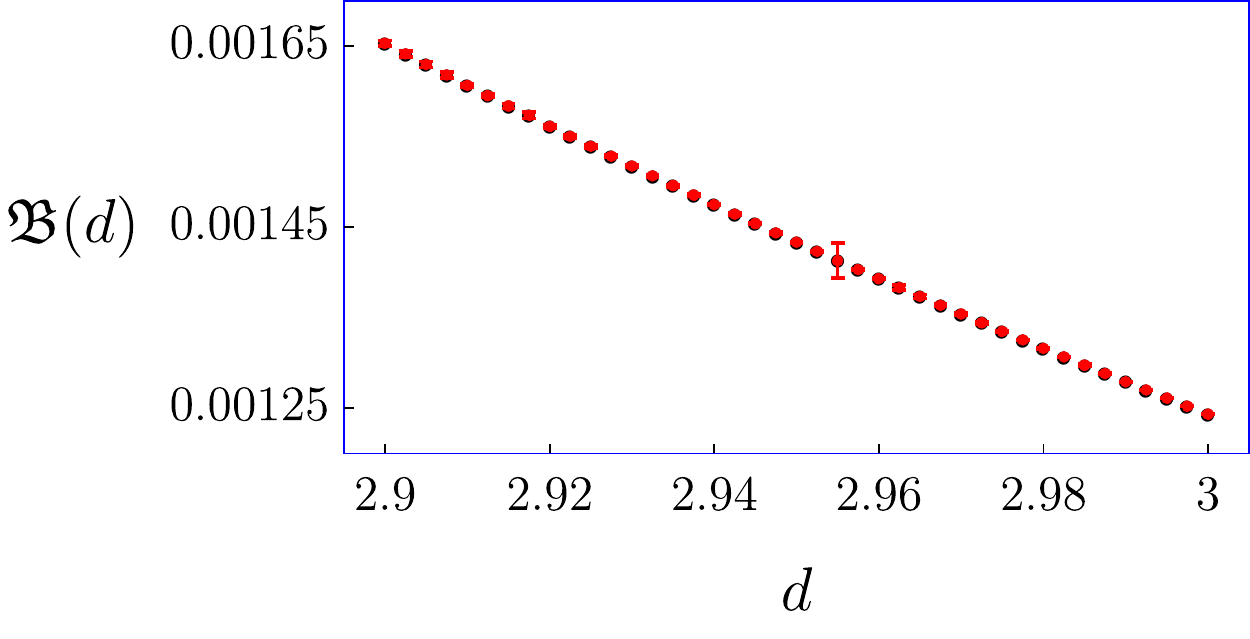}
				\caption{\hspace{0.5em}  $\mathfrak{B}(d)$  near $d=2$ (top) and $d=3$ (bottom).}
			\end{subfigure}	
			\caption{{({\color{blue}$a$})} The function $\mathfrak{B}(d)$.  The (black) dots correspond to the value of the numerical integration and the (red) error bars represent the numerical error in the computation. ({\color{blue}$b$}) Numerical evaluation near $d=2$ (top) and $d=3$ (bottom).\label{fig:B}}
		\end{figure*}

		\noindent We  proceed by scaling out $q_y$ from the above integral and introduce a two-variable Feynman parametrization that allows the explicit computation of the $k_{-}$ integration. 
Performing this integration yields
		\begin{align}\label{eq:Provi}
		\begin{split}
		\Delta\Pi^{\mathrm{2L}}_{1}(\vec{q}) &= -\frac{\beta^4_{d}(d-2)|q_y|^{d-1}w(v)}{64\pi N_cN_f}{\Gamma\left(\frac{d-2}{d-1}\right)}\Gamma\left(\frac{d}{d-1}\right)\textswab{S}(d-2;w(v))\int\limits_{\mathbb{R}^{d-1}}\frac{\dd \mathbf{K}}{(2\pi)^{d-1}} \int\limits_{\mathbb{R}^{d-1}} \frac{\dd\mathbf{P}}{(2\pi)^{d-1}}\\
		&\times\int\limits^{1}_{0}\dd x_1\int\limits^{1-x_1}_{0}\dd x_2 \frac{|\mathbf{P}|^{2-d}}{|\mathbf{K}||\mathbf{K}+\mathbf{P}|}\left[\frac{3A+4B\left[ \mathbf{K}\cdot(\mathbf{K}+\mathbf{P}) \left(3x_1+3x_2-2\right)\right]}{\left(\mathbf{K}^2-2 \mathbf{K}\cdot\mathbf{P} \left(x_1-1\right)-\mathbf{P}^2 \left(x_1-1\right)-\left(x_1+x_2\right){}^2+x_1+x_2\right)^\frac{5}{2}}\right],
		\end{split}
		\end{align}
		\noindent where
		\begin{align}
		\begin{split}
		A&=-16 \left(\mathbf{K}^2 \left(2 x_1+2 x_2-1\right) (\mathbf{K}+\mathbf{P})^2+\left(x_1+x_2-1\right) \mathbf{K}\cdot(\mathbf{K}+\mathbf{P}) \left((\mathbf{K}+\mathbf{P})^2-\left(x_1+x_2-1\right) \left(x_1+x_2\right)\right)\right),
		\end{split}\\
		\begin{split}
		B&=4 \left(\mathbf{K}^2-2 \mathbf{K}\cdot\mathbf{P} \left(x_1-1\right)-\mathbf{P}^2 \left(x_1-1\right)-\left(x_1+x_2\right){}^2+x_1+x_2\right).
		\end{split}
		\end{align}
Integrations over the remaining frequency and co-dimensional momentum components are done by introducing another two-variable Feynman parametrization. This yields the contribution from the $n=1$ hot spot to the two-loop boson self energy
	\begin{align}
\Delta\Pi^{\mathrm{2L}}_{1}(\vec{q}) &= \frac{2\beta^4_{d}|q_y|^{d-1}}{ N_cN_f}w(v)\mathfrak{B}(d)\textswab{S}(d-2;w(v))\widetilde{\textswab{S}}\left(3-d;\frac{|q_y|}{\Lambda}\right),
\end{align}
\noindent where
$ \mathfrak{B}(d) $ is a smooth function of $d$  (see {\color{blue}Fig.} \ref{fig:B})  defined by
\begin{align}
\begin{split}\label{eq:DefinitionofB}
\mathfrak{B}(d) &=\frac{ (d-2)\Gamma(3-d)}{3\times 2^{d+4}\pi^\frac{2d+3}{2}}\frac{\cos\left(\frac{\pi d}{2}\right)\Gamma\left(\frac{d-2}{d-1}\right)\Gamma\left(\frac{d}{d-1}\right)\Gamma\left(\frac{5-d}{2}\right)}{\Gamma\left(\frac{8-d}{2}\right)}\int\limits^{1}_{0}\dd x_1\int\limits^{1-x_1}_{0}\dd x_2\int\limits^{1}_{0}\dd y_1\int\limits^{1-y_1}_{0}\dd y_2\\
&\times \left[ \frac{(1-y_1-y_2)^\frac{3}{2}}{\sqrt{y_1}\sqrt{y_2}}\frac{(d-5) (d-3) C_3 D_1^2+D_2 \left((3-d) C_2 D_1+3 C_1 D_2\right)}{D^\frac{5}{2}_1D^\frac{7-d}{2}_2}\right],
\end{split}
\end{align}
\noindent with 
		\begin{align}
		\begin{split}\label{eq:D1}
		D_1&= -\left(x_1 \left(y_1+y_2-1\right)-y_1\right) \left(x_1 \left(y_1+y_2-1\right)-y_1+1\right),
		\end{split}\\
		\begin{split}\label{eq:D2}
		D_2&=\left(x_1+x_2-1\right) \left(x_1+x_2\right) \left(y_1+y_2-1\right),
		\end{split}\\
		\begin{split}\label{eq:C1}
		C_1 &=\left[-x_1^2 \left(y_1+y_2-1\right){}^2+x_1 \left(2 y_1-1\right) \left(y_1+y_2-1\right)-\left(y_1-1\right) y_1\right] \left\{\left(d^2-1\right) \left(-6 x_1-6 x_2+4\right) \right.\\
		&\times\left.  \left(-x_1^2 \left(y_1+y_2-1\right){}^2+x_1 \left(2
		y_1-1\right) \left(y_1+y_2-1\right)-\left(y_1-1\right) y_1\right)+(4-d) \left[(d-1) \left(-6 x_1^3 \left(y_1+y_2-1\right)\right.\right.\right. \\
		&\times\left.\left.\left.  \left(2 y_1+2 y_2-3\right)+x_1^2 \left(2 \left(4 \left(y_1+y_2\right) \left(4 y_1+y_2\right)-3 x_2
		\left(y_1+y_2-1\right) \left(2 y_1+2 y_2-3\right)\right)-62 y_1-32 y_2\right.\right.\right.\right.\\
		&+\left.\left.\left.27)+ x_1 \left(3 x_2 \left(2 y_1-1\right) \left(4 y_1+4 y_2-5\right)+y_1 \left(-28 y_1-16 y_2+39\right)+7 y_2-15\right)-6 x_2 \left(2
		\left(y_1-1\right) y_1+1\right)\right.\right.\right.\\
		&+\left.\left.\left. y_1 \left(8 y_1-7\right)+3\right)-2 \left(6 x_1^3 \left(y_1+y_2-1\right) \left(2 y_1+2 y_2-3\right)+x_1^2 \left(2 \left(3 x_2 \left(y_1+y_2-1\right) \left(2 y_1+2 y_2-3\right)\right.\right.\right.\right.\right.\\
		&-\left.\left.\left.\left.\left. 4
		\left(y_1+y_2\right) \left(4 y_1+y_2\right)\right)+62 y_1+32 y_2-27\right)+x_1 \left(-3 x_2 \left(2 y_1-1\right) \left(4 y_1+4 y_2-5\right)-7 y_2\right.\right.\right.\right.\\
		&+\left.\left.\left.\left. y_1 \left(28 y_1+16 y_2-39\right)+9\right)+y_1 \left(12 x_2
		\left(y_1-1\right)-8 y_1+7\right)\right)\right]\right\}-(d-6) (d-4) \left(x_1 \left(y_1+y_2-1\right)-y_1\right)\\
		&\times  \left(x_1 \left(y_1+y_2-1\right)-y_1+1\right) \left\{6 x_1^3 \left(y_1+y_2-2\right) \left(y_1+y_2-1\right)+x_1^2
		\left(2 \left(3 x_2 \left(y_1+y_2-2\right) \left(y_1+y_2-1\right)\right.\right.\right.\\
		&-\left.\left.\left. 2 \left(y_1+y_2\right) \left(4 y_1+y_2\right)\right)+36 y_1+18 y_2-17\right)+x_1 \left(-3 x_2 \left(2 y_1-1\right) \left(2 y_1+2 y_2-3\right)-3 y_2\right.\right.\\
		&+\left.\left. y_1
		\left(14 y_1+8 y_2-21\right)+5\right)+6 x_2 \left(y_1-1\right) y_1+\left(3-4 y_1\right) y_1\right\},
		\end{split}\\
		\begin{split}\label{eq:C2}
		C_2&=\left(-x_1^2 \left(y_1+y_2-1\right){}^2+x_1 \left(2 y_1-1\right) \left(y_1+y_2-1\right)-\left(y_1-1\right) y_1\right) \left[\left(d^2-1\right) \left(-6 x_1-6 x_2+4\right) \right.\\
		&\times\left.  \left(x_1^2+\left(2 x_2-1\right)
		x_1+\left(x_2-1\right) x_2\right) \left(y_1+y_2-1\right)+(4-d) (d-1) \left(-\left(x_1+x_2\right){}^2+x_1+x_2\right)\right]\\
		&+\left(x_1^2+\left(2 x_2-1\right) x_1+\left(x_2-1\right) x_2\right) \left(y_1+y_2-1\right)
		\left\{\left(d^2-1\right) \left(-6 x_1-6 x_2+4\right) \left(-x_1^2 \left(y_1+y_2-1\right){}^2\right.\right.\\
		&+\left.\left. x_1 \left(2 y_1-1\right) \left(y_1+y_2-1\right)-\left(y_1-1\right) y_1\right)+(4-d) \left[(d-1) \left(-6 x_1^3
		\left(y_1+y_2-1\right) \left(2 y_1+2 y_2-3\right)\right.\right.\right.\\
		&+\left.\left.\left. x_1^2 \left(2 \left(4 \left(y_1+y_2\right) \left(4 y_1+y_2\right)-3 x_2 \left(y_1+y_2-1\right) \left(2 y_1+2 y_2-3\right)\right)-62 y_1-32 y_2+27\right)\right.\right.\right.\\
		&+\left.\left.\left. x_1 \left(3 x_2
		\left(2 y_1-1\right) \left(4 y_1+4 y_2-5\right)+y_1 \left(-28 y_1-16 y_2+39\right)+7 y_2-15\right)-6 x_2 \left(2 \left(y_1-1\right) y_1+1\right)\right.\right.\right.\\
		&+\left.\left.\left. y_1 \left(8 y_1-7\right)+3\right)-2 \left\{6 x_1^3 \left(y_1+y_2-1\right)
		\left(2 y_1+2 y_2-3\right)+x_1^2 \left(2 \left(3 x_2 \left(y_1+y_2-1\right) \left(2 y_1+2 y_2-3\right)\right.\right.\right.\right.\right.\\
		&-\left.\left.\left.\left.\left. 4 \left(y_1+y_2\right) \left(4 y_1+y_2\right)\right)+62 y_1+32 y_2-27\right)+x_1 \left(-3 x_2 \left(2 y_1-1\right)
		\left(4 y_1+4 y_2-5\right)-7 y_2\right.\right.\right.\right.\\
		&+\left.\left.\left.\left. y_1 \left(28 y_1+16 y_2-39\right)+9\right)+y_1 \left(12 x_2 \left(y_1-1\right)-8 y_1+7\right)\right\}\right]\right\}-(d-6) (d-4) \left(x_1^2+\left(2 x_2-1\right) x_1\right.\\
		&+\left. \left(x_2-1\right)
		x_2\right) \left(x_1 \left(y_1+y_2-1\right)-y_1\right) \left(x_1 \left(y_1+y_2-1\right)-y_1+1\right),
		\end{split}\\
		\begin{split}\label{eq:C3}
		C_3&=\left(x_1^2+\left(2 x_2-1\right) x_1+\left(x_2-1\right) x_2\right) \left(y_1+y_2-1\right) \left\{\left(d^2-1\right) \left(-6 x_1-6 x_2+4\right) \left(x_1^2+\left(2 x_2-1\right) x_1\right.\right.\\
		&+\left.\left. \left(x_2-1\right) x_2\right) 
		\left(y_1+y_2-1\right)+(4-d) (d-1) \left(-\left(x_1+x_2\right){}^2+x_1+x_2\right)\right\},
		\end{split}
		\end{align}
and
\begin{align}
\begin{split}\label{eq:STilde}
\widetilde{\textswab{S}}\left(3-d;\frac{|q_y|}{\Lambda}\right)&\equiv -\frac{(d-2) \Gamma(3-d)\sin(\pi d)\csc\left(\frac{\pi d}{2}\right)}{6(2\pi)^{d+2}\mathfrak{B}(d)}{\Gamma\left(\frac{d-2}{d-1}\right)}\Gamma\left(\frac{d}{d-1}\right)\\
&\times \int\limits^{1}_{0}\dd x_1\int\limits^{1-x_1}_{0}\dd x_2\int\limits^{1}_{0}\dd y_1\int\limits^{1-y_1}_{0}\dd y_2 \frac{(1-y_1-y_2)^\frac{3}{2}}{\sqrt{y_1}\sqrt{y_2}} \int\limits^{\frac{\Lambda}{|q_y|}}_{0} \dd P\left[\frac{C_1 P^4+C_2P^2+C_3}{\left(D_1 P^2+D_2\right)^{\frac{8-d}{2}}}\right].
\end{split}
\end{align}
$\widetilde{\textswab{S}}\left(3-d;\frac{|q_y|}{\Lambda}\right)$ 
singles out the contribution that is divergent in the $d \rightarrow 3$ limit.
In the large $\Lambda/|q_y|$ limit, it satisfies the limits
		\begin{align}\label{eq:LimitsTilde}
			\lim_{d\rightarrow 3}\widetilde{\textswab{S}}\left(3-d;\frac{|q_y|}{\Lambda}\right) = -\log\left(\frac{|q_y|}{\Lambda}\right),\qquad \lim_{\frac{|q_y|}{\Lambda}\rightarrow 0}\widetilde{\textswab{S}}\left(3-d;\frac{|q_y|}{\Lambda}\right)= \frac{1}{3-d}.
		\end{align}

	\noindent  The contribution from the remaining hot spots are obtained by performing a $C_4$ transformation on the $n=1$ hot spot contribution. 
	Taking the contributions from all hot spots into account, Eq. (\ref{eq:ALLHOTS}) leads to
		\begin{align}\label{eq:TwoLoopSelfEnergyAnyDim}
	\Delta	\Pi^{\mathrm{2L}}(\mathbf{0},\vec{q}) = \frac{4\beta^4_{d}}{ N_cN_f}w(v)\mathfrak{B}(d)\textswab{S}(d-2;w(v))\left[|q_y|^{d-1}\widetilde{\textswab{S}}\left(3-d;\frac{|q_y|}{\Lambda}\right)+|q_x|^{d-1}\widetilde{\textswab{S}}\left(3-d;\frac{|q_x|}{\Lambda}\right)\right].
		\end{align}
		According to Eq. (\ref{eq:LimitsTilde}), the UV cutoff drops out in $d<3$ and we have
		\begin{align}
		\Delta\Pi^{\mathrm{2L}}(\mathbf{0},\vec{q}) = \frac{4\beta^4_{d}}{(3-d) N_cN_f}w(v)\mathfrak{B}(d)\textswab{S}(d-2;w(v))\left[|q_y|^{d-1}+|q_x|^{d-1}\right].
		\end{align}
		We note that Eq. (\ref{eq:DefinitionofB}) reproduces
		the result obtained in  Ref. \cite{SCHLIEF} in the $d \rightarrow 2$ limit and is consistent with the findings of Ref. \cite{LUNTS} close to three dimensions.

Now we show that {\color{blue}Fig.} \ref{fig:SDDiagramD} 
does not contribute to the momentum dependent self-energy.
{\color{blue}Fig.} \ref{fig:SDDiagramD} is written as
\begin{align}
\Upsilon^{\mathrm{2L}}(q) & \sim \frac{4(N^2_c-1)\beta^4_{d}v^2}{N_c N_f}\sum^{4}_{n=1}\int\dd k\int \dd p \mathrm{Tr}\left[G^{(0)}_{\overline{n}}(k+q)\gamma_{d-1}G^{(0)}_n(k)\gamma_{d-1}G^{(0)}_{\overline{n}}(k+p)\gamma_{d-1}G^{(0)}_{n}(k)\gamma_{d-1}\right]D(p).
\end{align}
Taking the trace over the spinor indices, making the change of variables $k_{+}=\varepsilon_{n}(\vec{k};v)$,  and ${k_{-}=\varepsilon_{\overline{n}}(\vec{k}+\vec{p};v)}$ and integrating over $k_{+}$ results in
\begin{align}
\hspace{-0.95cm}\Upsilon^{\mathrm{2L}}(q)&=\frac{2(N^2_c-1)\beta^4_{d}v}{N_c N_f}\sum^{4}_{n=1}\int\dd p\int\limits_{\mathbb{R}^{d-1}}\frac{\dd\mathbf{K}}{(2\pi)^{d-1}}\int\limits_{\mathbb{R}}\frac{\dd k_{-}}{(2\pi)}\left[\frac{((\mathbf{K}\cdot\mathbf{P})(\mathbf{K}\cdot\mathbf{Q})-\mathbf{K}^2(\mathbf{P}\cdot\mathbf{Q}))D(p)}{|\mathbf{K}|^{3}((\mathbf{K}+\mathbf{P})^2+k^2_{-})((\mathbf{K}+\mathbf{Q})^2+(k_{-}-\varepsilon_{\overline{n}}(\vec{p}-\vec{q};v))^2)}\right].
\end{align}
\noindent This expression vanishes when  $\mathbf{Q}=\mathbf{0}$ for any $v$,
and there is no spatial momentum dependent contribution in ${d>2}$. We note that this diagram is exactly zero in ${d=2}$ \cite{SCHLIEF,MAX1,ABANOV1,ABANOV2}.

	\subsection{One-loop fermion self-energy}\label{sec:ONELOOPFERMION}
	
		\noindent  The quantum correction in {\color{blue}Fig.} \ref{fig:FSE1} reads
		\begin{align}
		\delta\boldsymbol{\Gamma}^{(2,0)}_{\mathrm{1L}} = \sum^{4}_{n=1}\sum^{N_c}_{\sigma=1}\sum^{N_f}_{j=1}\int\dd q \overline{\Psi}_{n,\sigma,j}(q)\Sigma^{\mathrm{1L}}_{n}(q)\Psi_{n,\sigma,j}(q),
		\end{align}
		\noindent where the one-loop fermion self-energy is given by		
		\begin{align}\label{eq:Sigma1L}
		\Sigma^{\mathrm{1L}}_{n}(q) = \frac{2\beta^2_{d}(N^2_c-1)v}{N_cN_f}\int\dd k\gamma_{d-1}G^{(0)}_{\overline{n}}(k+q)\gamma_{d-1} D(k).
		\end{align}
		\noindent Here $G^{(0)}_{n}(k)$, $\beta_d$ and $D(k)$ are defined in Eqs. (\ref{eq:BareFermionPropagator}), (\ref{eq:Betad}) and (\ref{eq:BosonPropagator}), respectively. We will consider the part of the self-energy that depends on the spatial momentum and the one that depends on the frequency and co-dimensional momentum, separately. For this purpose we write
		\begin{align}
		\Sigma_{n}(\mathbf{Q},\vec{q}) = (\boldsymbol{\Gamma}\cdot\mathbf{Q})\overline{\Sigma}_{n,\mathrm{f}}(\mathbf{Q})+\gamma_{d-1}\Sigma_{n,\mathrm{s}}(\vec{q}),
		\end{align}
		with
				\begin{align}\label{eq:FrequencySpatialParts}
		\overline{\Sigma}_{n,\mathrm{f}}(\mathbf{Q})&=\frac{1}{2}\mathrm{Tr}\left[\frac{(\boldsymbol{\Gamma}\cdot\mathbf{Q})}{\mathbf{Q}^2}\Sigma_{n}(\mathbf{Q},\vec{0})\right],\qquad  
		\Sigma_{n,\mathrm{s}}(\vec{q})
=\frac{1}{2}\mathrm{Tr}\left[\gamma_{d-1}\Sigma_{n}(\mathbf{0},\vec{q})\right].
		\end{align}


		\subsubsection{ $\overline{\Sigma}_{n,\mathrm{f}}(\mathbf{Q})$ }
		
		\noindent We focus on the frequency and co-dimensional momentum component first,
				\begin{align}
		\overline{\Sigma}^{\mathrm{1L}}_{n,\mathrm{f}}(\mathbf{Q}) =\frac{ 2i\beta^2_d(N^2_c-1) v}{N_cN_f}\frac{1}{\mathbf{Q}^2}\int\dd k\left(\frac{\mathbf{Q}\cdot(\mathbf{K}+\mathbf{Q})}{(\mathbf{K}+\mathbf{Q})^2+\varepsilon_{\overline{n}}(\vec{k};v)^2}\right)D(k).
		\end{align}
		\noindent  For concreteness we consider the $n=1$ hot spot in the small $v$ limit. 
Performing the scaling $k_x\rightarrow k_x/c(v)$ yields 
		\begin{align}
		\overline{\Sigma}^{\mathrm{1L}}_{1,\mathrm{f}}(\mathbf{Q}) =\frac{ 2i\beta^2_d(N^2_c-1) w(v)}{N_c N_f}\frac{1}{\mathbf{Q}^2}\int\dd k \left(\frac{\mathbf{Q}\cdot(\mathbf{K}+\mathbf{Q})}{(\mathbf{K}+\mathbf{Q})^2+(w(v) k_x-k_y)^2}\right)\frac{1}{|\mathbf{K}|^{d-1}+|k_x|^{d-1}+c(v)^{d-1}|k_y|^{d-1}},
		\end{align}
		\noindent where $w(v)=v/c(v)$. 
The integration over $ {\vec{k}}$ gives 
		\begin{align}
	\hspace{-0.5cm}	\overline{\Sigma}^{\mathrm{1L}}_{1,\mathrm{f}}(\mathbf{Q}) = \frac{i \beta^2_{d}(N^2_c-1)w(v)}{N_c N_f\pi\mathbf{Q}^2}\Gamma\left(\frac{d}{d-1}\right)\int\limits_{\mathbb{R}^{d-1}}\frac{\dd\mathbf{K}}{(2\pi)^{d-1}}\left(\frac{\mathbf{Q}\cdot\left(\mathbf{K}+\mathbf{Q}\right)}{|\mathbf{K}+\mathbf{Q}|}\right)\left[|\mathbf{K}|^{2-d}\Gamma\left(\frac{d-2}{d-1}\right)-\frac{(c(v)\widetilde{\Lambda})^{2-d}(d-1)}{(d-2)\Gamma\left(\frac{1}{d-1}\right)}\right]
\label{eq36}
		\end{align}
in the small $c(v)$ limit.
		\noindent 
In $d>2$, the second term in the square brackets of Eq. (\ref{eq36}) can be dropped, 
and the first term gives rise to a logarithmically divergent contribution.
In $d=2$, the two terms in the square brackets combine to become a logarithm,
and the integration over ${\bf K}$ is finite.
In all cases, the logarithmically divergent contribution can be written as
		\begin{align}
		\overline{\Sigma}^{\mathrm{1L}}_{1,\mathrm{f}}(\mathbf{Q})=-\frac{(N^2_c-1)  \cos \left(\frac{\pi  d}{2}\right) \Gamma \left(\frac{2d-3}{d-1}\right) \Gamma \left(\frac{1}{d-1}\right) \Gamma
			\left(\frac{d-1}{2}\right)}{2^{3-d}N_c N_f\pi ^{3/2} \Gamma \left(\frac{d}{2}\right)} i w(v)\log\left(\frac{\Lambda}{|\mathbf{Q}|}\right).
		\end{align}
		\noindent Here we have used the fact that
$\Lambda \approx \widetilde{\Lambda}$
and the definition of $\beta_d$ in Eq. (\ref{eq:Betad}). 
Combining this result with the renormalization condition in Eq. (\ref{eq:RGConditionFermionFrequency1}) and the fact that the other three hot spots give the same contribution, $Z_1$ is fixed to be
		\begin{align}
		Z_1&= 1-\frac{(N^2_c-1) \zeta(d)}{N_c N_f}w(v)\log\left(\frac{\Lambda}{\mu}\right),
		\end{align}
		\noindent with $\zeta(d)$ defined in Eq. (\ref{eq:zetad}).
			
		\subsubsection{$\Sigma_{n,\mathrm{s}}(\vec{q})$}
		
		\noindent Now we turn our attention to the spatial part of the self-energy defined in Eq. (\ref{eq:FrequencySpatialParts}):
		\begin{align}
		\Sigma^{\mathrm{1L}}_{n,\mathrm{s}}(\vec{q}) &= -\frac{2(N^2_c-1)i\beta^2_{d}v}{N_c N_f}\int\dd k\left(\frac{\varepsilon_{\overline{n}}(\vec{k}+\vec{q};v)}{\varepsilon_{\overline{n}}(\vec{k}+\vec{q};v)^2+\mathbf{K}^2}\right)D(k). 
		\end{align}
			
		\noindent Without loss of generality we consider the contribution from the $n=1$ hot spot, 
		\begin{align}
		\Sigma^{\mathrm{1L}}_{1,\mathrm{s}}(\vec{q}) &= -\frac{2i(N^2_c-1)\beta^2_{d}v}{N_c N_f}\int\dd k\left(\frac{(vk_x-k_y+\varepsilon_{3}(\vec{q};v))}{(vk_x-k_y+\varepsilon_{3}(\vec{q};v))^2+\mathbf{K}^2}\right)\frac{1}{|\mathbf{K}|^{d-1}+c(v)^{d-1}(|k_x|^{d-1}+|k_y|^{d-1})}. 
		\end{align}
		
		\noindent 
 When  $v$ and $c(v)$ are small, the integration over $k_x$ yields 
		\begin{align}
		\begin{split}\label{eq:BeforeAD}
		\Sigma^{\mathrm{1L}}_{1,\mathrm{s}}(\vec{q}) = -\frac{2i(N^2_c-1)\beta^2_{d}}{(d-1)\pi N_c N_f}w(v)\int\limits_{\mathbb{R}^{d-1}}\frac{\dd\mathbf{K}}{(2\pi)^{d-1}}\int\limits_{\mathbb{R}}\frac{\dd k_y}{(2\pi)}\left[\left(\frac{(\varepsilon_{3}(\vec{q};v)-k_y)}{(\varepsilon_{3}(\vec{q};v)-k_y)^2+\mathbf{K}^2}\right)\right.\\
		\times\left[\Gamma\left(\frac{d-2}{d-1}\right)\Gamma\left(\frac{1}{d-1}\right)\frac{1}{\left(|\mathbf{K}|^{d-1}+c(v)^{d-1}|k_y|^{d-1}\right)^{\frac{d-2}{d-1}}}-\frac{c(v)^{2-d}\widetilde{\Lambda}^{2-d}(d-1)}{(d-2)}\right].
		\end{split}
		\end{align}
We drop the second term in the square brackets because the integrand is odd in $(\varepsilon_{3}(\vec{q};v)-k_y)$.
Focusing only on the first term, the remaining integrations are done by writing the expression as an antiderivative with respect to $c(v)$:
		\begin{align}
		\begin{split}
		\Sigma^{\mathrm{1L}}_{1,\mathrm{s}}(\vec{q}) = \frac{2(d-2)i(N^2_c-1)\beta^2_{d}}{(d-1)\pi N_c N_f}\Gamma\left(\frac{d-2}{d-1}\right)\Gamma\left(\frac{1}{d-1}\right)w(v)\int\limits^{c(v)}_{0}\dd \mathsf{c} \mathsf{c}^{d-2}\int\limits_{\mathbb{R}^{d-1}}\frac{\dd\mathbf{K}}{(2\pi)^{d-1}}\\
		\times \int\limits_{\mathbb{R}}\frac{\dd k_y}{(2\pi)}\left[\left(\frac{(\varepsilon_{3}(\vec{q};v)-k_y)}{(\varepsilon_{3}(\vec{q};v)-k_y)^2+\mathbf{K}^2}\right)\frac{|k_y|^{d-1}}{\left(|\mathbf{K}|^{d-1}+\mathsf{c}^{d-1}|k_y|^{d-1}\right)^{\frac{2d-3}{d-1}}}\right]. 
		\end{split}
		\end{align}

\noindent The lower limit of the integration over $\mathsf{c}$ is determined from the fact that the integration over $k_y$ in Eq. (\ref{eq:BeforeAD}) vanishes in the small $c(v)$ limit. 
The radial integration for $\mathbf{K}$ is divided into two regions: 
$K\equiv |\mathbf{K}| \in (0,{|\varepsilon_{3}(\vec{q};v)}-k_y|)$ and $K\in (|{\varepsilon_{3}(\vec{q};v)}-k_y|,\infty)$. 
In the first region, the fermionic contribution to  the integrand  varies slowly in $\mathbf{K}$ and can be Taylor expanded around the origin.
Only the zeroth order term in the expansion becomes IR divergent when $\mathsf{c}=0$, and thus, provides the leading order contribution to the integration in the small $c(v)$ limit.  
The contribution from the second region is regular and therefore is subleading in the small $c(v)$ limit.
Keeping only the leading contribution in the small $c(v)$ limit, we obtain
		\begin{align}\label{eq:2LoopProvi2}
	\hspace{-1.2cm}	\Sigma^{\mathrm{1L}}_{1,\mathrm{s}}(\vec{q}) = \frac{(d-2)i(N^2_c-1)\beta^2_{d}w(v)}{2^{d-2}\pi^\frac{d+1}{2} N_c N_f}\frac{\Gamma\left(\frac{d-2}{d-1}\right)\Gamma\left(\frac{1}{d-1}\right)}{\Gamma\left(\frac{d+1}{2}\right)}\int\limits^{c(v)}_{0}\dd \mathsf{c} \int\limits_{\mathbb{R}}\frac{\dd k_y}{(2\pi)}\frac{(\varepsilon_{3}(\vec{q};v)-k_y)}{(\varepsilon_{3}(\vec{q};v)-k_y)^2}|k_y|\overline{\textswab{S}}'\left(d-2;\mathsf{c};\frac{|\varepsilon_{3}(\vec{q};v)-k_y|}{|k_y|}\right),
			\end{align}
where
\begin{align}
|k_y|\overline{\textswab{S}}'\left(d-2;\mathsf{c};\frac{|\varepsilon_{3}(\vec{q};v)-k_y|}{|k_y|}\right)
\equiv 
		\int\limits^{|\varepsilon_{3}(\vec{q};v)-k_y|}_{0}\dd K	\frac{\mathsf{c}^{d-2}|k_y|^{d-1}K^{d-2}}{\left(K^{d-1}+\mathsf{c}^{d-1}|k_y|^{d-1}\right)^{\frac{2d-3}{d-1}}}.
\end{align}
While $\overline{\textswab{S}}'(d-2;\mathsf{c};|\varepsilon_3(\vec{q};v)-k_y|/|k_{y}|)$ depends on $k_y$ and $\varepsilon_3(\vec{q};v)$,
these dependences are suppressed in the $d\rightarrow 2$ or $\mathsf{c}\rightarrow 0$ limits. In either of these limits, $\overline{\textswab{S}}'(d-2;\mathsf{c};|\varepsilon_3(\vec{q};v)-k_y|/|k_{y}|)$ reduces to $\textswab{S}(d-2;\mathsf{c})$ defined in Eq. (\ref{eq:SFunction}). From now on, we replace $\overline{\textswab{S}}'(d-2;\mathsf{c};|\varepsilon_3(\vec{q};v)-k_y|/|k_{y}|)$  with $\textswab{S}(d-2;\mathsf{c})$  in Eq. (\ref{eq:2LoopProvi2}).
Integration over $\mathsf{c}$ can be done by using the following limits:
		\begin{align}
\lim_{\xi\rightarrow 0^{+}}	\int \dd a {\textswab{S}}(\xi;a) &= a-a\log(a) \stackrel{a\ll 1}{=} \lim_{\xi\rightarrow 0^{+}} a {\textswab{S}}(\xi;a),\\
\lim_{b\rightarrow 0}	\int \dd a {\textswab{S}}(\xi; ba) &= \frac{a}{\xi}  = a\lim_{a\rightarrow 0^{+}}{\textswab{S}}(\xi;a).
		\end{align}
This allows us to write Eq. (\ref{eq:2LoopProvi2}) as
		 	\begin{align}
		 \begin{split}
		 \Sigma^{\mathrm{1L}}_{1,\mathrm{s}}(\vec{q}) = \frac{(d-2)i(N^2_c-1)\beta^2_{d}}{2^{d-2}\pi^\frac{d+1}{2} N_c N_f}\frac{\Gamma\left(\frac{d-2}{d-1}\right)\Gamma\left(\frac{1}{d-1}\right)}{\Gamma\left(\frac{d+1}{2}\right)}v\textswab{S}\left(d-2;c(v)\right) \varepsilon_3(\vec{q};v)\int\limits_{\mathbb{R}}\frac{\dd k_y}{(2\pi)}\left(\frac{(1-k_y)}{(1-k_y)^2}\right)|k_y|.
		 \end{split}
		 \end{align}
		 \noindent Here we have scaled out the external momentum through the change of variables $k_y\rightarrow |{\varepsilon_3(\vec{q};v)}|k_y$. The integration over $k_y$ is UV divergent and we cut it off by $\widetilde{\Lambda}/|{\varepsilon_{3}(\vec{q};v)}|$. In the large ${\widetilde{\Lambda}/|\varepsilon_{3}(\vec{q};v)|}$ limit,
		 \begin{align}
		 \hspace{-0.5cm}\int\limits^{\frac{\widetilde{\Lambda}}{|\varepsilon_{3}(\vec{q};v)|}}_{-\frac{\widetilde{\Lambda}}{|\varepsilon_{3}(\vec{q};v)|}}\frac{\dd k_y}{(2\pi)} \left(\frac{1-k_y}{(1-k_y)^2}\right)|k_y| = \lim_{\delta\rightarrow 0}\left(\int\limits^{1-\delta}_{-\frac{\widetilde{\Lambda}}{|\varepsilon_{3}(\vec{q};v)|}}\frac{\dd k_y}{(2\pi)}+\int\limits^{\frac{\widetilde{\Lambda}}{|\varepsilon_{3}(\vec{q};v)|}}_{1+\delta}\frac{\dd k_y}{(2\pi)}\right)\left(\frac{1-k_y}{(1-k_y)^2}\right)|k_y|= \frac{1}{\pi}\log\left(\frac{|\varepsilon_{3}(\vec{q};v)|}{\widetilde{\Lambda}}\right).
		 \end{align}
Hence, the divergent contribution to the spatial  part of the one-loop fermion self-energy for the fermions at the $n=1$ hot spot is given by
		\begin{align}
		\begin{split}
		\Sigma^{\mathrm{1L}}_{1,\mathrm{s}}(\vec{q}) = -\frac{(d-2)i(N^2_c-1)\beta^2_{d}}{2^{d-2}\pi^\frac{d+3}{2} N_c N_f}\frac{\Gamma\left(\frac{d-2}{d-1}\right)\Gamma\left(\frac{1}{d-1}\right)}{\Gamma\left(\frac{d+1}{2}\right)}v\textswab{S}\left(d-2;c(v)\right) \varepsilon_3(\vec{q};v)\log\left(\frac{\widetilde{\Lambda}}{|\varepsilon_{3}(\vec{q};v)|}\right)
		\end{split}
		\end{align}
in the small $v$ and large $\widetilde{\Lambda}/|{\varepsilon_{3}(\vec{q};v)}|$ limits.
Introducing the value of $\beta_d$ defined in Eq. (\ref{eq:Betad}) and combining this expression with the renormalization conditions in Eqs. (\ref{eq:RGConditionKx}) and (\ref{eq:RGConditionKy}) fixes the counter term coefficients $A_2$ and $A_3$ to the one-loop order,
		\begin{align}
		A^{\mathrm{1L}}_2&= \frac{2(d-1)(N^2_c-1) \zeta(d)}{\pi N_cN_f}v\textswab{S}\left(d-2;c(v)\right) \log\left(\frac{\widetilde{\Lambda}}{\mu}\right),\label{eq:A21}\\
		A^{\mathrm{1L}}_3&=   -\frac{2(d-1)(N^2_c-1)\zeta(d)}{\pi N_c N_f}v\textswab{S}\left(d-2;c(v)\right) \log\left(\frac{\widetilde{\Lambda}}{\mu}\right)\label{eq:A31}
		\end{align}
	\noindent with $\zeta(d)$ defined in Eq. (\ref{eq:zetad}).

	\subsection{Two-loop fermion self-energy}\label{sec:TWOLOOPFERMION}
	
	\noindent We consider the  two-loop fermion self-energy depicted in {\color{blue}Fig.} \ref{fig:FSE2},
	\begin{align}
	\delta\boldsymbol{\Gamma}^{(2,0)}_{\mathrm{2L}} = \sum^{4}_{n=1}\sum^{N_c}_{\sigma=1}\sum^{N_f}_{j=1}\int\dd q \overline{\Psi}_{n,\sigma,j}(q)\Sigma^{\mathrm{2L}}_{n}(q)\Psi_{n,\sigma,j}(q),
	\end{align}
	\noindent where the two-loop fermion self-energy is given by
	\begin{align}
{	\Sigma^{\mathrm{2L}}_{n}(k) = \frac{4(N^2_c-1)\beta^4_d v^2}{N^2_c N^2_f}\int\dd q\int\dd p\left[\gamma_{d-1}G^{(0)}_{\overline{n}}(k+q)\gamma_{d-1}G^{(0)}_{n}(k+q+p)\gamma_{d-1}G^{(0)}_{\overline{n}}(k+p)\gamma_{d-1}\right]D(p)D(q).}
	\end{align}
Without loss of generality, we consider the $n=1$ hot spot contribution to the spatial piece of this quantum correction since its frequency part is strictly subleading with respect to the one-loop correction due to an additional factor of $w(v)=v/c(v)$. 
The self-energy at $\mathbf{K}=\mathbf{0}$ becomes
	\begin{align}
\hspace{-0.5cm}	{\Sigma^{\mathrm{2L}}_{1,\mathrm{s}}(\vec{k})} &{= -\frac{4i(N^2_c-1)\beta^4_d v^2}{N^2_c N^2_f}\int \dd q\int\dd p\left\{\frac{D(p)D(q)}{(\mathbf{P}^2+\varepsilon_3(\vec{k}+\vec{p};v)^2)(\mathbf{Q}^2+\varepsilon_3(\vec{k}+\vec{q};v)^2)((\mathbf{P}+\mathbf{Q})^2+\varepsilon_{1}(\vec{k}+\vec{q}+\vec{p};v)^2)}\right.}\\
\hspace{-0.5cm}	&\times\left.\left[(\mathbf{P}\cdot\mathbf{Q})\varepsilon_{1}(\vec{k}+\vec{p}+\vec{q};v)+\mathbf{Q}\cdot(\mathbf{P}+\mathbf{Q})\varepsilon_{3}(\vec{k}+\vec{p};v)+[\mathbf{P}\cdot(\mathbf{P}+\mathbf{Q})-\varepsilon_{1}(\vec{k}+\vec{p}+\vec{q};v)\varepsilon_{3}(\vec{k}+\vec{p};v)]\varepsilon_{3}(\vec{k}+\vec{q};v)\right]\right\}\notag.
	\end{align}

We proceed by performing the scaling $p_x\rightarrow p_x/v$ and $q_x\rightarrow q_x/v$
and dropping the dependences on $p_y$ and $q_y$ inside the boson propagators in the small $v$ limit.
In the small $c(v)$ limit, the integrations over $p_x$ and $q_x$ give
	\begin{align}
	\begin{split}
	{\Sigma^{\mathrm{2L}}_{1,\mathrm{s}}(\vec{k})}&{= -\frac{4(d-2)^2i(N^2_c-1)\beta^4_d w(v)^2}{\pi^2 N^2_c N^2_f}\Gamma\left(\frac{d-2}{d-1}\right)^2\Gamma\left(\frac{d}{d-1}\right)^2\textswab{S}\left(d-2;w(v)\right)^2\int\limits_{\mathbb{R}^{d-1}}\frac{\dd \mathbf{Q}}{(2\pi)^{d-1}}}\\
	&\times\int\limits_{\mathbb{R}^{d-1}}\frac{\dd \mathbf{P}}{(2\pi)^{d-1}}\int\limits_{\mathbb{R}}\frac{\dd q_y}{(2\pi)}\int\limits_{\mathbb{R}}\frac{\dd p_y}{(2\pi)}\left\{\frac{|\mathbf{P}|^{2-d}|\mathbf{Q}|^{2-d}}{(\mathbf{Q}^2+(\varepsilon_3(\vec{k};v)-q_y)^2)((\mathbf{P}+\mathbf{Q})^2+(\varepsilon_{1}(\vec{k};v)+p_y+q_y)^2)}\right.\\
	&\times \left.\frac{1}{(\mathbf{P}^2+(\varepsilon_3(\vec{k};v)-p_y)^2)}\left[(\mathbf{P}\cdot\mathbf{Q})(\varepsilon_{1}(\vec{k};v)+p_y+q_y)+\mathbf{Q}\cdot(\mathbf{P}+\mathbf{Q})(\varepsilon_{3}(\vec{k};v)-p_y)\right.\right.\\
	&\left.\left.+(\mathbf{P}\cdot(\mathbf{P}+\mathbf{Q})-(\varepsilon_{1}(\vec{k};v)+p_y+q_y)(\varepsilon_{3}(\vec{k};v)-p_y))(\varepsilon_{3}(\vec{k};v)-q_y)\right]\right\},
	\end{split}
	\end{align}
where $\textswab{S}(d-2;w(v))$ is defined in Eq. (\ref{eq:SFunction}). 
Here we ignore terms that are  subleading in $c(v)$.
	 We continue by making the change of variables $p_y\rightarrow p_y+\varepsilon_{3}(\vec{k};v)$ and $q_y\rightarrow q_y+\varepsilon_{3}(\vec{k};v)$ which makes the two-loop fermion self-energy depend on the external spatial momentum only through ${\delta(\vec{k};v)}\equiv\varepsilon_{1}(\vec{k};v)+2{\varepsilon_3(\vec{k};v)}=3vk_x-k_y$. 
  After an introduction of a single-variable Feynman parametrization, the integration over $p_y$ yields
	\begin{align}
	\hspace{-0.5cm}\begin{split}
	{\Sigma^{\mathrm{2L}}_{1,\mathrm{s}}(\vec{k})} &{= -\frac{i(d-2)^2(N^2_c-1)\beta^4_d w(v)^2}{\pi^2 N^2_c N^2_f}\Gamma\left(\frac{d-2}{d-1}\right)^2\Gamma\left(\frac{d}{d-1}\right)^2\textswab{S}(d-2;w(v))^2}\int\limits_{\mathbb{R}^{d-1}}\frac{\dd \mathbf{Q}}{(2\pi)^{d-1}}\int\limits_{\mathbb{R}^{d-1}}\frac{\dd \mathbf{P}}{(2\pi)^{d-1}}\int\limits^{1}_{0}\dd x\\
	&\times\int\limits_{\mathbb{R}}\frac{\dd q_y}{(2\pi)}|\mathbf{P}|^{2-d}|\mathbf{Q}|^{2-d}\left[\frac{\mathcal{A}-q_y(\mathbf{P}^2+2(1-x)\mathbf{P}\cdot\mathbf{Q}+(1-x)(\mathbf{Q}^2+x(q_y+\delta(\vec{k};v))^2))}{(\mathbf{P}^2+2(1-x)\mathbf{P}\cdot\mathbf{Q}+(1-x)(\mathbf{Q}^2+x(q_y+\delta(\vec{k};v))^2))^\frac{3}{2}(q^2_y+\mathbf{Q}^2)}\right]
	\end{split}
	\end{align}
with
	\begin{align}
	\mathcal{A}&= -q_y(\mathbf{P}\cdot(\mathbf{P}+\mathbf{Q}))+x(\mathbf{P}\cdot\mathbf{Q})(q_y+\delta(\vec{k};v))+(1-x)(q_y+\delta(\vec{k};v))(\mathbf{Q}\cdot(\mathbf{Q}+\mathbf{P})+xq_y(q_y+\delta(\vec{k};v))).
	\end{align}
By introducing a second single-variable Feynman parametrization, the integration over $q_y$ yields
	\begin{align}
	\begin{split}\label{eq:Sigma2Loop}
	{\Sigma^{\mathrm{2L}}_{1,\mathrm{s}}(\vec{k})} &{= -\frac{i(d-2)^2(N^2_c-1)\delta(\vec{k};v)\beta^4_d w(v)^2}{\pi^3 N^2_c N^2_f}\Gamma\left(\frac{d-2}{d-1}\right)^2\Gamma\left(\frac{d}{d-1}\right)^2\textswab{S}(d-2;w(v))^2}\\
	&\times\int\limits_{\mathbb{R}^{d-1}}\frac{\dd \mathbf{Q}}{(2\pi)^{d-1}}\int\limits_{\mathbb{R}^{d-1}}\frac{\dd \mathbf{P}}{(2\pi)^{d-1}}\int\limits^{1}_{0}\dd x\int\limits^{1}_{0}\dd y\frac{|\mathbf{P}|^{2-d}|\mathbf{Q}|^{2-d}}{\sqrt{y+x(1-x)(1-y)}}\left(\frac{\mathcal{C}_1}{\mathcal{D}^2_1}\right),
	\end{split}
	\end{align}
	\noindent where
	\begin{align}
	\begin{split}
	\mathcal{C}_1 &= \frac{\sqrt{1-y}}{x+x^2(y-1)+y-x y}\left[x y (\mathbf{P}\cdot\mathbf{Q})+(1-x)\left(x(1-y)\mathbf{P}\cdot(\mathbf{P}+\mathbf{Q})+y\mathbf{Q}\cdot(\mathbf{P}+\mathbf{Q})\right.\right.\\
	&+\left.\left. x(1-y)(\mathbf{P}^2+2(1-x)(\mathbf{P}\cdot\mathbf{Q})+(1-x)\mathbf{Q}^2)\right)\right],
	\end{split}\\
	\mathcal{D}_1&=(1-y)\mathbf{P}^2+2(1-x)(1-y)(\mathbf{P}\cdot\mathbf{Q})+(1-x(1-y))\mathbf{Q}^2+\frac{(1-x)(1-y)x y}{y+(1-x)(1-y)x}\delta(\vec{k};v)^2.
	\end{align}
\noindent Integration over $\mathbf{P}$ is done by introducing a third single-variable Feynman parametrization. This process yields a $\mathbf{Q}$-dependent integrand that can be cast in the rotationally invariant way,
	\begin{align}
	\begin{split}
	{\Sigma^{\mathrm{2L}}_{1,\mathrm{s}}(\vec{k}) }&{= \frac{i(d-2)^2(N^2_c-1)\delta(\vec{k};v)\beta^4_d w(v)^2}{2^{2d-2}\pi^{\frac{2d+3}{2}} N^2_c N^2_f}\frac{\Gamma\left(\frac{d-2}{d-1}\right)^2\Gamma\left(\frac{d}{d-1}\right)^2}{\Gamma\left(\frac{d-2}{2}\right)\Gamma\left(\frac{d-1}{2}\right)}\textswab{S}(d-2;w(v))^2}\\
	&\times\int\limits^{1}_{0}\dd x\int\limits^{1}_{0}\dd y\int\limits^{1}_{0}\dd z\frac{(1-z)z^\frac{d-4}{2}\sqrt{1-y}}{(y+x(1-x)(1-y))^\frac{3}{2}(1-y(1-z))^\frac{d+1}{2}}\int\limits^{\frac{\Lambda}{|\delta(\vec{k};v)|}}_{0}\dd Q\left[\frac{(\mathcal{E}_1Q^2+\mathcal{E}_2)}{(\mathcal{H}_1 Q^2+\mathcal{H}_2)^\frac{3}{2}}\right],
	\end{split}
	\end{align}
	\noindent where the integration over the angular components has been done, and the integration over $Q\equiv |\mathbf{Q}|$ has been cut off in the UV since it is logarithmically divergent.
The coefficients $\mathcal{E}_i$ and $\mathcal{H}_i$ are defined as follows:
		\begin{align}
	\mathcal{E}_1&=\frac{1}{1-y (1-z)}\left\{(x-1) \left[2 x^3 (y-1)^2 (z-1) (d (y-1) (z-1)-y z+y+2 z-1)\right.\right.\\
	&\left.\left.+x^2 (y-1) \left(-2 d (y-1) (z-1) (y (z-1)-2 z+1)+2
	y^2 (z-1)^2-y (9 z-4) (z-1)+z (8 z-9)+2\right)+\right.\right.\notag\\
	&\left.\left. x (y-1) \left(2 (d-1) y^2 (z-1)^2-y (z-1) (2 d (z-1)-3 z+2)+z (2 d
	(z-1)-4 z+3)\right)+y z (y (z-1)+1)\right]\right\}\notag,\\
	\mathcal{E}_2&=\frac{2 (d-1) (x-1)^2 x^2 (y-1)^2 y (z-1) }{(x-1) x (y-1)+y},\\
	\mathcal{H}_1&=-\frac{(z-1)  \left(x^2 (y-1)^2 (z-1)-x (y-1) (y (z-1)-2 z+1)+y (y-1) (z-1)+z\right)}{y (z-1)+1},\\
	\mathcal{H}_2&=-\frac{(1-x) x (1-y) y (1-z)}{((x-1) x (y-1)+y)}.
	\end{align}
	\noindent   For $\Lambda/|{\delta(\vec{k};v)}|\gg 1$, the divergent contribution to the two-loop fermion self-energy is given by
	\begin{align}\label{eq:TheLogDivergent}
		{\Sigma^{\mathrm{2L}}_{1,\mathrm{s}}(\vec{k}) }&{=-\frac{2i\beta^4_d(N^2_c-1)}{N^2_c N^2_f}\mathfrak{F}(d)w(v)^2\textswab{S}(d-2;w(v))^2\delta(\vec{k};v)\log\left(\frac{\Lambda}{|\delta(\vec{k};v)|}\right),}
	\end{align}
	\noindent where the positive function $\mathfrak{F}(d)$ is given by
	\begin{align}
\label{eq:FunctionF}
	\mathfrak{F}(d)& = \frac{(d-2)^2}{(2\pi)^{d+2}}\frac{\Gamma\left(\frac{d-2}{d-1}\right)^2\Gamma\left(\frac{d}{d-1}\right)^2}{\Gamma(d-2)}\int\limits^{1}_{0}\dd x\int\limits^{1}_{0}\dd y\int\limits^{1}_{0}\dd z \frac{(1-x)\sqrt{1-y}z^\frac{d-4}{2}}{(1 - y (1 - z))^\frac{d}{2}[y+(1 - x) x (1-y) ]^\frac{3}{2}\sqrt{1-z}}\\
	&\times\left. \frac{\left(2 x^3 (y-1)^2 (z-1) (d (y-1) (z-1)-yz+y+2 z-1)+\right.}{((x (1- y) (1 + y (-1 + z) - 2 z) + 
		x^2 (-1 + y)^2 (-1 + z) + (-1 + y) y (-1 + z) + z))^\frac{3}{2}}\right.\notag\\
	&+\left.\frac{x^2 (y-1) \left(-2 d (y-1) (z-1) (y (z-1)-2 z+1)+2 y^2 (z-1)^2-y (9 z-4) (z-1)+z (8 z-9)+2\right)}{((x (1- y) (1 + y (-1 + z) - 2 z) + 
		x^2 (-1 + y)^2 (-1 + z) + (-1 + y) y (-1 + z) + z))^\frac{3}{2}}\right.\notag\\
	&+\left.\frac{\left.x (y-1) \left(2 (d-1) y^2 (z-1)^2-y (z-1) (2 d (z-1)-3 z+2)+z (2 d (z-1)-4 z+3)\right)+y z (y (z-1)+1)\right)}{((x (1- y) (1 + y (-1 + z) - 2 z) + 
		x^2 (-1 + y)^2 (-1 + z) + (-1 + y) y (-1 + z) + z))^\frac{3}{2}}\right]\notag.
	\end{align}

	\begin{figure*}
		\centering
		
		\begin{subfigure}{0.45\linewidth}
			\centering
			\includegraphics[scale=0.65]{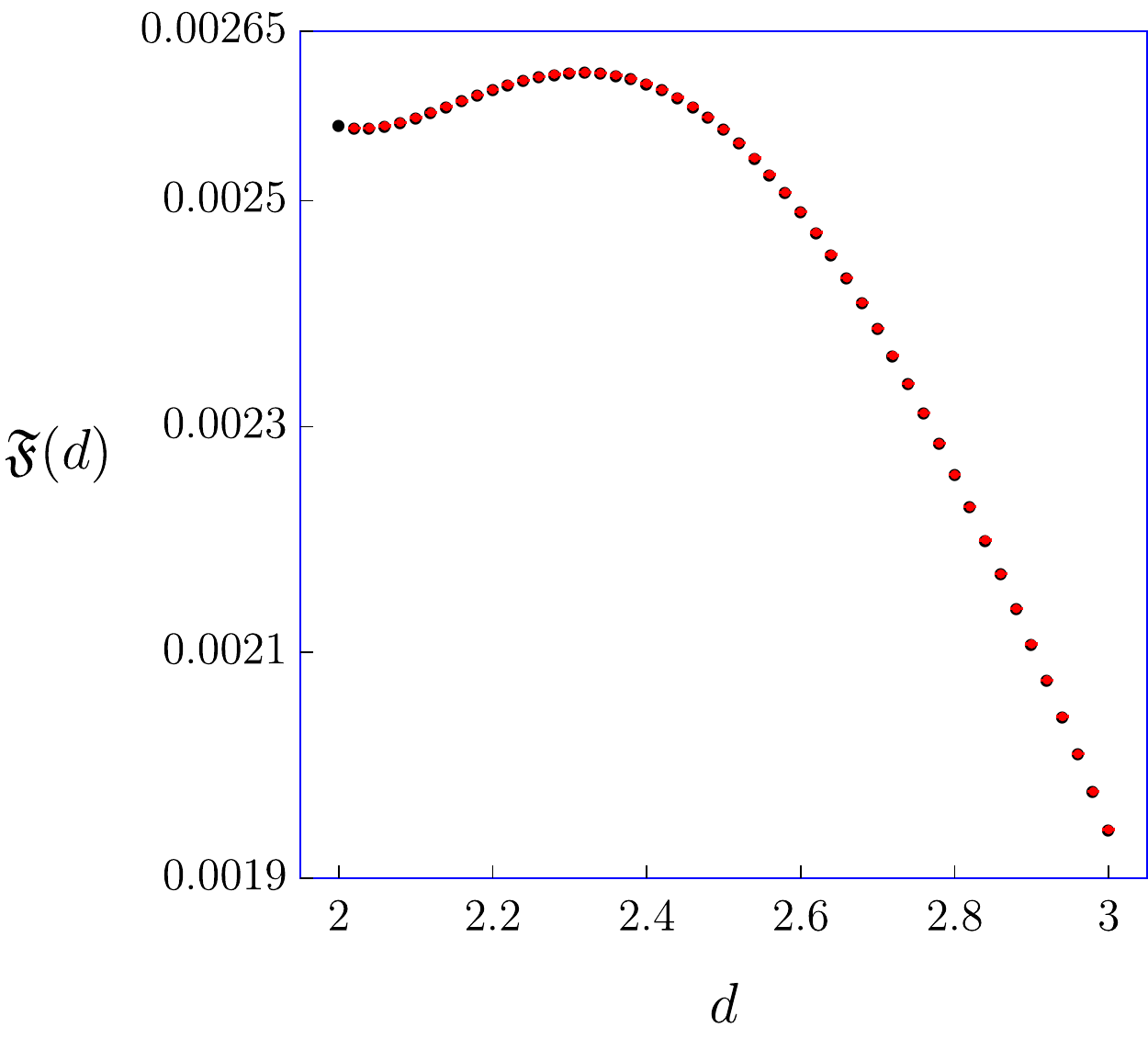}
			\caption{\hspace{0.5em}  $\mathfrak{F}(d)$ for $2\leq d\leq 3$.}
		\end{subfigure}
		\begin{subfigure}{0.45\linewidth}
			\includegraphics[scale=0.6]{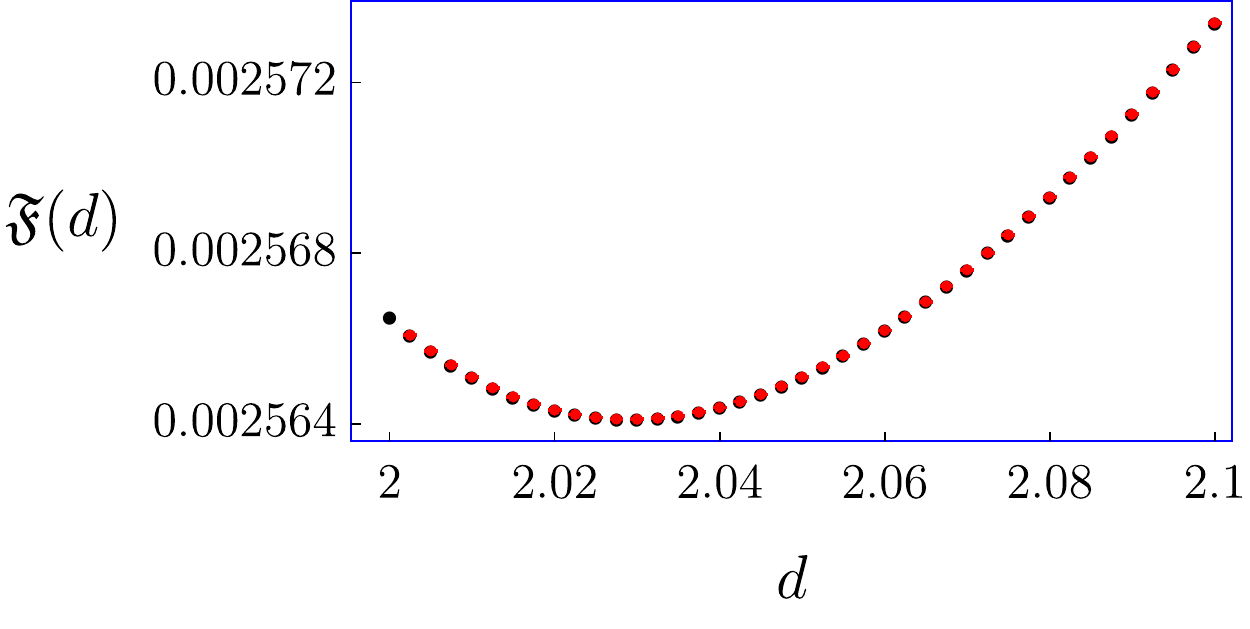}\\\vfill
			\includegraphics[scale=0.6]{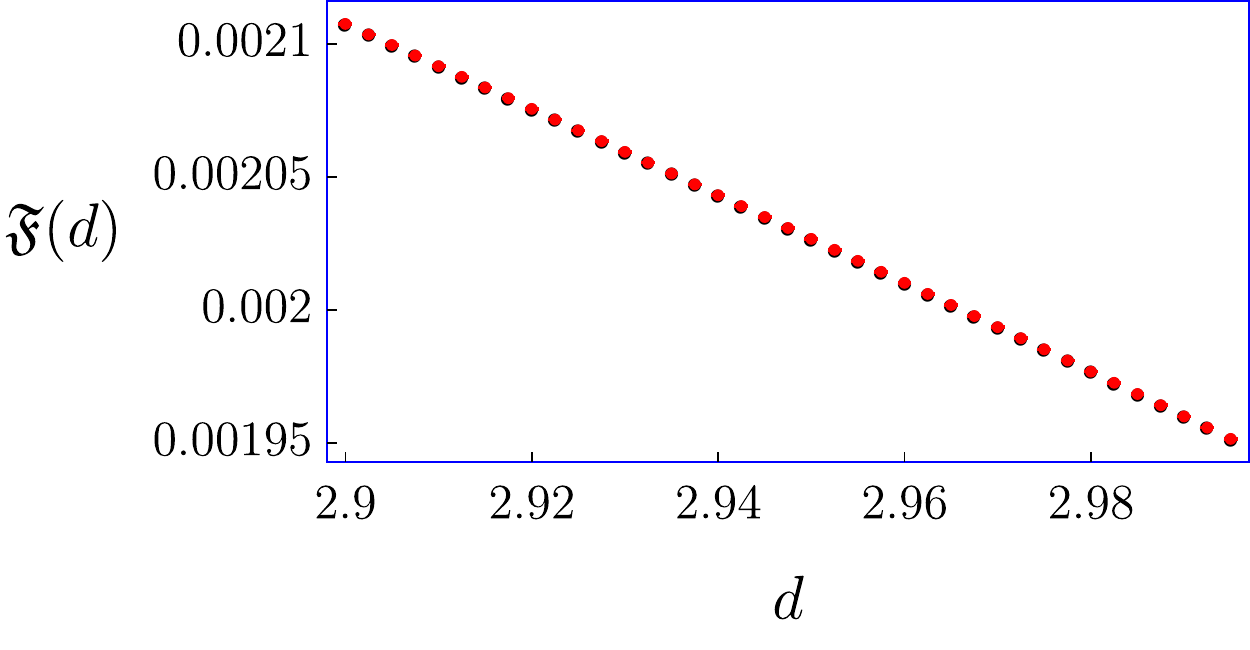}
			\caption{\hspace{0.5em}  $\mathfrak{F}(d)$  near $d=2$ (top) and $d=3$ (bottom).}
		\end{subfigure}

		\caption{{({\color{blue}$a$})} The function $\mathfrak{F}(d)$. Each point is computed numerically except for the one in $d=2$ where it can be determined analytically. The (black) dots correspond the value of the numerical integration and the (red) error bars represent the numerical error in the computation. ({\color{blue}$b$}) Numerical evaluation near $d=2$ (top) and $d=3$ (bottom).
\label{fig:functionF}}
	\end{figure*}
Despite the multiplicative factor  that vanishes in $d=2$,
Eq. (\ref{eq:FunctionF}) does not vanish because the integration over $z$ is divergent in $d=2$.
In this dimension, an explicit integration over the Feynman parameters gives rise to
	\begin{align}\label{eq:FunctionF2D}
	\mathfrak{F}(2) = \frac{1}{4\pi^4}\int\limits^{1}_{0}\dd x \int\limits^{1}_{0}\dd y \frac{(1-x)x}{(y+(1-y)(1-x)x)^2}= \frac{1}{4\pi^4}.
	\end{align} 
This agrees with the result obtained in Ref. \cite{SCHLIEF}. 
For $d>2$, the expression is computed numerically as shown in {\color{blue}Fig.} \ref{fig:functionF}.	
 From Eq. (\ref{eq:TheLogDivergent}) and the renormalization conditions in Eqs. (\ref{eq:RGConditionKx}) and (\ref{eq:RGConditionKy}) the two-loop counter term coefficients are determined to be
	\begin{align}
	{A^{\mathrm{2L}}_2 }&{= \frac{6(N^2_c-1)\beta^4_{d}}{ N^2_c N^2_f}\mathfrak{F}(d)w(v)^2\textswab{S}\left(d-2;w(v)\right)^2\log\left(\frac{\Lambda}{\mu}\right),\label{eq:A22}}\\
	{A^{\mathrm{2L}}_3 }&{=- \frac{2(N^2_c-1)\beta^4_{d}}{ N^2_c N^2_f}\mathfrak{F}(d)w(v)^2\textswab{S}\left(d-2;w(v)\right)^2\log\left(\frac{\Lambda}{\mu}\right).\label{eq:A32}}
	\end{align}
	
	\noindent Combining this result with Eqs. (\ref{eq:A21}) and (\ref{eq:A31}) it follows that the counter term coefficients $Z_2$ and $Z_3$ are given, to the leading order in $v$, by
	\begin{align}
Z_2&{= 1+\frac{(N^2_c-1)}{N_c N_f}\left[\frac{2(d-1)}{\pi}\zeta(d)v\textswab{S}(d-2;c(v))\log\left(\frac{\widetilde{\Lambda}}{\mu}\right)+\frac{6\beta^4_d}{N_cN_f}\mathfrak{F}(d)w(v)^2\textswab{S}\left(d-2;w(v)\right)^2\log\left(\frac{\Lambda}{\mu}\right)\right],}\label{eq:Z2Appendix}\\
Z_3&={ 1-\frac{(N^2_c-1)}{N_c N_f}\left[\frac{2(d-1)}{\pi}\zeta(d)v\textswab{S}(d-2;c(v))\log\left(\frac{\widetilde{\Lambda}}{\mu}\right)+\frac{2\beta^4_d}{N_cN_f}\mathfrak{F}(d)w(v)^2\textswab{S}\left(d-2;w(v)\right)^2\log\left(\frac{\Lambda}{\mu}\right)\right].}\label{eq:Z3Appendix}
	\end{align}

	\subsection{One-loop vertex correction}

		\noindent We consider the one-loop vertex correction in {\color{blue}Fig.} \ref{fig:YUK},
		\begin{align}
		\delta\boldsymbol{\Gamma}^{(2,1)}_{\mathrm{1L}}=\frac{i\beta_d\sqrt{v}}{\sqrt{N_f}}\sum^{4}_{n=1}\sum^{N_c}_{\sigma,\sigma'=1}\sum^{N_f}_{j=1}\int \dd k\int\dd q\overline{\Psi}_{\overline{n},\sigma,j}(k+q){\Phi}_{\sigma\sigma'}(q)\varGamma^{(2,1),\mathrm{1L}}_{n}(k,q)\Psi_{n,\sigma',j}(k),
		\end{align}
		\noindent where the one-loop vertex function is given by
		\begin{align}
		\varGamma^{(2,1),\mathrm{1L}}_{n}(k,q) &= \frac{2\beta^2_d v}{N_c N_f}\int\limits\dd p\left[\gamma_{d-1} G^{(0)}_{\overline{n}}(p+k+q)\gamma_{d-1} G^{(0)}_{n}(p+k)\gamma_{d-1}\right] D(p).
		\end{align}
In view of the renormalization condition in Eq. (\ref{eq:RGConditionVertex}), we consider the vertex function
at $k=0$ and $\vec q=0$,
		\begin{align}\label{eq:UpsilonYuk}
		\Upsilon^{\mathrm{1L}}_{n}(\mathbf{Q}) = \frac{1}{2}\mathrm{Tr}\left[\gamma_{d-1}\varGamma^{(2,1),\mathrm{1L}}_{n}(k,q)\right]\bigg|_{k=0,\vec{q}=\vec{0}}.
		\end{align}
For $n=1$, Eq. (\ref{eq:UpsilonYuk}) becomes
		\begin{align}
		\Upsilon^{\mathrm{1L}}_{1}(\mathbf{Q}) = \frac{2\beta^2_d v}{N_c N_f}\int\dd p \left(\frac{\mathbf{P}\cdot(\mathbf{P}+\mathbf{Q})-(v^2p^2_x-p^2_y)}{(\mathbf{P}^2+(vp_x+p_y)^2)((\mathbf{P}+\mathbf{Q})^2+(vp_x-p_y)^2)}\right)D(p).
		\end{align}
Following the same steps used in Secs. \ref{sec:TWOLOOPBOSON} and \ref{sec:TWOLOOPFERMION} of this appendix, we obtain
		\begin{align}\label{eq:Ups}
		\begin{split}
		\Upsilon^{\mathrm{1L}}_{1}(\mathbf{Q}) &= \frac{2(d-2)\beta^2_d w(v)}{\pi N_c N_f}\Gamma\left(\frac{d-2}{d-1}\right)\Gamma\left(\frac{d}{d-1}\right)\textswab{S}(d-2;w(v))\\
		&\times\int\limits_{\mathbb{R}^{d-1}}\frac{\dd \mathbf{P}}{(2\pi)^{d-1}}\int\limits_{\mathbb{R}}\frac{\dd p_y}{(2\pi)}|\mathbf{P}|^{2-d}\left(\frac{\mathbf{P}\cdot(\mathbf{P}+\mathbf{Q})+p^2_y}{(\mathbf{P}^2+p_y^2)((\mathbf{P}+\mathbf{Q})^2+p_y^2)}\right)
		\end{split}
		\end{align}
to leading order in the small $v$ limit.
Here $\textswab{S}(d-2;w(v))$ is defined in Eq. (\ref{eq:SFunction}). 
We introduce a two-variable Feynman parametrization that allows a straightforward integration over $p_y$, 
		\begin{align}
			\begin{split}
		\Upsilon^{\mathrm{1L}}_{1}(\mathbf{Q}) &= \frac{(d-2)\beta^2_d w(v)}{d\pi^\frac{3}{2} N_c N_f}\frac{\Gamma\left(\frac{d-2}{d-1}\right)\Gamma\left(\frac{d}{d-1}\right)\Gamma\left(\frac{d+2}{2}\right)\Gamma\left(\frac{d-1}{2}\right)}{\Gamma\left(\frac{d}{2}\right)\Gamma\left(\frac{d-2}{2}\right)}\textswab{S}(d-2;w(v))\int\limits^{1}_{0}\dd x_1\int\limits^{1-x_1}_{0}\dd x_2\\
		&\times\int\limits_{\mathbb{R}^{d-1}}\frac{\dd \mathbf{P}}{(2\pi)^{d-1}}\frac{(1-x_1-x_2)^{\frac{d-4}{2}}}{(x_1+x_2)^\frac{3}{2}}\frac{(\mathbf{P}^2-(d-1)(x_1+x_2)\mathbf{P}\cdot(\mathbf{P}+\mathbf{Q})+x_2\mathbf{Q}\cdot(\mathbf{Q}+2\mathbf{P}))}{(\mathbf{P}^2+2x_2\mathbf{P}\cdot\mathbf{Q}+x_2\mathbf{Q}^2)^\frac{d+1}{2}}.
		\end{split}
		\end{align}
		\noindent  The integration over $\mathbf{P}$ is logarithmically divergent, and in the large $\Lambda/|\mathbf{Q}|$ limit one obtains
		\begin{align}
		\Upsilon^{\mathrm{1L}}_{1}(\mathbf{Q}) &= \frac{(2-d)\beta^2_{d}w(v)}{2^{d-2}N_cN_f\pi^\frac{d+1}{2}}\frac{\Gamma\left(\frac{d-2}{d-1}\right)\Gamma\left(\frac{d}{d-1}\right)}{\Gamma\left(\frac{d-1}{2}\right)}\textswab{S}(d-2;w(v))\log\left(\frac{\Lambda}{|\mathbf{Q}|}\right).
		\end{align}
From Eq. (\ref{eq:Betad}) and the renormalization condition in Eq. (\ref{eq:RGConditionVertex}),
we obtain a counter term for the vertex with
		\begin{align}
		Z_6=1-\frac{(d-1)\zeta(d)}{ N_c N_f  }w(v)\textswab{S}\left(d-2;w(v)\right)\log\left(\frac{\Lambda}{\mu}\right),
		\end{align}
	\noindent with $\zeta(d)$ defined in Eq. (\ref{eq:zetad}).

	\section{\bf{Derivation of the Low-Energy Fixed Point} }\label{Sec:Fixed}


		\noindent The self-consistent equation for the dynamically generated boson velocity reads
		\begin{align}\label{eq:SelfConsistentEquationAppendix}
	c(v)^{d-1} = \frac{4\beta^4_d\mathfrak{B}(d)}{(3-d)N_c N_f}\frac{v}{c(v)}\textswab{S}
\left(d-2;\frac{v}{c(v)}\right),
		\end{align}
		\noindent where $\textswab{S}(d-2;v/c(v))$ its given in Eq. (\ref{eq:SFunction}). 
It is easy to see that 
	\begin{align}\label{eq:CVAppendix}
	c(v) = \left(\frac{4\beta^4_d\mathfrak{B}(d)}{(3-d)N_c N_f}\right)^\frac{1}{d}v^\frac{1}{d}\textswab{S}\left(d-2;v^{\frac{(d-1)}{d}}\right)^\frac{1}{d}
	\end{align}
solves Eq. (\ref{eq:SelfConsistentEquationAppendix}) to the leading order in $v/c(v)$
both in the $v/c(v) \rightarrow 0$ limit with $d>2$ 
and in the $d \rightarrow 2$ limit.
This follows from the fact that
${\textswab{S}\left(d-2;{v}/{c(v)}\right) \sim 1/(d-2)}$ in the $v/c(v) \rightarrow 0$ limit with $d>2$,
and 
${\textswab{S}\left(d-2;{v}/{c(v)}\right) \sim -\log ({v}/{c(v)})}$ in  $d=2$.


Now we compute the beta function for $v$ in $2\leq d<3$. 
Eqs. (\ref{eq:RenormalizedBare}) and  (\ref{eq:Betav}), and the fact that the bare velocity is independent of the running energy scale $\mu$, yield the beta function for $v$  as a solution to the equation
	\begin{align}\label{eq:ToBetav}
	\beta_v = v\left(\frac{1}{Z_3}\frac{\partial Z_3}{\partial\log\mu}-\frac{1}{Z_2}\frac{\partial Z_2}{\partial\log\mu}\right)+v\beta_{v}\left(\frac{1}{Z_3}\frac{\partial Z_3}{\partial v}-\frac{1}{Z_2}\frac{\partial Z_2}{\partial v}\right).
	\end{align}
From the counter-term coefficients $Z_2$ and $Z_3$ that are obtained 
to the leading order in the small $v$ limit, 
the beta function becomes
	\begin{align}\label{eq:BetavFistStep}
	{\beta_v 
	\equiv \frac{ \partial v}{\partial \ln \mu}
	= \frac{4(N^2_c-1)}{N_cN_f}v\left[\frac{(d-1)}{\pi}\zeta(d)v\textswab{S}(d-2;c(v))+\frac{2\beta^4_{d}}{N_cN_f}\mathfrak{F}(d)w(v)^2\textswab{S}\left(d-2;w(v)\right)^2\right].}
	\end{align}
In any $2 \leq d < 3$, $\beta_v > 0$ for $v \ll 1$.
This implies that $v$ decreases as energy is lowered
once the bare value of $v$ is small.
Since our calculation is controlled in the small $v$ limit,
we conclude that $v \rightarrow 0$ limit is 
a stable fixed point 
with a finite basin of attraction.

Factoring out the first term in the square brackets  we can use
	\begin{align}\label{eq:NiceProperty}
	\begin{split}
	\frac{2\pi\beta^{4}_{d}\mathfrak{F}(d)}{N_cN_f(d-1)\zeta(d)}\frac{w(v)^2}{v}\frac{\textswab{S}\left(d-2;w(v)\right)^2}{\textswab{S}\left(d-2;c(v)\right)}&\approx  v^{\frac{(d-2)}{d}}\textswab{S}\left(d-2;v^\frac{d-1}{d}\right)^\frac{(d-2)}{d},
	\end{split}
	\end{align}
close to $d=2$, to rewrite the beta function as
	\begin{align}
	\beta_v = \frac{4(d-1)(N^2_c-1)}{\pi N_c N_f}\zeta(d)v^2\textswab{S}\left(d-2;v^\frac{d-1}{d}\right)\left[1+v^{\frac{(d-2)}{d}}\textswab{S}\left(d-2;v^\frac{d-1}{d}\right)^\frac{(d-2)}{d}\right].
	\end{align}
\noindent Eq. (\ref{eq:NiceProperty}) follows from Eq. (\ref{eq:SelfConsistentEquationAppendix}) and  the fact that $\textswab{S}(d-2;c(v))\approx \textswab{S}(d-2;w(v)) \approx  \textswab{S}\left(d-2;v^\frac{d-1}{d}\right)$ close to $d=2$. In Eq. (\ref{eq:NiceProperty}), we use the numerical coefficient evaluated at $d=2$ because the second term in the square brackets of Eq. (\ref{eq:BetavFistStep}) is only important in the low-energy limit in $d=2$. 

We define the logarithmic length scale $l=\log(\Lambda/\mu)$, where $\Lambda$ is a UV energy scale. 
The IR beta function 
that describes the flow of $v$ with increasing length scale
 can be rewritten as
	\begin{align}\label{eq:BetavDG2}
	\frac{\dd v(l)}{\dd l} = -\frac{4(d-1)(N^2_c-1)}{\pi N_c N_f}\zeta(d)v(l)^2\textswab{S}\left(d-2;v(l)^\frac{d-1}{d}\right)\left[1+v(l)^{\frac{(d-2)}{d}}\textswab{S}\left(d-2;v(l)^\frac{d-1}{d}\right)^\frac{(d-2)}{d}\right]
	\end{align}
with a boundary condition $v(0)=v_0$. We note that the term in square brackets is merely a constant in the small $v$ limit in $d\geq 2$. On the other hand, $\textswab{S}\left(d-2;v(l)^{\frac{d-1}{d}}\right)$ provides, at most, a logarithmic correction in $d=2$. With the $l$-dependence of $v(l)$ ignored inside logarithms, the solution to Eq. (\ref{eq:BetavDG2}) can be cast in the following implicit form:
\begin{align}\label{eq:vl}
\begin{split}
v(l) &= \frac{1}{\frac{1}{v_0} + \mathbb{F}(v(l))l},\\
 \mathbb{F}(v(l)) &= \frac{4(d-1)(N^2_c-1)\zeta(d)}{\pi N_c N_f}\textswab{S}\left(d-2;v(l)^\frac{d-1}{d}\right)\left[1+v(l)^{\frac{(d-2)}{d}}\textswab{S}\left(d-2;v(l)^\frac{d-1}{d}\right)^\frac{(d-2)}{d}\right].
 \end{split}
\end{align}
\noindent This equation for $v(l)$ can be solved iteratively in the low-energy limit. The initial condition naturally provides the logarithmic length scale $l^{-1}_0 = v_0\mathbb{F}(v_0)\sim \frac{(N^2_c-1)}{N_cN_f}v_0\textswab{S}\left(d-2;v^\frac{d-1}{d}_0\right)$  below which the solution in Eq. (\ref{eq:vl}) becomes independent of $v_0$ and reduces to the universal form:
	\begin{align}\label{eq:VOFLSimplified}
	v(l) = \frac{\pi N_c N_f}{4(d-1)(N^2_c-1)}\frac{1}{\zeta(d)}\frac{1}{l}\frac{1}{\textswab{S}\left(d-2;l^{-\frac{(d-1)}{d}}\right)}\left[\frac{1}{1+l^{-\frac{(d-2)}{d}}}\right].
	\end{align}
	\noindent Eq. (\ref{eq:VOFLSimplified}) continuously interpolates the form of $v(l)$ found in two and close to three dimensions\cite{SCHLIEF,LUNTS}. In obtaining Eq. (\ref{eq:VOFLSimplified}) we used the limiting forms of Eq. (\ref{eq:SFunction}) repeatedly to discard subleading terms in the $l\gg l_0$ limit.

	\section{\bf{Critical Exponents and Physical Observables}}\label{Sec:UNIV}

	\noindent From Eqs. (\ref{eq:z1}) and (\ref{eq:etapsiphi}), the dynamical critical exponent and the anomalous dimensions of the fields, defined as the deviations from the interaction-driven scaling, are given by
	\begin{align}
	z &= 1-\frac{\dd}{\dd\log\mu}\left(\log Z_3-\log Z_1\right),\\
	\eta_\Psi&= \frac{1}{2}\frac{\dd}{\dd\log\mu}\left(d\log Z_3-(d-1)\log Z_1\right),\\
	\eta_\Phi &= \frac{1}{2}\frac{\dd}{\dd\log\mu}\left(2\log Z_6-2(d-1)\log Z_1-\log Z_2+\left(2d-3\right)\log Z_3\right).
	\end{align}
	\noindent At low energies, $v$ flows to zero faster than the ratio $w(v)=v/c(v)$. 
The dominant contributions to $z$ and $\eta_\Psi$ come from $Z_1$ in the small $v$ limit,
while ${\eta_{\Phi}}$ is dominated by $Z_6$ and $Z_1$. 
To the leading order in $v$, the dynamical critical exponent and the anomalous scaling dimensions are given by
	\begin{align}
	z &= 1+\frac{(N^2_c-1)}{N_cN_f}\zeta(d)w(v), \label{eq:z} \\
	\eta_\Psi &= -\frac{(N^2_c-1)}{N_cN_f}\frac{(d-1)\zeta(d)}{2}w(v),  \label{eq:eta_Psi} \\
	\eta_{\Phi} & =\frac{(d-1)\zeta(d)}{N_cN_f} \left[\textswab{S}\left(d-2;w(v)\right)-(N^2_c-1)\right]w(v).
\label{eq:eta_Phi}
	\end{align}

The scaling forms of the two-point functions 
are dictated by the renormalization group equation,
	\begin{align}\label{eq:RG2Point}
	\begin{split}
	\left[\left(z\mathbf{K}\cdot\nabla_{\mathbf{K}}+\vec{k}\cdot\nabla_{\vec{k}}\right)-\beta_v\frac{\partial}{\partial v}+D_{\mathsf{a}}\right]	\varGamma^{(2)}_{\mathsf{a}}(k, v;\mu)=0.
	\end{split}
	\end{align}
	\noindent 
Here $\mathsf{a}=\mathsf{b,f}$ labels the bosonic and fermionic cases, respectively.
$D_{\mathsf{a}}$ denotes the total scaling dimension of the operator given by
	\begin{align}
	D_{\mathsf{f}}&= 2\eta_\Psi+z(d-1)-d,\\
	D_{\mathsf{b}}&= 2\eta_\Phi+z(d-1)-2(d-1).
\label{eq:Da}
	\end{align}
Eq. (\ref{eq:RG2Point}) can be written as
	\begin{align}\label{eq:RG2Point2}
\left[	\mathbf{K}\cdot\nabla_{\mathbf{K}}+\frac{\vec{k}}{z(l)}\cdot\nabla_{\vec{k}}+\frac{\dd}{\dd l}+\frac{D_{\mathsf{a}}(l)}{z(l)}\right]\varGamma^{(2)}_{\mathsf{a}}(k,v(l))=0,
	\end{align}
where $v(l)$ is the solution of 
	\begin{align}\label{eq:boundaryProblem}
	\frac{\dd v(l)}{\dd l} = -\frac{\beta_v}{z(l)},\qquad v(0)=v_0,
	\end{align}
and $l=l(\Lambda/\mu)$ is a logarithmic length scale.
The solution to Eq. (\ref{eq:RG2Point2}) can be written as
	\begin{align}
		\label{eq:SolRGEquation}
	\varGamma^{(2)}_{\mathsf{a}}(\mathbf{K},\vec{k},v_0) = \exp\left(\int\limits^{l}_{0}\dd\ell\frac{D_{\mathsf{a}}(\ell)}{z(\ell)}\right) \varGamma^{(2)}_{\mathsf{a}}\left(e^{l}\mathbf{K},\exp\left(\int\limits^{l}_{0}\frac{\dd\ell}{z(\ell)}\right)\vec{k},v(l)\right).
	\end{align}
Here $z(\ell)$ and $D_{\mathsf{a}}(\ell)$ should be viewed as functions of the logarithmic length scale,
where $w(v)$ in 
Eqs. 
(\ref{eq:z}),
(\ref{eq:eta_Psi}),
(\ref{eq:eta_Phi})
and
(\ref{eq:Da})
are replaced by
\begin{align}
w(l) &=\frac{\pi N_cN_f}{4(d-1)\left(\zeta(d)(N^2_c-1)\right)^\frac{d-1}{d}}\left[\frac{(3-d)(d-1)	}{\pi\beta^4_d\mathfrak{B}(d)}\right]^\frac{1}{d} \left[\frac{1}{1+l^{-\frac{(d-2)}{d}}}\right]^\frac{d-1}{d}\frac{1}{l^{\frac{d-1}{d}}}\frac{1}{\textswab{S}\left(d-2;l^{-\frac{d-1}{d}}\right)},\label{eq:WOFL}
\end{align}
in the large $l$ limit.
Similarly, the scale-dependent velocity of the collective mode is given by
\begin{align}
c(l) &=\left(\frac{\pi\beta^4_d\mathfrak{B}(d)}{(3-d)(d-1)(N^2_c-1)\zeta(d)}\right)^\frac{1}{d}\left[\frac{1}{1+l^{-\frac{(d-2)}{d}}}\right]^\frac{1}{d}\frac{1}{l^\frac{1}{d}}.\label{eq:COFL}
\end{align}


In order to compute 
Eq. (\ref{eq:SolRGEquation}) 
for the fermionic two-point function, 
we consider
	\begin{align}
	\frac{D_\mathsf{f}(l)}{z(l)} = \frac{2\eta_\Psi(l)+z(l)(d-1)-d}{z(l)} = -\frac{1}{z(l)}+\frac{2\eta_{\Psi}(l)+(d-1)(z(l)-1)}{z(l)},
	\end{align}
where the contribution from the dynamical critical exponent 
and that from the net anomalous scaling dimension of the fermion field are separated. 
The crossover function  in Eq. (\ref{eq:SolRGEquation}) is determined by
	\begin{align}
	I_{z}(l) &= \int\limits^{l}_{0}\frac{\dd\ell}{z(\ell)},\qquad \text{and} \qquad 	
I_\Psi(l) = \int\limits^{l}_{0}\dd\ell\left(\frac{2\eta_\Psi(\ell)+(d-1)(z(\ell)-1)}{z(\ell)}\right).
	\end{align} 

	\noindent  Since the critical exponents are controlled by $w(v)=v/c(v)\gg v$ it follows that, to the leading order in $v$,  the contribution from the dynamical critical exponent is dominated by the counter-term coefficient $Z_1$ and, consequently,
	\begin{align}
	\begin{split}\label{eq:Iz}
	I_{z}(l) &=\int\limits^{l}_{0}\dd\ell\hspace{0.5em}\left(1-\frac{(N^2_c-1)}{N_c N_f}\zeta(d) w(\ell)\right)= l - \mathfrak{F}_z(d)(N^2_c-1)^\frac{1}{d} \left[\frac{1}{1+l^{-\frac{(d-2)}{d}}}\right]^\frac{d-1}{d}\frac{l^\frac{1}{d}}{\textswab{S}\left(d-2;l^{-\frac{(d-1)}{d}}\right)},
	\end{split}
	\end{align}
	\noindent where $\textswab{S}(\xi;a)$ is defined in Eq. (\ref{eq:SFunction}), $\mathfrak{F}_z(d)$ is defined in Eq. (\ref{eq:Fzz}) and the last equality follows from taking 
the $l \rightarrow \infty$ limit. 
Similarly, the contribution from the net anomalous scaling dimension is dominated by $Z_3$ at low energies,
	\begin{align}
	\begin{split}\label{eq:IPsi0}
	I_{\Psi}(l)	&=\frac{(N^2_c-1)}{N_cN_f} \int\limits^{l}_{0}\dd\ell \left[\frac{2(d-1)}{\pi}\zeta(d)v(\ell)\textswab{S}(d-2;c(\ell))+\frac{2\beta^4_d}{N_cN_f}\mathfrak{F}(d)w(\ell)^2\textswab{S}\left(d-2;w(\ell)\right)^2\right].
	\end{split}
	\end{align}
From Eq. (\ref{eq:VOFLSimplified}) we use
	\begin{align}\label{eq:NiceProperty2}
	v(l)\textswab{S}\left(d-2;v(l)^{\frac{d-1}{d}}\right)\approx \frac{\pi N_cN_f}{4(d-1)(N^2_c-1)}\frac{1}{\zeta(d)}\left[\frac{1}{1+l^{-\frac{(d-2)}{d}}}\right]\frac{1}{l}
	\end{align}
to write
	\begin{align}\label{eq:IPsi}
	I_{\Psi}(l)&{=} \frac{1}{2}\left[\frac{1}{1+l^{-\frac{(d-2)}{d}}}\right]\left(\log l+\textswab{S}\left(d-2;l^{-\frac{1}{d}}\right) \right).
	\end{align}
in the large $l$ limit. In obtaining Eq. (\ref{eq:IPsi}) from Eq. (\ref{eq:IPsi0}) one has to use the expression for $v(l)$ without dropping the term depending on $v_0$ prior to the integration. Only after the integration is done, the terms depending on $v_0$ can be thrown away safely.
Since the fermion two-point function reduces to the bare one in the small $v$ limit,
the two-point function for nonzero $v$ is given by
	\begin{align}
\varGamma^{(2,0)}_{n}(\mathbf{K},\vec{k})=\varGamma^{(2)}_{\mathsf{f}}(\mathbf{K},\vec{k})={F_\Psi(|\mathbf{K}|)}{\left(iF_z(|\mathbf{K}|)\boldsymbol{\Gamma}\cdot\mathbf{K}+i\gamma_{d-1}\varepsilon_{n}(\vec{k};v_{|\mathbf{K}|} )\right)}
	\end{align}
for $e^{I_z( \log(\Lambda/|\mathbf{K}|) )}\vec{k}$ fixed.
Here, $v_{ |\mathbf{K}| } = v \left( \log(\Lambda/|\mathbf{K}|) \right)$. 
The universal functions,
		\begin{align}
	F_z(|\mathbf{K}|) &=\exp\left(\mathfrak{F}_z(d)(N^2_c-1)^\frac{1}{d} \left[\frac{1}{1+(\log(\Lambda/|\mathbf{K}|))^{-\frac{(d-2)}{d}}}\right]^\frac{d-1}{d}\frac{\log(\Lambda/|\mathbf{K}|)^\frac{1}{d}}{\textswab{S}\left(d-2;\log(\Lambda/|\mathbf{K}|)^{-\frac{(d-1)}{d}}\right)}\right),\label{eq:FzAppendix}\\
	F_{\Psi}(|\mathbf{K}|)&=\exp\left(\frac{1}{2}\left[\frac{1}{1+(\log(\Lambda/|\mathbf{K}|))^{-\frac{(d-2)}{d}}}\right]\left(\log \log(\Lambda/|\mathbf{K}|)+\textswab{S}\left(d-2;\log(\Lambda/|\mathbf{K}|)^{-\frac{1}{d}}\right) \right)\right),
	\end{align}
capture the deviations of the dynamical critical exponent and the anomalous scaling dimension of the fermion field from their values at the low-energy fixed point.\\

For the bosonic two-point function in Eq. (\ref{eq:SolRGEquation}),
we consider
	\begin{align}
	\frac{D_{\mathsf{b}}(l)}{z(l)} = \frac{2\eta_{\Phi}(l)+z(l)(d-1)-2(d-1)}{z(l)}  = -\frac{(d-1)}{z(l)}+\frac{2\eta_{\Phi}(l)+(z(l)-1)(d-1)}{z(l)},
	\end{align}
	\noindent where we have separated the contribution from the dynamical critical exponent and 
the net anomalous dimension of the bosonic field. 
The latter contribution to the two-point function is captured by
	\begin{align}
	I_{\Phi}(l)  = \int\limits^l_{0}\dd\ell\frac{(2\eta_{\Phi}(\ell)+(d-1)(z(\ell)-1))}{z(\ell)}.
	\end{align}
$I_\Phi(l)$ is dominated by the counter-terms $Z_6$ and $Z_1$ in the small $v$ limit,
and we can write
	\begin{align}
	\begin{split}
	I_\Phi(l)&= \frac{\mathfrak{F}_{\Phi}(d)}{(N^2_c-1)^\frac{d-1}{d}}\left[\frac{1}{1+l^{-\frac{(d-2)}{d}}}\right]^\frac{d-1}{d}l^\frac{1}{d}\left(1-\frac{(N^2_c-1)}{2\textswab{S}\left(d-2;l^{-\frac{(d-1)}{d}}\right)}\right),
	\end{split}
	\end{align}
	\noindent where $\mathfrak{F}_{\Phi}(d)$ is defined in Eq. (\ref{eq:Fphi}). 
Using Eqs. (\ref{eq:SolRGEquation}) and (\ref{eq:Iz}) and taking into account the fact that the bosonic 
two-point vertex function reduces to Eq. (\ref{eq:BosonPropagator}) for $v\ll 1$, 
we obtain
	\begin{align}
	\varGamma^{(0,2)}(\mathbf{Q},\vec{q}) =  \varGamma^{(2)}_{\mathsf{b}}(\mathbf{Q},\vec{q})={F_\Phi(|\mathbf{Q}|)}{ \left({F}_{z}(|\mathbf{Q}|)^{d-1}|\mathbf{Q}|^{d-1}+c_{|\mathbf{Q}|}^{d-1}(|q_x|^{d-1}+|q_y|^{d-1})\right)}
\end{align}
for $e^{I_z( \log(\Lambda/|\mathbf{Q}|) )}\vec{q}$ fixed.
Here 
\begin{align}
\begin{split}
F_{\Phi}(|\mathbf{Q}|) &=\exp\left[ \frac{\mathfrak{F}_{\Phi}(d)\log(\Lambda/|\mathbf{Q}|)^\frac{1}{d}}{(N^2_c-1)^\frac{d-1}{d}}\left[\frac{1}{1+(\log(\Lambda/|\mathbf{Q}|))^{-\frac{(d-2)}{d}}}\right]^\frac{d-1}{d}\left[1-\frac{(N^2_c-1)}{2\textswab{S}\left(d-2;\log(\Lambda/|\mathbf{Q}|)^{-\frac{(d-1)}{d}}\right)}\right]\right],
\end{split}
\end{align}
and $c_{|\mathbf{Q}|} =  c \left( \log(\Lambda/|\mathbf{Q}|) \right)$
capture the scale-dependent anomalous dimension and the velocity of the collective mode.

	\end{document}